\documentclass[11pt,a4paper]{article}

\pdfoutput=1 

\usepackage{jcappub} 

\usepackage[T1]{fontenc} 

\usepackage{amsmath,amssymb}
\usepackage{epsfig,graphicx}
\usepackage{subfigure}
\usepackage{graphicx}
\usepackage{soul}
\usepackage{hyperref}
\usepackage{rotating}
\usepackage{cancel}
\usepackage{bm}
\usepackage{xcolor}
\usepackage[font={small}]{caption}
\usepackage{comment}
\usepackage{cite}
\usepackage{psfrag}
\usepackage[numbers,sort&compress]{natbib}
\usepackage{physics}
\usepackage{siunitx}
\usepackage{multirow}

\newcommand{\eV}{\mathrm{\,eV}}
\newcommand{\GeV}{\mathrm{\,GeV}}

\newcommand{\ak}{\hat{a}_{\vec{k}}}
\newcommand{\akd}{\hat{a}_{\vec{k}}^{\dagger}}

\newcommand{\kv}{\vec{k}}
\newcommand{\kpv}{\vec{k'}}

\newcommand{\xv}{\vec{x}}
\newcommand{\xpv}{\vec{x'}}

\newcommand{\ttilde}{\tilde{t}}
\newcommand{\ktilde}{\tilde{k}}

\newcommand{\ttosc}{\tilde{t}_{\mathrm{osc}}}
\newcommand{\Vol}{\mathcal{V}}
\newcommand{\eq}{\mathrm{EQ}}
\newcommand{\Mpl}{M_{\mathrm{pl}}}

\newcommand{\ta}{\mathrm{ta}}
\newcommand{\ini}{\mathrm{in}}
\newcommand{\vir}{\mathrm{vir}}
\newcommand{\coll}{\mathrm{coll}}
\newcommand{\osc}{\mathrm{osc}}
\newcommand{\os}{\mathbb{OSC}}

\newcommand{\crit}{\mathrm{crit}}
\newcommand{\eff}{\mathrm{eff}}
\newcommand{\cl}{\mathrm{cl}}
\newcommand{\form}{\mathrm{form}}

\renewcommand{\eqref}[1]{eq.~(\ref{#1})}

\newcommand{\Eqref}[1]{Eq.~(\ref{#1})}

\newcommand{\secref}[1]{section~\ref{#1}}

\newcommand{\figref}[1]{figure~\ref{#1}}
\newcommand{\figsref}[1]{figures~\ref{#1}}

\newcommand{\Figref}[1]{Figure~\ref{#1}}

\newcommand{\citre}[1]{ref.~\cite{#1}}
\newcommand{\citres}[1]{refs.~\cite{#1}}
\newcommand{\Citre}[1]{Ref.~\cite{#1}}
\newcommand{\Citres}[1]{Refs.~\cite{#1}}

\newcommand{\beq}{\begin{equation}}
\newcommand{\eeq}{\end{equation}}

\definecolor{tab:blue}{HTML}{1f77b4}
\definecolor{tab:purple}{HTML}{9467bd}
\definecolor{tab:red}{HTML}{d62728}
\definecolor{tab:green}{HTML}{2ca02c}
\definecolor{goldenrod}{HTML}{daa520}

\title{ALP dark matter with non-periodic potentials: parametric resonance, halo formation and gravitational signatures}

\author[a]{Aleksandr Chatrchyan,}
\author[a,b]{Cem Eröncel,}
\author[a,c]{Matthias Koschnitzke,}
\author[a,c]{Géraldine Servant}

\affiliation[a]{Deutsches Elektronen-Synchrotron DESY, Notkestr. 85, 22607 Hamburg, Germany}
\affiliation[b]{Istanbul Technical University, Department of Physics, 34469 Maslak, Istanbul, Turkey}
\affiliation[c]{II. Institut für Theoretische Physik, Universit\"{a}t Hamburg, Luruper Chaussee 149, 22761 Hamburg, Germany}

\emailAdd{aleksandr.chatrchyan@desy.de}
\emailAdd{cem.eroncel@itu.edu.tr}
\emailAdd{matthias.koschnitzke@desy.de}
\emailAdd{geraldine.servant@desy.de}

\abstract{Axion-like particles (ALPs) are leading candidates to explain the dark matter in the universe. Their production via the misalignment mechanism has been extensively studied for cosine potentials characteristic of pseudo-Nambu-Goldstone bosons. In this work we investigate ALPs with non-periodic potentials, which allow for large misalignment of the field from the minimum. As a result, the ALP can match the relic density of dark matter in a large part of the parameter space. Such potentials give rise to self-interactions which can trigger an exponential growth of fluctuations in the ALP field via parametric resonance, leading to the fragmentation of the field. We study these effects with both Floquet analysis and lattice simulations. Using the Press-Schechter formalism, we predict the halo mass function and halo spectrum arising from ALP dark matter. These halos can be dense enough to produce observable gravitational effects such as astrometric lensing, diffraction of gravitational wave signals from black hole mergers, photometric microlensing of highly magnified stars, perturbations of stars in the galactic disk or stellar streams. These effects would provide a probe of dark matter even if it does not couple to the Standard Model. They would not be observable for halos predicted for standard cold dark matter and for ALP dark matter in the standard misalignment mechanism. We determine the relevant regions of parameter space in the (ALP mass, decay constant)-plane and compare predictions in different axion fragmentation models.
}

\begin{document}

\begin{flushright}
	\footnotesize
	DESY-23-060 \\
\end{flushright}
\color{black}

\maketitle
\flushbottom

\section{Introduction}\label{sec:introduction}
One prominent idea to explain dark matter is to introduce a light scalar particle called axion \cite{Weinberg:1977ma_axion2,Wilczek:1977pj}, which arises as the pseudo-Nambu-Goldstone boson \cite{PhysRev.117.648_Nambu,Goldstone:1961eq,PhysRev.127.965_broken_sym} of a new spontaneously broken global $\mathrm{U}(1)$-symmetry. The latter was first introduced as extension of the Standard Model by Peccei and Quinn \cite{Peccei:1977hh} not to solve the puzzle of dark matter, but the strong CP-problem in QCD. In this work, we will focus on a more general class of particles, usually referred to as \textit{axion-like particles (ALPs)}, which do not generally solve the strong CP-problem, but are still highly motivated, as they arise naturally in many Standard-Model extensions, in particular in string theory \cite{Svrcek:2006yi_string_axions3,Arvanitaki:2009fg_String_Axiverse,Demirtas:2018akl}.

While interactions between axions and other particles are assumed to be too weak for axions to get into thermal equilibrium with the rest of the early-universe plasma,  the most promising cosmological mechanism to produce axions or ALPs is the \textit{vacuum misalignment} (or \textit{vacuum realignment}) mechanism in which the  axion or the ALP field, modelled as classical scalar field due to its bosonic nature and high occupation numbers, has a non-zero initial field value and non-zero potential energy in the early universe, leading to oscillations of the field that let it act as (dark) matter component \cite{Preskill:1982cy_axion_DM_1,Abbott:1982af_axion_DM_2,Dine:1982ah_axion_DM_3}. 

In most of the models, axions and ALPs have a periodic potential, which arises due to non-perturbative instanton effects \cite{RevModPhys.53.43}. In this work we focus on an alternative type of potentials, which are non-periodic and allow larger displacements of the ALP field from the minimum. Such potentials can be generated from interactions with strongly coupled Yang-Mills gauge fields \cite{Witten:1980sp_Large_N_1,Witten:1998uka_Large_N_2,Nomura:2017zqj,Nomura:2017ehb_pure_natural_inflation}. Non-periodic potentials are also motivated by \textit{axion-monodromy} \cite{Silverstein:2008sg, McAllister:2008hb, Dong:2010in_Monodromy3}. Oscillations in such potentials can lead to a strong growth of fluctuations of the field due to \textit{parametric resonance} \cite{Amin:2014eta_floquet, Olle:2019kbo}. This process can lead to \textit{fragmentation} of the field in which all of the energy density is transferred to the fluctuations \cite{Fonseca:2019ypl_fragmentation}.

Recently, there has been a lot of interest in the study of parametric resonance in ALP models. \Citres{Greene:1998pb,Arvanitaki:2019rax,Zhang:2017dpp,Cedeno:2017sou,LinaresCedeno:2020dte,LinaresCedeno:2021sws} show that parametric resonance can be effective also for a periodic potential if the initial angle is very close to the top of the potential. Furthermore, \Citres{Fonseca:2019ypl_fragmentation,Eroncel:2022vjg} demonstrate that the parametric resonance is very efficient if the ALP field has a large initial kinetic energy as in the Kinetic Misalignment Mechanism \cite{Co:2019jts,Chang:2019tvx}. Similar effects have also been observed in the axiverse models where two ALP fields have similar masses \cite{Cyncynates:2021xzw,Cyncynates:2022wlq}, when the discrete symmetry of the ALPs is broken by a quadratic monomial \cite{Jaeckel:2016qjp, Berges:2019dgr,Chatrchyan:2020pzh}, when the ALP has an $\alpha$-attractor-type potential \cite{Kitajima:2018zco}, and also in models where the ALP potential is temperature-dependent, such as the QCD axion \cite{Sikivie:2021trt,Kitajima:2021inh}. Very recently, \Citre{Brandenberger:2023idg} claims that oscillations of the Hubble parameter induced by the ALP oscillations can also cause fragmentation of the low-wavelength ALP modes.

The growth of fluctuations has major consequences. Once gravitational interactions become important, the overdense regions, corresponding to the fluctuations, collapse to halos. This process can be described by the Press-Schechter formalism \cite{Press:1973iz} and its modifications. This is similar to formation of miniclusters in the post-inflationary scenario, studied in \citres{Hogan:1988mp,Kolb:1993zz_early_axion_stars_1,Kolb:1993hw_early_axion_stars2,Kolb:1994fi,Hardy:2016mns,Enander:2017ogx,Fairbairn:2017dmf,Fairbairn:2017sil,Bertschinger:1985pd,Zurek:2006sy,Ricotti:2009bs,Gosenca:2017ybi,Delos:2017thv,Vaquero:2018tib,Buschmann:2019icd,Eggemeier:2019khm,OHare:2021zrq,Niemeyer:2019aqm_virial_theorem_FDM,Xiao:2021nkb,Barman:2021rdr}. For the amplified scales, the halos can differ significantly from the generic prediction for cold dark matter (CDM) halos at the respective size and could lead to observable effects purely caused by gravitational interactions, similar to what was found in~\citre{Arvanitaki:2019rax} for an ALP field with a cosine potential, and in~\citre{Eroncel:2022efc} for an ALP field with initial kinetic energy.
\Citre{Arvanitaki:2019rax} also considered the transfer function for non-periodic potentials, however, only in a linear analysis. Our non-linear lattice analysis shows that, while the onset of oscillations is more delayed the larger the initial field value gets, the densest halos are only formed close to the critical field value, defined as the initial field value that separates the linear regime from the non-linear one. This detailed analysis allowed us to identify the parameter regions in the axion mass $m_a$, decay constant $f_a$-plane where observable signatures are expected.

\begin{figure}[t!]
    \centering
    \includegraphics[width=\textwidth]{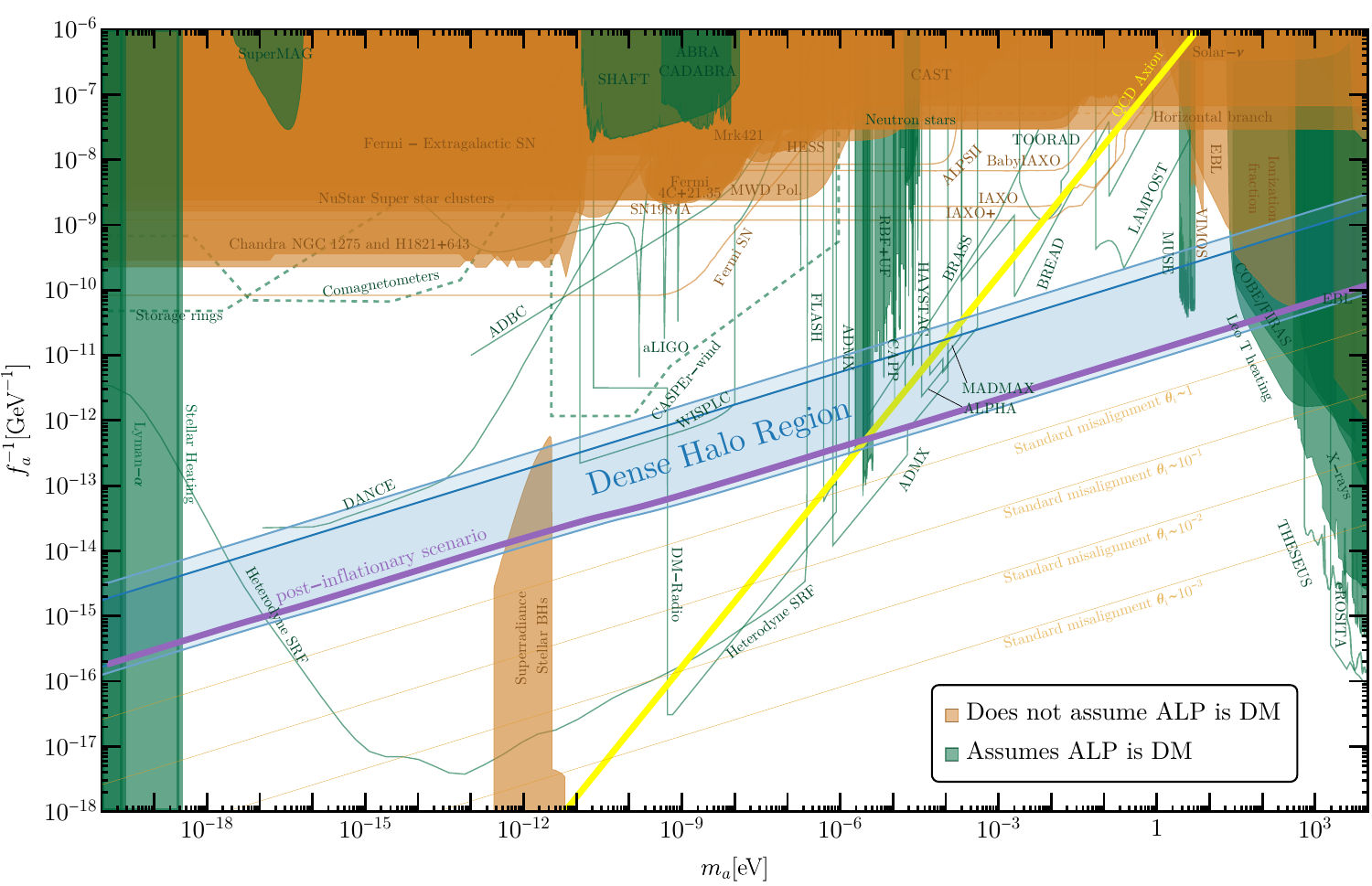}
    \caption{A sketch of the region in the ALP parameter space where dense halos are expected to form, together with all the experimental constraints and projections on ALPs assuming a coupling to the electromagnetic field with \eqref{eq:KSVZ_coupling}. We have obtained this region by combining the regions where dense halos are expected from the Kinetic Misalignment Mechanism and from ALPs with non-periodic potentials considered in this work. For caveats about this plot, see \secref{sec:comparison}. All the data for the constraints and projections are compiled from \citre{AxionLimits}.}
    \label{fig:dense-halo-region}
\end{figure}

The outline of this work is the following: The next section comprises a detailed investigation of the vacuum misalignment mechanism, where we focus on the evolution of fluctuations of the ALP field during this process. Using the semi-analytic Floquet analysis we explain why parametric resonance is inefficient in the case of the cosine potential, which serves as a motivation for our studies of non-periodic potentials. In \secref{sec:the_spectra} we compute the energy density power spectrum of the ALP field, resulting from parametric resonance and fragmentation, using both linear evolution of the modes, as well as fully non-linear lattice simulations. In \secref{sec:minihalos} we take the results of the preceding chapter and examine how the fluctuations decouple from the Hubble flow and evolve to virialised halos after gravitational interaction becomes important. The main result, to be discussed in \secref{sec:obs_prospects}, is the identification of the parts of the parameter space of ALP mass $m_a$ and symmetry breaking scale $f_a$ in which we predict halos that have observable gravitational effects, assuming that the ALPs make up all of dark matter. Finally, we compare this scenario with others that lead to fragmentation of the ALP field in \secref{sec:comparison}. There, we estimate that all the scenarios we have considered predict dense halos in a roughly similar region in the $(m_a,f_a)$-parameter space which we show in \figref{fig:dense-halo-region} with a blue band labelled as "Dense Halo Region". In this plot, we also show the experimental constraints and projections assuming that the ALP field has a KSVZ-like coupling \cite{Kim:1979if_KSVZ_1,Shifman:1979if_KSVZ_2} to the photon and the neutron. For simplicity, we consider ALPs with constant mass. The different predictions for axion fragmentation due to a temperature-dependent ALP mass were discussed in \citre{Eroncel:2022vjg}.

Throughout this work, unless otherwise noted, we work in units where $c=\hbar=1$, use the mostly minus convention for $g_{\mu\nu}$, the reduced Planck mass $\Mpl=\sqrt{1/(8\pi G)}$ and $H_0=h 100\,\mathrm{km}\,\mathrm{s}^{-1}\,\mathrm{Mpc}^{-1}$ with $h=0.68$.

\section{From misalignment mechanism to parametric resonance}

In this section, we investigate the significance of parametric resonance for ALP dark matter produced from the vacuum misalignment mechanism. Both periodic, as well as non-periodic ALP potentials are considered. 

After a pedagogical review of the essentials of the mechanism, as well as estimations of the relic density in \secref{ssec:essentials_of_mm}, we focus on the dynamics of ALP fluctuations in the subsequent subsections. The corresponding equations of motion are presented in \secref{ssec:fluctuations}. We then discuss the growth of fluctuations in a periodic, ``cosine'' potential, and in a non-periodic, ``power-law'' potential in sections~\ref{ssec:resonance_periodic} and \ref{ssec:resonance_non-periodic}, respectively, reviewing and using the well-established method of Floquet analysis. 
We demonstrate why in the first case Hubble friction prevents the fluctuation modes from experiencing an amplification unless the misalignment field value is very close to the top of the potential. Then, we show how this constraint is avoided in the non-periodic case.

\subsection{ALP dark matter from the misalignment mechanism}
\label{ssec:essentials_of_mm}

Ignoring interactions with all other particles and assuming a minimal coupling to gravity, the action of an ALP field is given by
\begin{equation}\label{action_ALP_general}
S_{\phi}=\int d^4x\sqrt{-g}\left[\frac{1}{2}\partial_\mu \phi g^{\mu\nu}\partial_\nu\phi-V(\phi)\right]\, ,
\end{equation}
where $g$ is the determinant of the FLRW metric $g_{\mu\nu}$. 

We assume that the ALPs were present during inflation. In this case, similar to the conventional misalignment mechanism for the QCD axion, inflation leaves the ALP field with an almost homogeneous value inside our Hubble patch, which is essentially frozen before $H\sim m_a$. 
We denote this initial misalignment value by $\phi_{i}$. 
Throughout this work, we assume $m_a=\,$cst and $\dot{\phi}_i=0$.
The field obeys the classical equation of motion (EOM):
\begin{equation}\label{full_eom_axion_unperturbed}
\ddot{\phi}+3H\dot{\phi}-\frac{1}{a^2}\nabla^2\phi+V'(\phi)=0\,,
\end{equation}
where we denoted $\partial V/\partial\phi$ as $V'(\phi)$. In the radiation-dominated era, once the Hubble friction term becomes subdominant, the field rolls to the minimum of the potential and oscillates around it at later times. Near the minimum of the potential, where $V \approx \frac{1}{2}m_a^2\phi^2$, the equation of state of such an oscillating ALP field averages to $w=0$ so that it behaves as a (dark) matter component.

Assuming that oscillation starts in the radiation era, and $m_a=\,$cst, a general expression for the relic energy density of ALPs today is given by 
\begin{equation}
\boxed{
\label{eq:relic_density_general}
\Omega_{a,0}=\frac{1}{3}(\Omega_{r,0})^{3/4}\frac{g_s(T_0)}{g_s(T_\osc)}\left(\frac{g_\rho(T_\osc)}{g_\rho(T_0)}\right)^{3/4}\left(\frac{1}{H_\osc}\right)^{3/2}\left(\frac{1}{H_0}\right)^{1/2}\left(\frac{1}{\Mpl}\right)^{2}V(\phi_i)\mathcal{Z}, }  
\end{equation}
where $H_{\osc}$ is the Hubble scale at the onset of oscillation, $g_s$ and $g_\rho$ are the effective degrees of freedom in entropy and energy, respectively, while $\Omega_{r,0}$ is today's density parameter of radiation. For this estimate one uses that the energy density scales approximately as $\rho_a=\rho_{a,\,\osc}\left({a_{\osc}}/{a}\right)^3$ for $a>a_{\osc}$, where $\rho_{a,\,\osc} = V(\phi_i)$ is the energy density of the field before the onset of oscillation. $\mathcal{Z}$ incorporates corrections to this estimate. For a harmonic potential, or close to the minimum of a general potential, $\mathcal{Z}$ can be found to be
\begin{equation}\label{eq:z1-analytical}
    \mathcal{Z}=\frac{8}{\pi}\qty(\Gamma(5/4))^2\simeq 2.1,
\end{equation}
when setting $H_\osc=m_a$.
\Eqref{eq:relic_density_general} can differ strongly from the harmonic estimate due to anharmonic effects, as it will be discussed in the next subsections. 

\subsubsection{Relic density for a periodic potential}\label{sec:relic_density_LMM}

In this section we consider the usual periodic axion potential
\begin{equation}\label{eq:cosine_potential}
V(\phi)=m_a^2f_a^2[1-\cos(\phi/f_a)]\,,
\end{equation}
where $f_a$ is the symmetry breaking scale, also referred to as the decay constant. For small initial field values, the evolution of the field in this potential defines the well-studied 'standard misalignment mechanism'. For small initial field values, $\phi_i/f_a\lesssim1$, the potential looks almost harmonic and the relic density can be found from \eqref{eq:relic_density_general} and \eqref{eq:z1-analytical}. Since the axion mass $m_a$ and the initial field value determine when the axion field begins to oscillate and decay, in a scenario with small initial field value, the axion can only match the relic density for large decay constants $f_a$. To explain dark matter from the axion-like-particle in a larger part of the parameter space, i.e. also in the region of small $f_a$, where dark matter from an ALP with small initial field value would be underproduced, the onset of oscillations of the homogeneous mode (or zero mode) has to be delayed, such that more of the maximal energy budget of $m_a^2f_a^2$ can be used.

\begin{figure}[t!]
    \centering
    \includegraphics[width=0.8\textwidth]{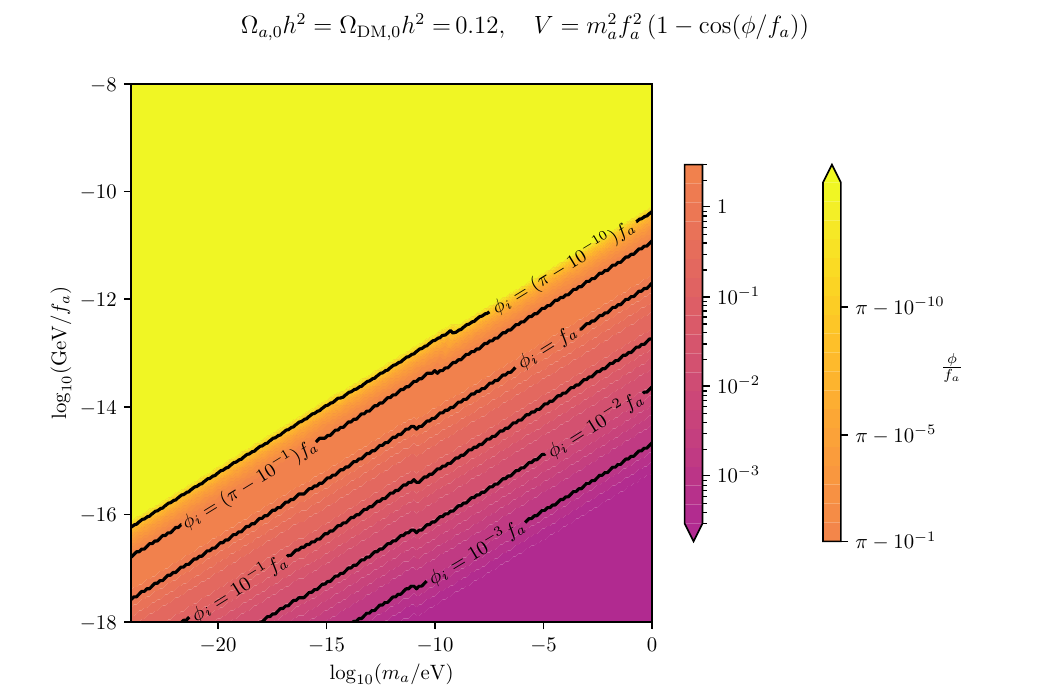}
    \caption{The required initial field value $\phi_i/f_a$ to match the dark matter relic density at each point in the $(m_a,f_a)$-plane for ALPs with a cosine potential. The relic density is given by \eqref{eq:relic_density_general} with the cosine potential and $\mathcal{Z}=2.1$  for small initial field values and \eqref{eq:relic_density_cosine} for initial field values close to the top.}
    \label{fig:cosine_relic_plot}
\end{figure}

In a model with the cosine potential, the onset of oscillations is delayed if the axion field starts close to top of the potential. This mechanism is called \em Large Misalignment Mechanism \em or \em extreme axion \em and was described in \citre{Arvanitaki:2019rax} and \cite{Zhang:2017dpp}. Examples on how such an initial condition could be realised can be found in \citres{Co:2018mho,Takahashi:2019pqf}. Defining $$\theta=\phi/f_a,$$
\Citre{Arvanitaki:2019rax} found as empirical expression for the onset of oscillation for values close to the top with $\left|\pi-\theta_i\right|<10^{-2}$, 
\begin{equation}
    t_{\osc}=\frac{1}{m_a}\ln\left(\frac{1}{\pi-\left|\theta_i\right|}\frac{2^{1/4}\pi^{1/2}}{\Gamma(5/4)}\right),
\end{equation}
which we confirmed numerically for tunings up to $\abs{\pi-\theta_i}=10^{-12}$.
Using $H_{\osc}=1/(2t_{\osc})$, we can plug this into \eqref{eq:relic_density_general} and find for the relic density:
\begin{equation}
\boxed{
\label{eq:relic_density_cosine}
\begin{split}
    \Omega_{a,0}=&~\frac{2^{3/2}}{3}(\Omega_{r,0})^{3/4}\frac{g_s(T_0)}{g_s(T_\osc)}\left(\frac{g_\rho(T_\osc)}{g_\rho(T_0)}\right)^{3/4}\left(\frac{m_a}{H_0}\right)^{1/2}\left(\frac{f_a}{\Mpl}\right)^{2}\\
    &\times\left[\ln\left(\frac{1}{\pi-\left|\theta_i\right|}\frac{2^{1/4}\pi^{1/2}}{\Gamma(5/4)}\right)\right]^{3/2}\left[1-\cos(\theta_i)\right] .
\end{split}
}
\end{equation}
We show the parameter region where an axion-like particle with a cosine potential, initially at rest, can make up all of the DM, $\Omega_{a, 0}h^2 = \Omega_{\mathrm{DM, 0}}h^2 = 0.12$, in \figref{fig:cosine_relic_plot}.

\subsubsection{Relic density for a non-periodic potential}\label{sec:relic_density_monodromy} 

\begin{figure}[t!]
    \centering
    \includegraphics[width=0.6\textwidth]{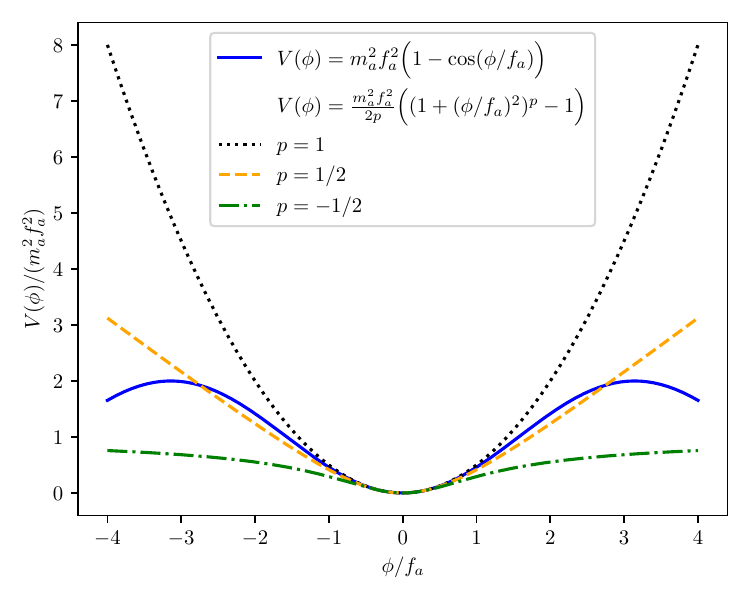}
    \caption{Comparison of the cosine potential (blue solid line) with the non-periodic potential \eqref{eq:monodromy_potential} for different $p$ values. $p=1$ corresponds to a harmonic potential (black dotted line), while the potential is flatter for $p=1/2$ (orange dashed line) and for negative values for $p$, such as $p=-1/2$ (green dashed-dotted line), approaches a plateau for large values of $\phi/f_a$.}
    \label{fig:comparison_potential}
\end{figure}

Following \citres{Olle:2019kbo,Nomura:2017ehb_pure_natural_inflation} we take as a generic non-periodic potential for the ALP field:
\begin{equation}
\boxed{
\label{eq:monodromy_potential}
V(\phi)=\frac{m_a^2f_a^2}{2p}\left[\left(1+\frac{\phi^2}{f_a^2}\right)^p-1\right]. } 
\end{equation}
As in the previous case, $m_a$ is the mass near the minimum, while the decay constant $f_a$ is supposed to make a link to the UV completion of the ALP. The parameter $p$ can be adjusted to incorporate different models. For $p=1$ we recover the generic quadratic potential. The values $0<p<1$ lead to axion-monodromy potentials \cite{Silverstein:2008sg, McAllister:2008hb, Dong:2010in_Monodromy3}, while the values $p<0$ represent potentials with a plateau at large field values~\cite{Kitajima:2018zco}. The latter can arise when the axion is coupled
to pure strongly coupled Yang-Mills gauge fields, which has been shown by considering a $SU(N)$ gauge theory in the large $N$ limit \cite{Witten:1980sp_Large_N_1,Witten:1998uka_Large_N_2,Nomura:2017zqj,Nomura:2017ehb_pure_natural_inflation}. For more details on possible UV-completions see appendix~\ref{App.UV}.
A comparison of the potentials for different $p$-values is shown in \figref{fig:comparison_potential}.

A good estimate for the beginning of oscillations is given by \cite{Kitajima:2018zco}
\begin{equation}\label{H_oscillation}
H_{\osc}\cong\sqrt{\Bigl|\frac{V'(\phi_i)}{\phi_i}\Bigr|}\,.
\end{equation}
The right hand side reduces to $m_a$ in the case of a quadratic potential, but is smaller for $p<1$, meaning that the onset of oscillations is delayed in this case.

Considering axion masses such that oscillation starts in radiation era, hence starting from \eqref{eq:relic_density_general} and using \eqref{H_oscillation}, we find:
\begin{equation}
\boxed{
\label{eq:relic_density_monodromy}
\Omega_{a,0}=\frac{1}{6p}(\Omega_{r,0})^{3/4}\frac{g_s(T_0)}{g_s(T_\osc)}\left(\frac{g_\rho(T_\osc)}{g_\rho(T_0)}\right)^{3/4}\left(\frac{m_a}{H_0}\right)^{1/2}\left(\frac{f_a}{\Mpl}\right)^{2}\frac{(1+\theta_i^2)^p-1}{(1+\theta_i^2)^{3(p-1)/4}}\mathcal{Z}_p(\theta). } 
\end{equation}
Here $\mathcal{Z}_p(\theta)$ contains the corrections due to the anharmonicity of the potential, which leads to a delayed relaxation of the average equation of state to $w=0$. As a fit function to numerical results we used $\mathcal{Z}_p(\theta)=\mathcal{Z}_1[1 + \beta_p((1+\theta^2)^{\alpha_p}-1)]$ where we found $\alpha_{1/2}=0.25$, $\beta_{1/2}=1.33/\mathcal{Z}_1$, $\alpha_{-1/2}=0.26$, $\beta_{-1/2}=4.18/\mathcal{Z}_1$, and $\mathcal{Z}_1$ is given by~\eqref{eq:z1-analytical}.  The delayed onset of oscillations is accounted for by using \eqref{H_oscillation}. 

\begin{figure}[t!]
	\centering
	\includegraphics[width=\textwidth]{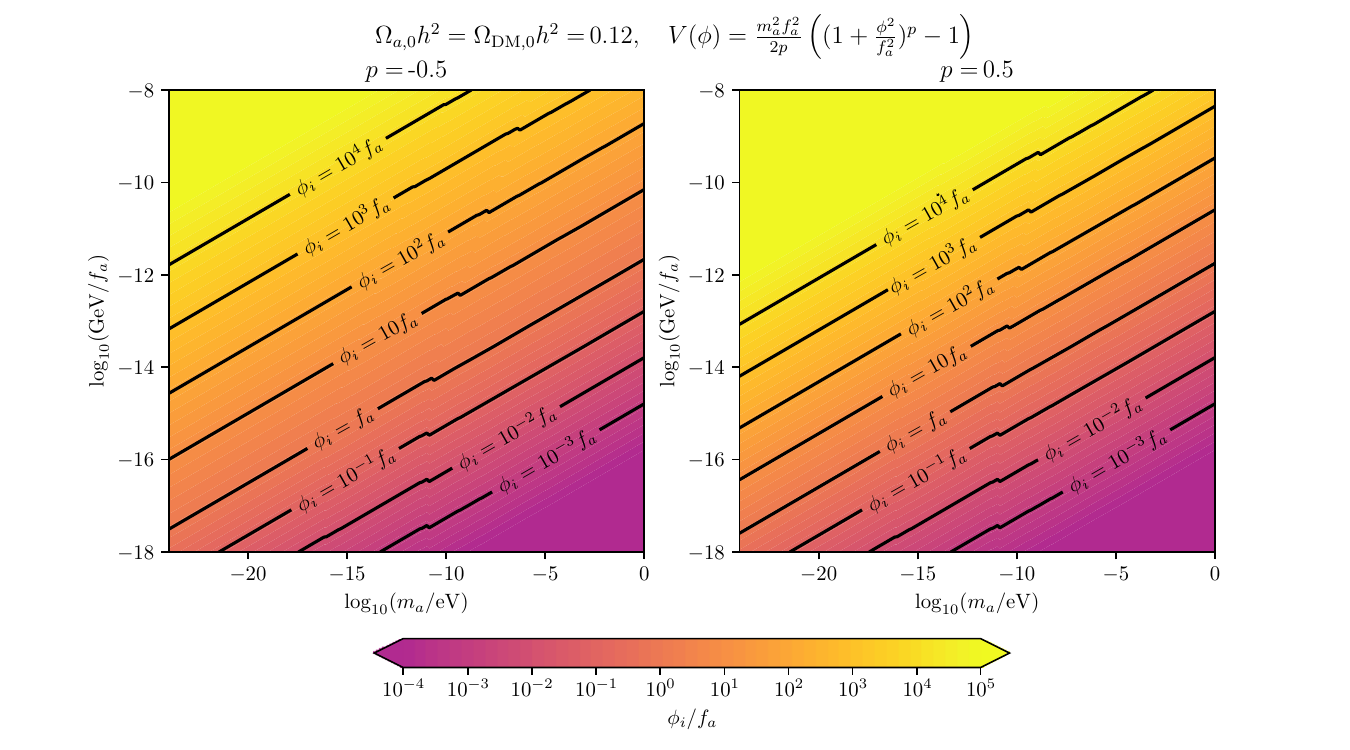}
	\caption{The required initial field value $\phi_i/f_a$ to match the dark matter relic density at each point in the $(m_a,f_a)$-plane according to \eqref{eq:relic_density_monodromy} for ALPs with the non-periodic potential. We set $p=-1/2$ and $p=1/2$ in the left and the right panels, respectively. Compare with \figref{fig:cosine_relic_plot}, which shows the same parameter space for ALPs with the cosine potential.}
	\label{fig:monodromy_relic_plots}
\end{figure}

For $p\neq1$ and large $\theta_i$, this can significantly change the $(m_a,f_a)$-parameter space for ALP DM with respect to the case of periodic potentials. 
We find the initial field values to match the ALP relic density to the one of dark matter in the $(m_a,f_a)$-plane and plot the result for different values of $p$ in \figref{fig:monodromy_relic_plots}. In contrast to the cosine potential, large initial field values make it possible to cover also the parameter space of smaller $m_a$ and $f_a$.

As it will be discussed in the next sections, when the field undergoes fragmentation, a significant fraction of its energy is transferred from the homogeneous mode into initially relativistic fluctuations. This effects leads to an additional correction to the prefactor $\mathcal{Z}$ since the equation of state parameter shifts for some time towards $w=1/3$, as one would expect for a gas of relativistic particles. As we clarify in section~\ref{ssec:obs_regions_parameter_space}, in the interesting part of the parameter region this correction is of $\mathcal{O}(1)$ \cite{Berges:2019dgr,Eroncel:2022vjg}. Therefore, for simplicity, we do not include this correction in our calculations.

\subsection{Fluctuations of the field}
\label{ssec:fluctuations}

It is convenient to separate the field into its homogeneous mode and fluctuations, 
\begin{equation}\label{field_with_fluctuations}
\hat \phi(t,\mathbf{x})=\phi(t)+\hat{\delta\phi}(t,\mathbf{x})\,.
\end{equation}
The fluctuation field can be expressed in terms of time dependent mode functions $u_{\kv}(t)$ for the individual Fourier modes~\cite{Fonseca:2019ypl_fragmentation}:
\begin{equation}\label{quantum_fluctuations_fourier}
\hat{\delta\phi}(t,\xv)=\int\frac{d^3k}{(2\pi)^3}\left[\ak u_{\kv}(t)e^{i\vec{k}\cdot\vec{x}}+\akd u_{\kv}^*(t)e^{-i\vec{k}\cdot\vec{x}}\right]\,,
\end{equation}
where the creation and annihilation operators $\ak$ and $\akd$ obey the usual commutation relations of scalar fields~\cite{Fonseca:2019ypl_fragmentation}. Since absolute direction of the momenta is not important when assuming homogeneity and isotropy of the ALP field, we are only interested in absolute values of $\kv$ and will from now on mostly use $k\equiv\lvert{\kv}\rvert$, especially for indices.

Due to the small mass of the considered ALPs, the fluctuations are usually characterised by large occupation numbers. As a consequence, these fluctuations can be treated as classical(-statistical) ones. Having this in mind, in the following, we will drop the hats on $\delta\phi$.

To separate equations of motions for the homogeneous mode and the small fluctuations, $\phi(t)\gg\delta\phi(t,\xv)$, we expand the potential with respect to the fluctuations and write
\begin{equation}\label{expansion of potential}
V(\phi+\delta\phi)=V(\phi)+V'(\phi)\delta\phi+\frac{1}{2}V''(\phi)\delta\phi^2+\mathcal{O}(\delta\phi^3)\,.
\end{equation}
Keeping only terms up to $\mathcal{O}(\delta\phi)$ in the potential and using~\eqref{quantum_fluctuations_fourier}, the result is an equation that decomposes into an integral part including fluctuations and a non-integral part only containing the homogeneous mode. This leads to the general EOMs for the homogeneous mode as well as the mode functions,
\begin{gather}\label{eom_zero_general}
\ddot{\phi}(t)+3H\dot{\phi}(t)+V'(\phi)=0\,,\\
\ddot{u}_k(t)+3H\dot{u}_k(t)+\left(\frac{\vec{k}^2}{a^2}+V''(\phi)\right)u_k(t)=0\,.\label{eom_fluctuations_general}
\end{gather}

Similarly, the total energy density separates into a part depending on the homogeneous mode and one part that depends on the fluctuations,
\begin{equation}\label{energy_density_general_allmodes}
\langle {\rho} \rangle =\frac{1}{2}\dot\phi(t)^2+V(\phi)\\
+\frac{1}{2}\int\frac{d^3k}{(2\pi)^3}\left[\lvert \dot{u}_k(t)\rvert^2+\left(\frac{\vec{k}^2}{a^2}+V''(\phi)\right)\lvert u_k(t)\rvert^2\right]\,,
\end{equation}

The evolution of the mode functions is trivial in the case of a quadratic potential, where $V''(\phi)$ is simply the mass squared. On the other hand, the anharmonicities of the potential can introduce instabilities for the fluctuations, which are discussed next.

\subsection{Growth of fluctuations in a periodic potential}
\label{ssec:resonance_periodic}

In this section we want to find out when the exponential growth of fluctuations is possible for the case of the cosine potential from \eqref{eq:cosine_potential}.

To be able to work with the EOMs numerically and to make the equations independent of the axion mass, we introduce a dimensionless time- and a dimensionless momentum-variable (as in \citre{Arvanitaki:2019rax,Eroncel:2022vjg}):
\begin{equation}
\label{eq_define_tilde}
\ttilde=m_at\,,\quad\quad\quad\tilde{k}^2=\frac{k^2}{2m_aa^2H}\,.
\end{equation}
From now on, the dot above a function will denote derivation with respect to the time variable given in the argument. During radiation era we have $a\sim t^{1/2}$ and $H=\frac{1}{2t}$, therefore $a^2H$ stays constant during this epoch. From these definitions we can see that $\tilde{k}$ corresponds to the ratio of the physical momentum to the axion mass at $\ttilde=1$, i.e.~how relativistic this mode is at this time, which is roughly when the homogeneous mode begins to oscillate. 

We can reformulate the equations of motion in terms of the dimensionless quantities, valid during radiation era:
\begin{gather}\label{eom_zero_dimless_cosine_rad_era}
\ddot{\theta}(\ttilde)+\frac{3}{2\ttilde}\dot{\theta}(\ttilde)+\sin(\theta(\ttilde))=0\,,\\
\ddot{u}_{\ktilde}(\ttilde)+\frac{3}{2\ttilde}\dot{u}_{\ktilde}(\ttilde)+\left(\frac{\ktilde^2}{\ttilde}+\cos(\theta(\ttilde))\right)u_{\ktilde}(\ttilde)=0\,.\label{eom_fluc_dimless_cosine_rad_era}
\end{gather}
Here we have used that $H=m_a/(2\ttilde)$ and $k^2/a^2=\ktilde^2m_a^2/\ttilde$ in radiation era. 

\subsubsection{Floquet analysis}\label{sec:floquet_cosine}
Let us ignore expansion for a moment, i.e.~set $H=0$ and $a=1$. This implies that the momenta do not redshift and that the corresponding term in the EOM is $k^2$ instead of $k^2/a^2$. One then arrives at the following equations for the homogeneous mode and the fluctuations:
\begin{gather}
\ddot{\theta}(t)+m^2\sin(\theta)=0\,,\\
\ddot{u}_k(t)+\left[k^2+m^2\cos(\theta)\right]u_k(t)=0\,.
\end{gather}
Since $\theta$ is periodic, this equation is a second-order differential equation with periodic coefficients, so it has the form of the Hill's equation~\cite{Hill1966,10.1007/BF02417081_Hills_equation}. According to the Floquet theorem \cite{ASENS_1883_2_12__47_0_Floquet}, the solutions are of the form
\begin{equation}
    u_k(t)=u_+(k;t)e^{\mu_k t}+u_-(k;t) e^{-\mu_k t},
\end{equation}
where $u_{\pm}$'s are periodic functions in time, and $\mu_k$'s are in general complex coefficients. A necessary, but not sufficient, condition for parametric resonance is that the Floquet exponent has non-zero real part $\Re{\mu_k}>0$. So determining whether parametric resonance occurs amounts to calculating the Floquet exponents. For more details on Floquet theory see appendix~\ref{App.1}. 

\begin{figure}[t]
    \centering
    \includegraphics[width=0.65\textwidth]{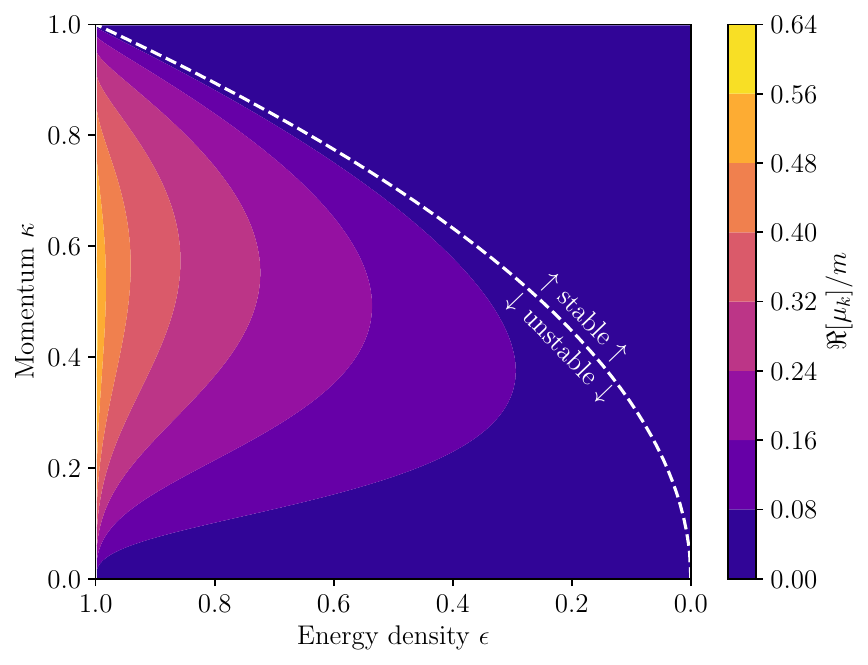}
    \caption{The contour of the real part of the Floquet exponents for the cosine potential calculated via~\eqref{floquet}. Larger values of $\mathfrak{R}[\mu_k]/m$ (warmer colours) indicate stronger parametric resonance. The dashed line separates the stable and the unstable regions as determined by~\eqref{instability-band}.}
    \label{fig:cosine-instability-bands}
\end{figure}

For the cosine potential, these exponents are calculated in~\citre{Greene:1998pb}:
\begin{equation}
\label{floquet}
    \mu_k=\frac{m \mathcal{I}}{\mathrm{K}(\sqrt{\epsilon})}\sqrt{2\kappa^2\qty(\epsilon-\kappa^2)\qty(1-\epsilon+\kappa^2)}
\end{equation}
where $$\kappa\equiv k/m. $$ 
$\mathrm{K}$ is the complete elliptic integral of the first kind, and $\mathcal{I}$ denotes the following integral:
\begin{equation}
    \mathcal{I}=\int_0^{\pi/2}\frac{\dd{\vartheta}}{1+\qty(1-2\epsilon+2\kappa^2)\sin^2\vartheta}\frac{2\sin^2\vartheta}{\sqrt{\qty(1+\sin^2\vartheta)\qty[1+(1-2\epsilon)\sin^2\vartheta]}}.
\end{equation}
In these expressions $\epsilon$ denotes the ratio of the homogeneous mode energy density to the height of the potential barrier $2m_a^2 f_a^2$, so it satisfies $0<\epsilon<1$. From the Floquet exponent~\eqref{floquet} we can immediately read off the condition for $\Re{\mu_k}>0$ as
\begin{equation}
    \label{instability-band}
    0<\qty(\frac{k}{m})^2<\epsilon.
\end{equation}
The modes for which the above condition is true are referred to as the modes inside the instability band. We see that the width of the instability band depends on the energy in the homogeneous mode. If $\epsilon$ is large, the homogeneous mode oscillates with a larger amplitude which allows the ALP field to probe the non-quadratic parts of the potential. This enhances the parametric resonance effect as we can see in \figref{fig:cosine-instability-bands}, where we show a plot of the Floquet exponents for the periodic potential without the Hubble expansion. We see that a large $\epsilon$ not only increases the width of the instability bands, but also renders the parametric resonance more efficient.

In the case of non-zero $H$, the evolution of the mode functions can be approximated by~\cite{Eroncel:2022vjg}
\begin{equation}
    u_k(t)\sim u_k(t_i)A_k(t)N_k(t),
\end{equation}
where $u_k(t_i)$ is the initial condition before parametric resonance becomes effective, $A_k(t)$ encompasses the decay of the mode amplitude due to the redshift which can be estimated using the WKB approximation as
\begin{equation}
    A_k(t)\propto \omega_k^{1/2}(t)a^{-3/2}(t)\qq{where}\omega_k(t)=\sqrt{\frac{k^2}{a^2(t)}+m_a^2},
\end{equation}
and $N_k(t)$ is the total growth of the mode due to parametric resonance which is obtained by integrating the Floquet exponent over time:
\begin{equation}
\label{eq:total-growth}
    N_k(t)=\exp\qty(\int_{t_i}^t\dd t' \mu_k\qty(\kappa(t'),\epsilon(t'))),\quad \kappa(t)=\frac{k}{m_a a(t)}.
\end{equation}
In terms of dimensionless variables
\begin{equation}
    \kappa_{\rm osc}\equiv \frac{k}{m_a a_{\rm osc}},\quad \tau\equiv 2 H_{\rm osc}t,\quad \tilde{u}_k\equiv \frac{\mu_k}{m_a},
\end{equation}
\eqref{eq:total-growth} reads
\begin{equation}
\boxed{
    N_k(\tau)=\exp\qty(\frac{m_a}{2 H_{\rm osc}}\int_{\tau_i}^\tau d\tau'\tilde{\mu}_{\kappa}\qty(\kappa(\tau'),\epsilon(\tau')))\equiv \exp\qty(\frac{m_a}{2 H_{\rm osc}}\mathcal{B}_{\kappa}(\tau)).} 
\end{equation}
By calculating $\mathcal{B}_{\kappa}$ numerically, we find that it becomes at most $\mathcal{O}(0.5)$. Therefore, the efficiency of the fragmentation is determined primarily by the hierarchy between the ALP mass and the Hubble scale at the onset of oscillations. This estimate shows that in the standard misalignment mechanism with a small initial field value $\phi_i\lesssim f_a$, the parametric resonance is not effective. However, in the Large Misalignment Mechanism, the tiny potential gradient at the top of the cosine potential delays the onset of oscillations, see \secref{sec:relic_density_LMM}, so that a hierarchy between $m_a$ and $H_{\rm osc}$ renders the parametric resonance effective. Such a delay also occurs in the Kinetic Misalignment Mechanism, where the ALP field has a large initial kinetic energy~\cite{Co:2019jts,Chang:2019tvx} which yields very efficient parametric resonance~\cite{Eroncel:2022vjg}.

\subsubsection{Tachyonic instability}\label{sec:tachyonic_instability} 
Let us again briefly ignore the Hubble expansion, but also assume that the homogeneous mode is still frozen at the initial field value $\theta_i$, although this is not actually possible when $H=0$. If the initial field value would be close to $-\pi$, the second derivative of the potential, $\cos(\theta_i)$, gets close to $-1$. A negative value of the second derivative implies a tachyonic instability \cite{Felder:2001kt_Tachyonic_instability}, which leads to growth of fluctuations, i.e.~particle production. This can be seen by assuming $\cos(\theta_i)=-1$, with which \eqref{eom_fluc_dimless_cosine_rad_era} becomes
\begin{equation}
\ddot{u}_k(\ttilde)+\left(\tilde{k}^2-1\right)u_k(\ttilde)=0\,.
\end{equation}
The solution to this is
\begin{equation}
u_k(\ttilde)=u_{k}(\ttilde_0)e^{\pm i\sqrt{\tilde{k}^2-1}(\ttilde-\ttilde_0)}\,.
\end{equation}
For $\tilde{k}>1$, i.e.~for relativistic modes, this results in oscillation of the fluctuation. This stays true even if the second derivative of the potential would be closer to zero, i.e.~if the initial field value would be smaller. For $\tilde{k}<1$, the term in the root is negative, which implies that the mode grows exponentially.

It is obvious that this analysis does not give a realistic result for the expanding universe. To find out if this exponential growth could have cosmological implications, we have to compare this (dimensionless) imaginary frequency $\sqrt{\tilde{k}^2-1}$ with the dimensionless Hubble-factor $H/m_a$, and check if the latter is large compared to the former:
\begin{equation}
\frac{\sqrt{\tilde{k}^2-1}\,m_a}{H(\ttilde)}=\sqrt{\tilde{k}^2-1}\cdot{2\ttilde}=\mathcal{O}(1)\cdot\ttilde\stackrel{!}{>}1\,.
\end{equation}
This shows that the tachyonic instability can only be cosmologically relevant if $\ttilde>1$. We saw that $\ttilde=1$ is approximately the dimensionless time at which the homogeneous mode begins to oscillate in the standard misalignment mechanism. This means, $\cos(\theta)$ does not stay negative and exponential growth of the fluctuations will only be possible under certain conditions, which we will investigate in the next subsection.

\begin{figure}[t!]
	\centering
	\includegraphics[width=0.6\textwidth]{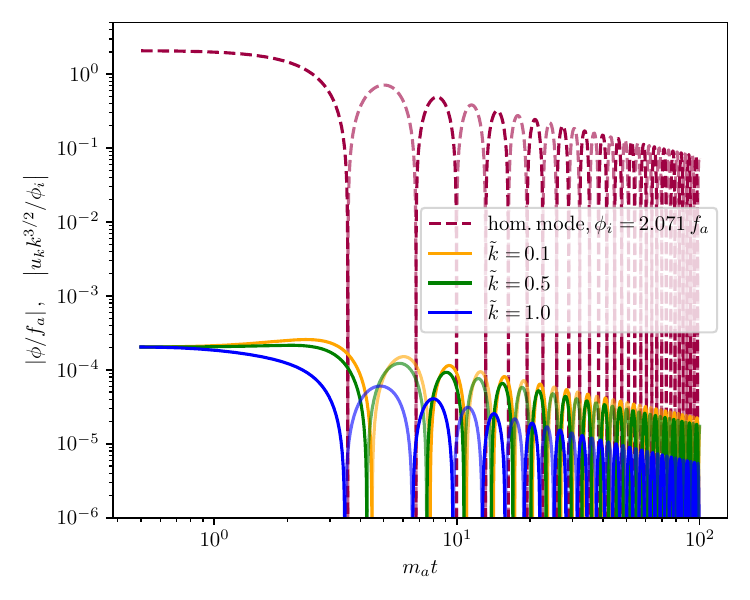}
	\caption{Numerical solutions to \eqref{eom_zero_dimless_cosine_rad_era} (homogeneous mode) with $\theta_i=\pi/2+0.5$, for which the second derivative of the potential is negative, and to \eqref{eom_fluc_dimless_cosine_rad_era} (fluctuations) for different values of $\tilde{k}$. 
 Naively one would expect exponential growth for $\ktilde=0.1$ and $0.5$ due to the tachyonic instability, but this would only apply if the homogeneous mode would stay frozen. Since it starts to oscillate before $\ttilde=1$, we see no growth of fluctuations. After the onset of oscillation all modes decay due to Hubble friction.}
	\label{fig:Fluctuations_multiple_plot1}
\end{figure}

To summarise, when the homogeneous mode oscillates harmonically near the minimum and is damped due to the expansion, no exponential growth of fluctuations due to the tachyonic instability is possible. \Figref{fig:Fluctuations_multiple_plot1} shows a numerical solution of the equations of motions, with no growth of fluctuations, since the homogeneous mode does not stay frozen long enough. This prediction changes if the oscillations start closer to the top of the potential. 

\subsubsection{Growth of fluctuations in the Large Misalignment Mechanism}

\begin{figure}[t!]
	\centering
	\includegraphics[width=0.6\textwidth]{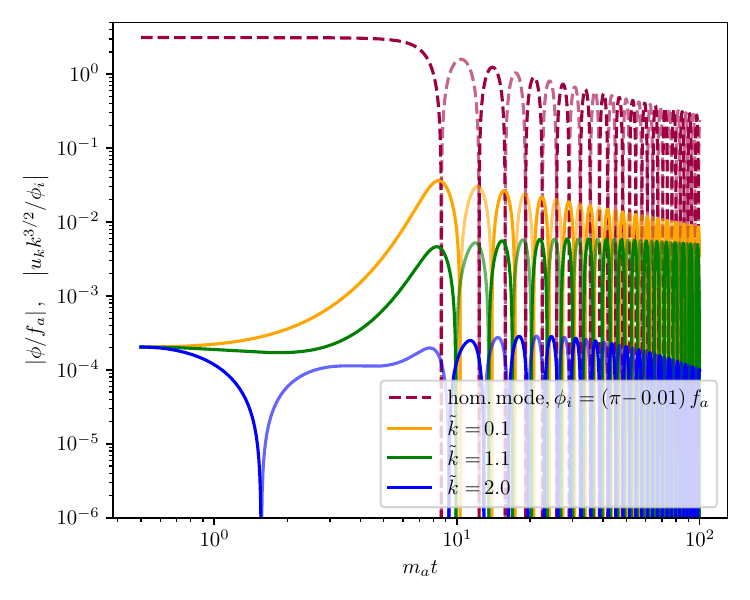}
	\caption{Homogeneous mode and mode functions with similar normalisation to \figref{fig:Fluctuations_multiple_plot1}. Oscillation of the homogeneous mode is delayed beyond $m_at>1$ due to the large misalignment $\theta_i=\pi-0.01$. The large misalignment makes it possible for fluctuations to grow exponentially due to the tachyonic instability already before they start to oscillate. This differs from \figref{fig:Fluctuations_multiple_plot1}, where no growth at early times could happen because oscillation of the homogeneous mode started too early.}
	\label{fig:Fluctuations_multiple_plot}
\end{figure}

In the Large Misalignment Mechanism introduced in \secref{sec:relic_density_LMM}, for an initial value of $\theta_i$ very close to $\pi$, like $\theta_i=\pi-0.01$, the damping term in \eqref{eom_zero_dimless_cosine_rad_era} stays dominant for longer and the homogeneous mode can still be frozen for $\ttilde>1$. Since $\cos(\theta_i)\approx -1$, \eqref{eom_fluc_dimless_cosine_rad_era} approximately becomes 
\begin{equation}
\ddot{u}_k(\ttilde)+\frac{3}{2\ttilde}\dot{u}_k(\ttilde)+\left(\frac{\tilde{k}^2}{\ttilde}-1\right)u_k(\ttilde)=0\,.
\end{equation}
When the damping term gets unimportant, i.e.~$\ttilde>1$, a mode begins to oscillate with frequency $\sqrt{{\tilde{k}^2}/{\ttilde}-1}$ if the square root is real. If this becomes imaginary, i.e.~$\tilde{k}^2/\ttilde<1$, modes show exponential growth. Because of the redshift of physical modes, also modes with $\tilde{k}>1$ can grow if the oscillations are delayed long enough. We show this in \figref{fig:Fluctuations_multiple_plot}. 

Due to the large amplitude of the oscillation of the homogeneous mode, which makes the higher order terms in the expansion of the cosine potential become important, at later times also parametric resonance leads to growth of fluctuations, which was analysed in \secref{sec:floquet_cosine}.

\subsection{Growth of fluctuations in a non-periodic potential}
\label{ssec:resonance_non-periodic}

Now we consider the potential from \eqref{eq:monodromy_potential} for which, due to the non-periodicity, it is possible to have large initial field values for the field, i.e.~$\phi_i \gtrsim f_a$. We are particularly interested in this regime, since it can involve an efficient parametric resonance and fragmentation of the ALP field. The motivation for large field values depends on the details of the actual UV-completion and on pre-inflationary physics. We discuss some of them in appendix~\ref{App.UV}. Generally, the ALP field can be driven to large values as a result of a random-walk behaviour during inflation, originating from the quantum fluctuations of the ALP field, stretched to superhorizon scales during inflation. These have an amplitude of the order of the inflationary Hubble scale $H_I$ when they leave the horizon.\footnote{When the field is in equilibrium, the variance of the field is approximately $H_I^2/m_a$. This shows that for small masses initial field values of $\phi_i\gg f_a$ originating from this process should be possible even under our assumption that $H_I\ll f_a$. \cite{Graham:2018jyp_initial_field_values}} Note, however, that such displacements during inflation can be noticeable only in the case of a very long duration of inflation, typically more than $10^{20}$ $e$-folds. 

\subsubsection{Floquet analysis}\label{sec:Floquet_non-periodic}
Analogous to \secref{sec:floquet_cosine}, for the analytical analysis of parametric resonance with the given non-periodic potential we will first ignore the expansion and set $H=0$ and $a=1$. The resulting EOMs for the dimensionless homogeneous mode and the mode functions are
\begin{equation}
\begin{split}
0&=\ddot{\theta}(\ttilde)+\frac{V'(\phi)}{m_a^2f_a}\\
\iff 0&=\ddot{\theta}(\ttilde)+\theta(\ttilde)\left[1+\theta^2(\ttilde)\right]^{p-1}\,\\
\end{split}
\end{equation}
\begin{equation}\label{eom_floquet_fluctuations_no_expansion}
\begin{split}
0&=\ddot{u}_k(\ttilde)+\left[\frac{k^2}{m_a^2}+\frac{V''(\phi)}{m_a^2}\right]u_k(\ttilde)\\
\iff 0&=\ddot{u}_k(\ttilde)+\left[\frac{k^2}{m_a^2}+(1+\theta(\ttilde)^2)^{p-2}\left(1+\theta(\ttilde)^2(2p-1)\right)\right]u_k(\ttilde)\,.
\end{split}
\end{equation}
Since the homogeneous mode is oscillating periodically (with period depending on the initial amplitude), the second derivative of the potential in \eqref{eom_floquet_fluctuations_no_expansion} is periodic in time. Therefore, we can again investigate the evolution of the fluctuations with the help of the Floquet theorem. (see appendix~\ref{App.1}).

\begin{figure}[t!]
	\centering
	\hspace*{-1cm}       
	\includegraphics[width=1.1\textwidth]{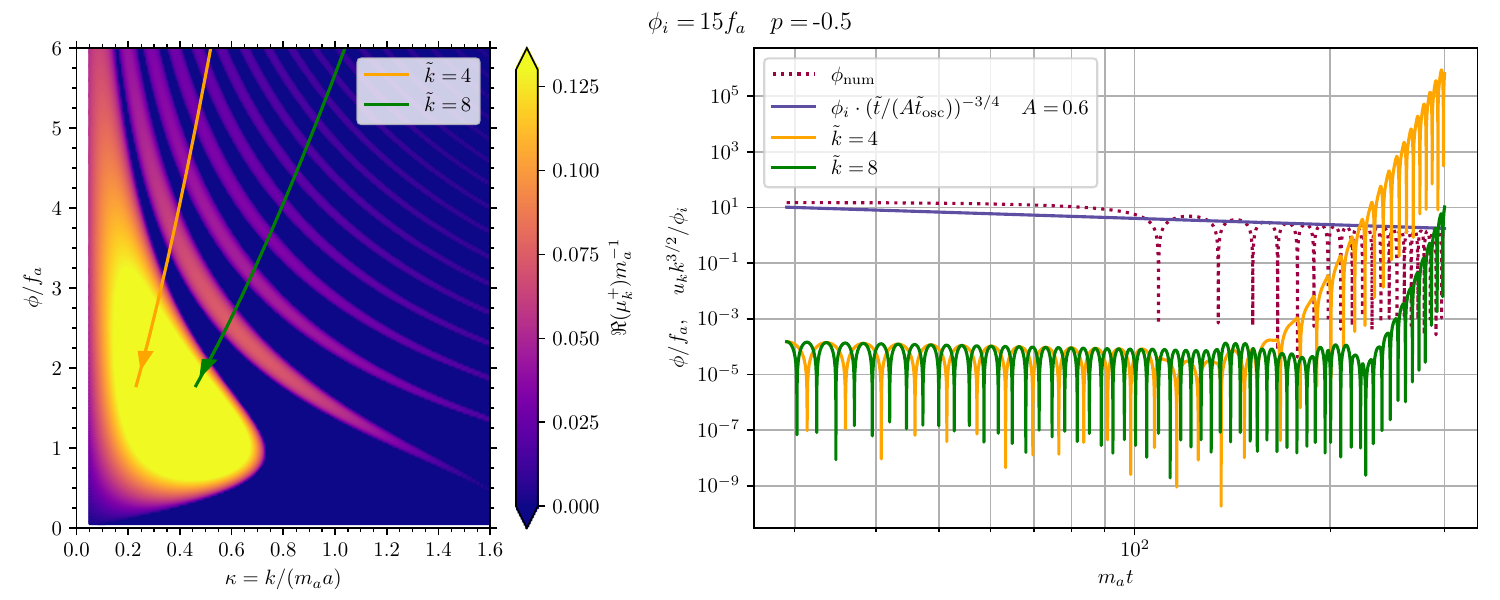}
	
	\caption{Parametric resonance in the ALP model with non-periodic potential from \eqref{eq:monodromy_potential} with $p=-1/2$ and $\phi_i=15f_a$. The left plot shows the real part of the Floquet exponents and how the two modes move through the $(\kappa$,$\phi)$-plane as they redshift and the homogeneous mode gets damped. The right plot shows numerical results for the absolute value of the homogeneous mode as a dotted line, and an analytical fit of the envelope in blue. We also show the absolute values of exponentially growing modes with $\tilde{k}=4$ and $8$, respectively. Details of this plot are discussed in the text of \secref{sec:Floquet_non-periodic}.}
	\label{fig:monodromy_potential_understanding_growth1}
\end{figure}

We show the real part of the Floquet exponents for different initial field values and momenta in the left panel of \figref{fig:monodromy_potential_understanding_growth1} for $p=-1/2$. The instability regions show for which values  of $\phi_i/f_a$ and $k/m_a$ exponential growth of the modes happens.

It goes beyond the scope of this analysis to attempt a rigorous Floquet analysis for $H\neq0$, that involves a time dependent physical momentum in the equations of motion. Nevertheless, we can use the analysis from the last subsection to analytically infer which modes will show exponential growth. Since the field behaves as matter component after the onset of oscillation, the homogeneous mode roughly dilutes as $\phi\sim t^{-3/4}$, as soon as it begins to oscillate. To find the beginning of oscillation, we again use \eqref{H_oscillation}.
For the non-periodic potential this results in $\ttosc=\left(2\sqrt{(1+\phi_{i}^2)^{p-1}}\right)^{-1}$. Hence, we conclude that the amplitude of the oscillation of the homogeneous mode behaves as $\phi_{\mathrm{ampl.}}({\ttilde})=\phi_{i}\left(\ttilde/(A\ttosc)\right)^{-3/4}$ for $\ttilde>A\ttosc$, where $A=\mathcal{O}(1)$ is introduced to compensate for the anharmonicity of the oscillation and fit the numerical result for the homogeneous mode (compare to \secref{sec:relic_density_monodromy}).
It is obvious that in radiation era, where $a\sim t^{1/2}$, the dimensionless physical momentum $\kappa\equiv k/(m_aa)$ of one given mode dilutes as $\kappa\sim t^{-1/2}$. Together with $\phi\propto a^{-3}$ this results in a curve in the $(\kappa$,$\phi)$-plane on which the different modes move towards smaller values of the homogeneous mode and physical momentum during the expansion.
From \figref{fig:monodromy_potential_understanding_growth1}, we can estimate which modes will grow and how strongly as different modes probe different parts of the instability region. We can compare this to numerical solutions for the homogeneous mode and fluctuations \eqref{eom_zero_general} and \eqref{eom_fluctuations_general}, which is shown for $p=-1/2$ for two exemplary modes in \figref{fig:monodromy_potential_understanding_growth1}. The left plot shows the real part of the Floquet exponent in the $(\kappa$,$\phi)$-plane and analytically derived curves that illustrate how modes ``move'' through this plane, i.e.~what the value of the amplitude of the oscillation of the homogeneous mode is when a mode has reached a certain physical momentum. These curves are outside the shown region at $\ttosc$ (at $\phi/f_a=15$ and $\kappa=\tilde{k}/\ttosc$) and move towards the origin. The curves end when $\ttilde$ reaches the end of the numerical simulation. The exponential growth in the modes that is seen on the right hand side corresponds to the respective time these modes enter the instability region that is depicted on the left hand side. 

\subsubsection{Tachyonic instabilities}

Tachyonic instabilities lead to growth of fluctuations when the homogeneous mode is still frozen and sits at a point in the potential at which the second derivative is negative. In the Large Misalignment Mechanism this was an important effect, since the top of the potential, where the field is fine-tuned to, has the largest negative value of the second derivative of the cosine potential. In the non-periodic potential, depending on the value of $p$, there may be no region of tachyonic instabilities, for example for $p=0.5$, while for $p=-0.5$ the second derivative is negative for most field values, but asymptotes to zero from below for larger ones, rendering the tachyonic instability increasingly unimportant for large initial field values.

\subsection{Regions of linear and non-linear growth}\label{sec:regions_non_linear}

\Eqref{eom_fluctuations_general} is based on linearization of the equations of motion, assuming small fluctuations. This neglects the interaction between different modes and the backreaction of the modes onto the homogeneous mode. Ignoring these effects is justified only as long as the fluctuations remain small relative to the homogeneous mode, which is the case for sufficiently small initial field values $\phi_i/f_a$. In contrast, for larger values of $\phi_i/f_a$, non-linear effects and the backreaction onto the homogeneous mode eventually become significant. This usually causes the rapid, non-perturbative process of fragmentation~\cite{Fonseca:2019ypl_fragmentation}, during which the energy from the homogeneous mode is completely transferred into fluctuations. For the cosine potential, numerical simulations show that this only happens for initial field values tuned close to the top or if the field initially carries enough kinetic energy to cross several maxima of the potential \cite{Arvanitaki:2019rax,Eroncel:2022vjg}.

To determine the transition between the two regimes, we track the components of the scalar field energy density in the homogeneous mode and in fluctuations.\footnote{Since the ratio of these two quantities, $\rho_{\mathrm{fluc}}/\rho_{0}$, is independent of $m_a$ and $f_a$, we only have to evaluate them for different initial values $\theta_i=\phi_i/f_a$ of the homogeneous mode.}
\begin{gather}
\rho_{0}=\frac{1}{2}\dot\phi(t)^2+V(\phi)\,,\\
\langle \rho_{\mathrm{fluc}} \rangle =\frac{1}{2}\int\frac{d^3k}{(2\pi)^3}\left[\lvert \dot{u}_k(t)\rvert^2+\left(\frac{\vec{k}^2}{a^2}+V''(\phi)\right)\lvert u_k(t)\rvert^2\right]\,.
\end{gather}
Note that the mode functions grow in the unstable phase, whereas at sufficiently late times, after leaving the instability regions and becoming non-relativistic, they behave as matter and decay in the same way as the homogeneous mode. As initial conditions for the mode functions at the onset of oscillation we take
\begin{equation}\label{initial_conditions_mode_func_no_curv}
    u_k(\ttosc)=\frac{\sqrt{2\pi^2\, 2.1\cdot10^{-9}}}{k^{3/2}}\phi_i\,,
\end{equation}
which we will justify in section~\ref{ssec:curvature}. As low momentum cutoff, we take the mode that enters the horizon at the end of the simulation and the large momentum cutoff at $k/a_{\osc}=5m_a$, since larger momenta will not be affected by parametric resonance.
We stop the simulation when $\rho_{\mathrm{fluc}}/\rho_{0}$ grows to unity since we can be sure that the linear description fails in that case. This is labelled as \textit{non-linear regime}. If $\rho_{\mathrm{fluc}}/\rho_{0}<1$ at sufficiently late times of the simulation, we are in the linear regime. The initial field value that separates the linear regime from the non-linear one is referred to as the \textit{critical field value}, $\theta_{\crit}=\phi_{\crit}/f_a$.

\begin{figure}[t!]
    \centering
    \includegraphics[width=0.8\textwidth]{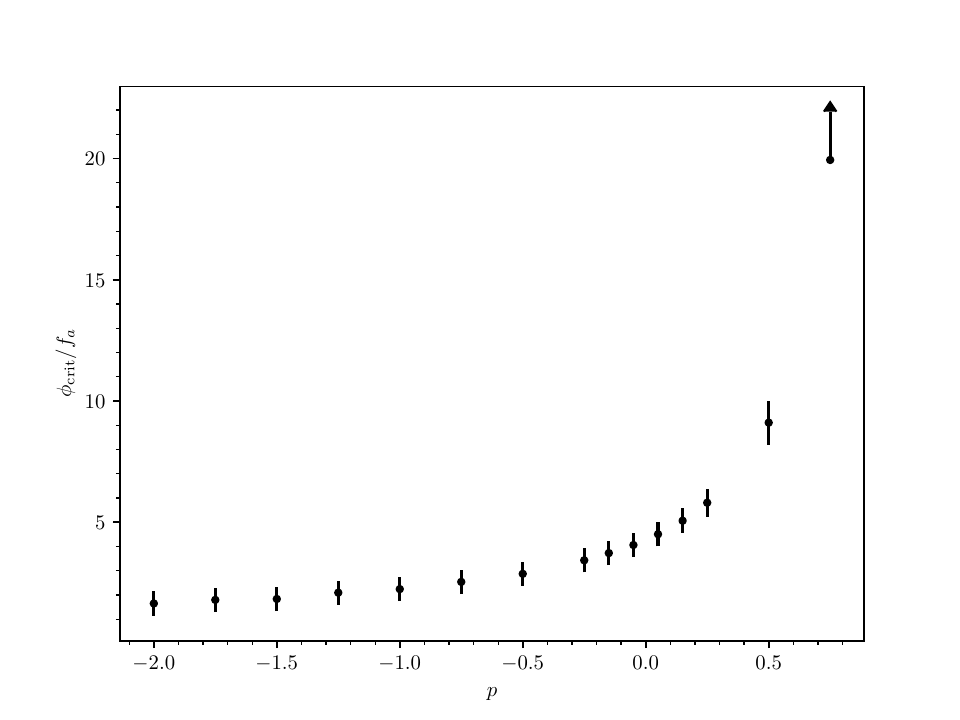}
    \caption{Critical field value $\phi_{\crit}$  at which $\rho_{\mathrm{fluc}}/\rho_0$ becomes larger than 1 for all $\phi_i>\phi_{\crit}$, with the initial conditions given in \secref{sec:regions_non_linear}, for different $p$ values in the non-periodic potential. The error-bars reflect the resolution of the numerical simulation. As indicated by the arrow, the rightmost point only gives a lower bound on the actual critical value.}
    \label{fig:phi_crit_no_curv_interp}
\end{figure}

If we numerically evaluate $\theta_{\crit}$ for several $p$, we find that it decreases for flatter potentials, i.e.~as $p$ gets more negative, while it diverges if $p$ gets close to unity, which corresponds to a quadratic potential where no parametric resonance is possible. The result is shown in \figref{fig:phi_crit_no_curv_interp}.

When we repeat this analysis for the Large Misalignment Mechanism and investigate for which parts of the parameter space in the $(m_a,f_a)$-plane fragmentation is expected, we find that we need a fine-tuning stronger than $\abs{\pi-\theta_i}\approx10^{-9}$.

Our goal is to investigate the amplification of ALP fluctuations and determine whether this can lead to observable effects. The crucial quantity that we will focus on is the energy density power spectrum, which we will discuss in the next section. It will quantify the inhomogeneities of the field and feed into the formalism that captures gravitational collapse of the overdense regions, which we discuss in \secref{sec:minihalos}.

\section{The power spectrum}
\label{sec:the_spectra}

We start this section by specifying the initial conditions for the fluctuations of the field. In order to properly describe the horizon entry of the curvature perturbations, we include the corresponding source terms in the linearized evolution equations. This is explained in \secref{ssec:curvature}, where both adiabatic and isocurvature fluctuations are discussed. In \secref{ssec:field_ps} we follow the subsequent evolution of ALP fluctuations and compute the resulting energy density power spectra. In the cases of efficient parametric resonance, the linear approximation eventually breaks down and we ``glue'' the linear evolution to a non-linear lattice simulation of the field dynamics, as it was done in~\cite{Berges:2019dgr, Chatrchyan:2020pzh}. We close with \secref{ssec:oscillons}, where the role of oscillons that are formed in the non-linear regime is explained.

\subsection{Initial conditions for the fluctuations}
\label{ssec:curvature}

The ALP field can have fluctuations of both adiabatic and isocurvature type, which are discussed in this subsection.

Adiabatic fluctuations of ALPs are sourced by curvature perturbations. This process can be described by including scalar metric perturbations in the field evolution equations. In the Newtonian gauge, the perturbed metric has the form 
\begin{equation}\label{pert_metric}
ds^2=[1+2\Phi(t,\xv)]dt^2-a^2(t)[1-2\Phi(t,\xv)]d\xv^2,
\end{equation}
where $\Phi(t,\xv)$ denotes the scalar metric perturbation field. In the radiation-dominated era, the Fourier modes $\Phi_{\kv}$ of the scalar perturbations are related, in the super-horizon limit, to those of curvature perturbations $R_{\kv}$ via $\Phi_{\kv} = - (2/3) R_{\kv}$. Inflation predicts a power-law form for the power spectrum of the curvature perturbations~\cite{Baumann:2022mni,Arvanitaki:2019rax}, 
\begin{equation}\label{initial_conditions_curvature_perturbations}
\langle R_{\kv} R^{\star}_{\kpv} \rangle =(2\pi)^3 \delta^{(3)}(\kv-\kpv) \frac{2\pi^2}{k^3} \Delta^2_{R}(k), \: \: \: \: \: \: \: \: \: \Delta^2_{R}(k) = A_s \Bigl( \frac{k}{k_{\star}} \Bigr)^{n_s - 1} \,.
\end{equation}
The amplitude $A_s =2.1\cdot10^{-9}$ at $k_{\star} = 0.05/\rm{Mpc}$ and the spectral tilt $n_s = 0.965$ are inferred from the PLANCK CMB-measurements~\cite{Planck:2018vyg_Curvature_pert_initial_cond}. For simplicity, the slight spectral tilt can be ignored in our calculations. The solution for the dynamics of curvature perturbations in the radiation era, derived in the linear theory, has the form\footnote{Note that the backreaction of the ALP field onto the curvature perturbations is negligible.}
\begin{equation}\label{curv_pert_in_rad_era}
\Phi_{\kv}(t_k)=3\Phi_{\kv,i}\left(-\frac{\cos(t_k)}{t_k^2}+\frac{\sin(t_k)}{t_k^3}\right)\,,
\end{equation}
where $\Phi_{\kv,i}$ denotes the super-horizon value of the mode function and $t_k$ is a momentum dependent, dimensionless time variable defined as $t_k \equiv k/(\sqrt{3}aH)$~\cite{Zhang:2017flu_solution_curvature_pert_rad_era}. As can be seen, the modes $\Phi_{\kv}$ decrease as $a^{-2}$ once they are deep inside the horizon, $t_k\gg1$.

Returning to the ALP field, due to the smallness of the fluctuations $\Phi$ and $\delta \varphi$, at least at early times, we keep only terms that are of zeroth or of linear order in any of these perturbations in the field evolution equations. In this way one arrives at~\cite{Zhang:2017dpp, Arvanitaki:2019rax}
\begin{equation}
\boxed{
\label{eom_with_curv_pert_fourier}
\ddot{u}_k(t)+3H\dot{u}_k(t)+\left(\frac{k^2}{a^2}+V''(\phi)\right)u_k(t)=4\dot{\phi}(t)\dot{\Phi}_k(t)-2V'(\phi)\Phi_k(t).} 
\end{equation}
The left hand side of the above equation is the same as the one in non-perturbed spacetime in \eqref{eom_fluctuations_general}, while the right hand side introduces source terms from the curvature perturbations. These terms can lead to fluctuations in the ALP field, even if the latter are initially zero. The sourcing mostly occurs close to the time at which the corresponding curvature mode enters the horizon, i.e.~when $H \sim k/a$. The simulation should therefore be started earlier than $\ttosc$, such that all modes of interest are superhorizon. 

Given that the ALP field was present during inflation, it unavoidably carries isocurvature fluctuations of the size of the inflationary Hubble scale $H_I$ (see \citre{Marsh:2015xka}). More specifically, the scale-invariant power spectrum of the fluctuations can be determined from the Bunch-Davies mode functions and is given by \begin{equation}
    \Delta^2_{\delta \phi}(k) = \left(\frac{H_I}{2\pi}\right)^2,
\end{equation}
in the superhorizon limit.

In the following it will be assumed that the isocurvature fluctuations are smaller compared to the adiabatic ones and, thus, they will be neglected. In this case one can work with mode functions $u_{k}$ that are initially zero. It was also verified that similar results are obtained if the source terms in \eqref{eom_with_curv_pert_fourier} are set to zero and, instead, reasonable initial values for the ALP mode functions are chosen, as it was done in \secref{ssec:resonance_non-periodic}.

\subsection{Growth of fluctuations and the density power spectrum}
\label{ssec:field_ps}

In this section we examine the amplification of ALP fluctuations. The discussion is split into two parts. In \secref{sssec:linear_regime},  we consider the case of $\theta<\theta_{\rm crit}$, where a linear analysis of the dynamics is performed. The non-linear regime for  $\theta>\theta_{\rm crit}$ is investigated in section~\ref{sssec:nonlinear_regime}.

To link the evolution of the fluctuations to observable quantities it is convenient to compute the \textit{energy density (contrast) power spectrum}.  
The density contrast is defined as 
\begin{equation}
    \delta(\vec{x})\equiv\frac{\rho(\vec{x})-\overline{\rho}}{\overline{\rho}}\,,
\end{equation}
where all quantities are time-dependent. $\overline{\rho}$ is the averaged energy density and in the quantum picture we identify this as $\expval{\rho}$. The Fourier transform of $\delta(\vec{x})$ is $\delta_{\kv}=\rho_{\kv}/\overline{\rho}$, where $\rho_{\kv}=\int d^3x\,\rho(x)e^{i\vec{x}\cdot\vec{k}}$ is the Fourier transform of $\rho(\vec{x})$ and we demand $k\neq0$. Again we will assume isotropy in momentum space, i.e.~care only for the magnitude of $\kv$.
The dimensionful density contrast power spectrum is defined as \cite{Enander:2017ogx}
\begin{equation}
    P_{\delta}(k)\equiv\frac{\expval{\delta_k^2}}{\Vol}=\frac{\expval{\abs{\rho_k}^2}}{\Vol\expval{\rho}^2}\,,
\end{equation}
where $\Vol$ is a large volume over which we integrate to keep the quantities finite. We will see that it drops out in the end. We also introduce the \textit{dimensionless density contrast power spectrum} which is defined as
\begin{equation}\label{dimless_density_contrast_power_spectrum}
    \Delta_{\delta}^2(k)=\frac{k^3}{2\pi^2}P_{\delta}(k)\,.
\end{equation}

\subsubsection{Linear regime}
\label{sssec:linear_regime}

Numerically calculating \eqref{eom_with_curv_pert_fourier} for a range of momenta allows one to track the  density power spectrum $\Delta_{\delta}^2(k)$. Note that the term $\dot{\Phi}$ in \eqref{eom_with_curv_pert_fourier} can be re-expressed as $m_a\frac{t_k}{\ttilde}\frac{d\Phi_{\kv}}{dt_k}$, where the derivative with respect to $t_k$ can be found from \eqref{curv_pert_in_rad_era}. This makes it possible to express \eqref{eom_with_curv_pert_fourier} fully in terms of dimensionless quantities and arrive at an EOM that is independent of $m_a$ and $f_a$.\footnote{Note that $f_a$ is again a common factor in all terms of \eqref{eom_with_curv_pert_fourier}.}  The enhancement of the fluctuations thus depends only on the values of $p$ and $\theta_i$. 

\begin{figure}[t!]
    \centering
    \includegraphics[width=0.82\textwidth]{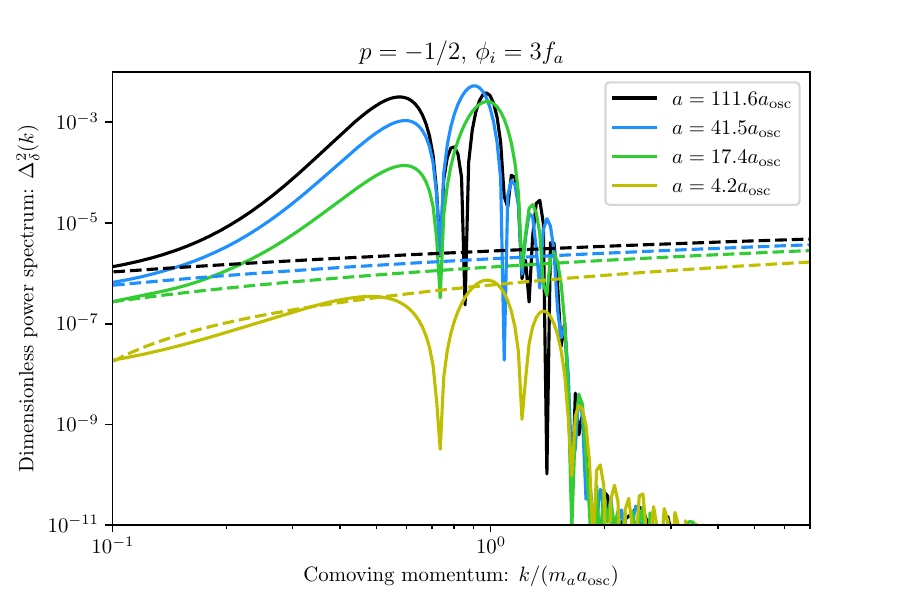}
    \caption{Dimensionless density contrast power spectra $\Delta_{\delta}^2(t,k)$ for ALP dark matter with $\phi_i=3f_a$ and $p=-1/2$ (solid lines) and for standard CDM (dashed lines), shown at different values of $a$, considering radiation dominated era, $a\sim t^{(1/2)}$, and adiabatic initial conditions. The horizontal axis shows the comoving momentum $k/(m_a a_{\rm osc})$. The curves converge at low momenta, while at high momenta the ALP spectrum is suppressed. }
    \label{fig:power spectrum}
\end{figure}

In the linear regime the density contrast perturbation can be computed using the following expression~\cite{Arvanitaki:2019rax},
\begin{equation}\label{density_contrast_power_spectrum_w_curvature_pert_linear}
    \delta_{\vec{k}, \mathrm{lin.,\,pert.} }=\frac{\dot{\phi}\dot{u}_{\vec{k}}+V'(\phi(t))u_{\vec{k}}-\dot{\phi}^2\Phi_{\kv}}{\frac{1}{2}\dot{\phi}^2+V(\phi(t))}\,.
\end{equation}
In~\figref{fig:power spectrum} we illustrate the dimensionless power spectrum, computed using the expression above, at several scale factors for $\theta_i=3$ and $p=-1/2$. The spectra are shown with solid lines. On the horizontal axis is the comoving momentum, in units of the ALP mass multiplied by the scale factor at $H_{\osc}$. As expected, the spectrum is peaked around the momenta corresponding to the most unstable modes. 

It is instructive to compare the evolution of the density power spectrum to the one of standard (pressureless) CDM. The latter has the following analytical form during the radiation-dominated era~\cite{Blinov:2021axd}
\beq\label{CDM_solution}
\delta_{k, \rm CDM}= C_2 + 9 \Phi_{k, i} \ln[ t_k ],
\eeq
where $C_2 = 3\Phi_{k,i}/2$ for adiabatic fluctuations. As can be seen, the density contrast grows logarithmically with the scale factor. 

The solution (\eqref{CDM_solution}) is shown in \figref{fig:power spectrum} with dashed lines at the same scale factors as the solid lines of the corresponding colour. One can distinguish three momentum ranges. At low momenta the ALP power spectrum converges to the one for CDM. At intermediate momenta $k\sim m_a a_{\osc}$, the ALP power spectrum is enhanced due to parametric resonance. Finally, at high momenta one observes a strong suppression of the ALP power spectrum. More details about the evolution of density perturbations for wave dark matter can be found in~\citre{Blinov:2021axd}.

\subsubsection{Nonlinear regime}
\label{sssec:nonlinear_regime}

If $\theta_i > \theta_{\rm crit}$, the linear approximation is expected to be valid only at early times. Once the fluctuations become large, their backreaction onto the homogeneous mode becomes important, which slows down their growth and, in most cases, leads to the fragmentation of the field.

In order to extend the discussion to this non-linear regime, we ``glue'' the linearized evolution to the lattice simulation that solves the full EOM~\eqref{full_eom_axion_unperturbed}. More specifically, the linear simulation is stopped when the fluctuations are still small, $\langle \delta\theta^2 \rangle = 10^{-4}$. The values of the field and its derivative, as well as the power spectra computed using the mode functions, are then extracted and used to initialise a Gaussian random field on the lattice. More details about the lattice setup can be found in appendix~\ref{App.Lattice} as well as in~\citre{Berges:2019dgr}. The field is evolved on a spatial lattice with $512^3$ grid points, periodic boundary conditions and a fixed comoving volume, using a standard leap-frog algorithm. In principle, it is possible to incorporate metric perturbations into the lattice simulation, in the form of an additional scalar field. However this would require the Fourier transformation of the field after each time step, to evolve the mode functions in momentum space according to~\eqref{curv_pert_in_rad_era}. On the other hand, in the regime when fragmentation takes place, the metric perturbations become unimportant as the field is no longer correlated with the metric perturbations. To this end we switch off the metric perturbations once the simulation starts.

The ALP dynamics after fragmentation usually involves the formation of long-lived compact objects, called \textit{oscillons}~\cite{Gleiser:1993pt, Copeland:1995fq}. These are coherent and localised configurations of the scalar field, held together by the attractive self-interactions. They can be very long-lived, even though there is no conserved charge that would ensure their stability and prevent them from decaying (see also \citre{Kasuya:2002zs} for the approximate conservation of the adiabatic invariant of oscillons). 
As a consequence, some fraction of the total energy, denoted by $F_{\os}$, is contained at late times in oscillons, while the rest is in the form of unbound fluctuations.\footnote{The subscript $\os$ will refer to oscillons, and should not be confused with the subscript $\osc$, referring to the onset of oscillations.}  The value of $F_{\os}$ depends on $p$, as well as on the initial field value $\theta_i$, and in some cases can reach around $70\%$~\cite{Kawasaki:2020jnw}. 

The separation into the contributions mentioned above can be observed in the field power spectrum. At sufficiently late times, the spectrum splits into two parts: one at high momenta, due to oscillons, and one at lower momenta due to fluctuations. The oscillon peak "blue-shifts" with time to higher momenta, as the oscillons maintain fixed physical size. In contrast, the low-momentum contribution from the fluctuations eventually becomes time-independent, maintaining constant comoving momenta. Such behaviour is expected due to the Hubble expansion since the self-interactions become weaker as the ALP field is driven to the vicinity of the quadratic minimum of the potential. The time for this ``freeze-out'' depends on the initial field value, i.e.~it is larger for larger values of $\theta_i$. 

\begin{figure}[t!]
	\centering
	\includegraphics[width=0.92\textwidth]{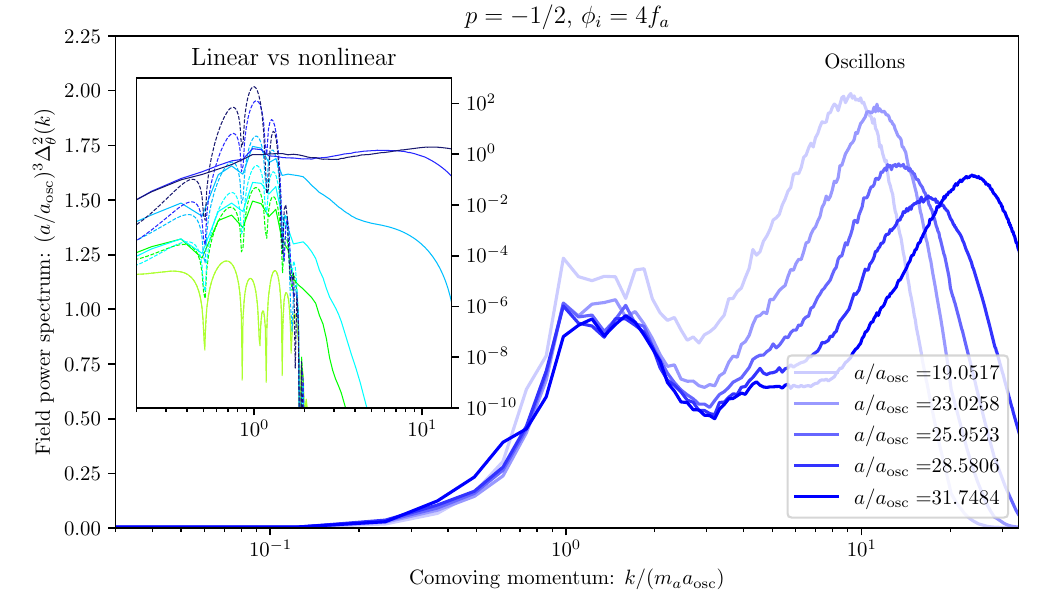}
	\caption{Snapshots of the dimensionless field power spectrum  $\Delta^2_{\theta}(t,k)$ for $\theta_i=4$ and $p=-1/2$, demonstrating the oscillon peak at high momenta and the fluctuation peak at small momenta. The colours from light to dark blue correspond to scale factors $a/a_{\osc}$ ranging from $19$ to $32$. The inset compares the linear (dashed lines) and the lattice (solid lines) evolution at scale factors $a/a_{\osc}$ between $4$ and $24$, illustrating how non-linear effects slow down and spread out the growth of low-momentum fluctuations.}
	\label{fig:field_ps}
\end{figure}

As a result of this behaviour, the oscillon peak gets well separated from the low-momentum fluctuations at late times. This is illustrated in \figref{fig:field_ps}, where several snapshots of the dimensionless field power spectrum $\Delta^2_{\theta}(k)$ are shown. The multiplication by the scale factor $a^3$ is done for convenience, to scale out the overall dilution due to the Hubble expansion. The curves are shown at scale factors in the range from $19a_{\osc}$ to $32a_{\osc}$. $\theta_i=4$ and $p=-1/2$ were used for this plot, although the pattern is rather generic. The height of the oscillon peak shows some oscillatory behaviour but on average does not change much within the simulation time. The relative strength of the two contributions depends on $F_{\os}$.

\begin{figure}[t!]
	\centering
	\includegraphics[width=0.82\textwidth]{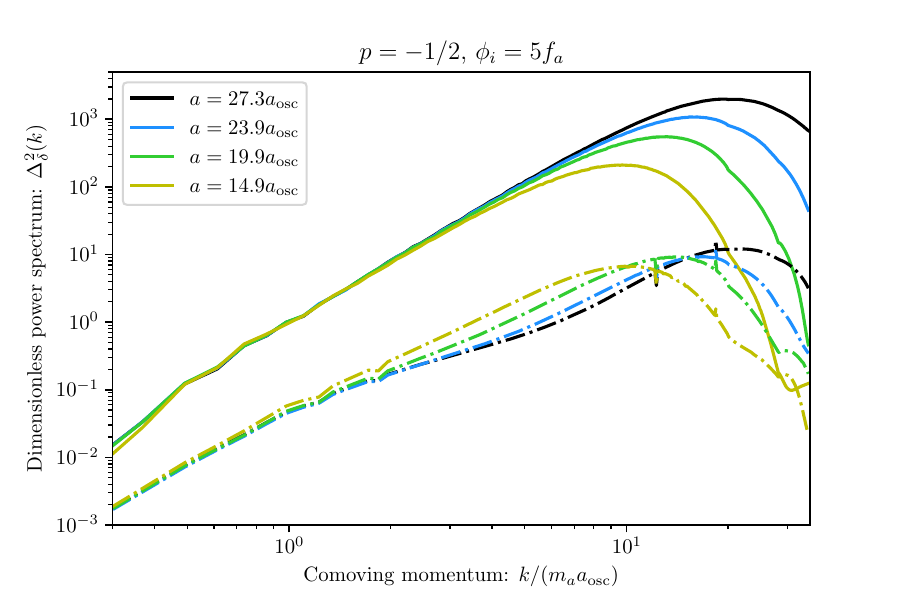}
	\caption{Snapshots of the energy density contrast power spectrum  $\Delta^2_{\delta}(t,k)$ for $\theta_i=5$ and $p=-1/2$ at the scale factors when oscillons have already formed, ranging from $a = 14a_{\osc}$ to $a = 28a_{\osc}$. The solid lines show the full (non-linear) power spectrum, dominated by oscilllon contribution. The dashed lines correspond to the Gaussian approximation for the power spectrum, according to \eqref{density contrast power spectrum2} which, at small momenta, is dominated by the contribution from fluctuations.}
	\label{fig:energy_ps_WKB}
\end{figure}

As in the linear regime, we are interested in the energy density power spectrum. Here it can be extracted from the lattice simulation by explicitly computing the energy density of the scalar field and evaluating its correlation functions. It is observed that the density power spectrum is dominated by the contribution from oscillons, in the form of a strong peak in momentum space. This is illustrated in \figref{fig:energy_ps_WKB}, where the solid lines show the dimensionless density contrast power spectrum at scale factors from $14a_{\osc}$ to $a=28a_{\osc}$ for $\theta_i=5$ and $p=-1/2$. A semi-analytical expression for this peak was found in \citre{Kawasaki:2020jnw}, where it was parametrised in terms of an extended Poisson distribution,
\begin{equation}
P_{\delta}(k; a)=\frac{F_{\os}^2}{ n_{\os} } \frac{ \langle M_{\os}^2 \exp\{- r_{\os}^2k^2 / (2a^2)\} \rangle }{\langle M_{\os} \rangle^2 }K(k/k_c)\,
\end{equation}
with $K(x) = 1 - ( 2/x )^2 \sin^2 ( x/2 )$. Here $n_{\os}$, $r_{\os}$ and $M_{\os}$ are the average number density (per comoving volume), the (physical) radius and the mass of the oscillons. $k_c$ is a cut-off scale for the oscillon power spectrum, which is determined by fitting the numerical data. It is related to the horizon scale at the oscillon formation. 

\subsection{The role of oscillon decay}
\label{ssec:oscillons}

The lifecycle of an oscillon can be separated into a long phase of slow ``evaporation'', via the emission of scalar radiation, and a very rapid ``decay'' in the end, when the oscillon energy is converted into a burst of scalar waves~\cite{Imagawa:2021sxt}. Power-law potentials with $p<1$, considered in this work, are known to support very long-lived oscillons. In \citres{Olle:2019kbo, Olle:2020qqy}, the lifetime was estimated to be $ \tau_{\os} \sim 10^{10}m_a^{-1}$ for $p=-1/2$,  $\tau_{\os} \sim10^8m_a^{-1}$ for $p=1/2$, and even longer for $p\rightarrow 1$. In contrast, in the case of the QCD axion the corresponding objects, often referred to as axitons, can survive only at most for $\sim 10^3 m_a^{-1}$~\cite{Vaquero:2018tib}. The very long lifetimes of oscillons make it impossible to extend the lattice simulations to their decay stage. Nevertheless, they do eventually decay and, in a large region of the parameter space given by
\beq\label{eq:osc_decay_before_eq}
m_{a} > 10^{-19}\, \mathrm{eV} \left( \frac{ \widetilde{\tau}_{\os} }{10^8} \right),
\eeq
with $\tilde{\tau}_{\os}=m\tau_{\os}$, this happens before the epoch of matter and radiation equality. It is thus important to understand how the decay is expected to impact the power spectra. This is discussed below, with a more detailed analysis of the decay process to be included in a future publication.

For simplicity we ignore the slow evaporation of oscillons, assuming that their energy is radiated away in the very end. In this last stage the oscillon peak in the field power spectrum will already be separated by several orders of magnitude from the low-momentum fluctuation modes. The decay of oscillons usually produces ALPs with wavelengths $(k/a)\gtrsim 0.1m_a$, which simply free-stream away~\cite{Imagawa:2021sxt}. It is thus expected that this process does not affect the field power spectrum at much smaller momenta.

We want to estimate the energy density power spectrum at late times, after the decay. At this stage the self-interactions are unimportant and, assuming that the field power spectrum is nearly Gaussian, it can be expressed  as~\cite{Vaquero:2018tib}
\begin{equation}\label{density contrast power spectrum2}
P_{\delta}(q)=2\frac{\int\frac{d^3k}{(2\pi)^3}n_k n_{|k+q|}}{\left(\int\frac{d^3k}{(2\pi)^3}n_k\right)^2}\,,
\end{equation}
in the deep non-relativistic limit. Here
\beq
\label{occup_mode}
n_k + \frac{1}{2} = \frac{ |\dot{u}_{k}|^2 +  \omega_k^2 |u_{k}|^2 }{2\omega_k} 
\eeq
denotes the occupation numbers. We discuss this in more detail in appendix~\ref{App.2}, where we demonstrate that the numerically extracted power spectrum after the decay of oscillons is indeed well approximated by (\ref{density contrast power spectrum2}).

In \figref{fig:energy_ps_WKB} we show several power spectra, computed using the above Gaussian expression, in dashed lines. The solid lines in that figure correspond to the full non-linear power spectra from the same simulation. As can be observed, $P_{\delta}(q)$ at small momenta becomes constant after some time. From the form of \eqref{density contrast power spectrum2} we can conclude that, by these times, it is dominated by the contribution from fluctuations at small $k$'s, rather than by the time-dependent oscillon contribution. This is expected to hold also after the decay of oscillons and, therefore, $P_{\delta}(q)$ should not change much any further. This observation allows us to estimate the power spectrum at small momenta and late times by its time-independent value from the simulation. 

To summarise, even though our simulations do not resolve the dynamics of the oscillons until their late decay, we can make predictions for the low momentum part of the energy power spectrum after their decay. 

\section{ALP halos}
\label{sec:minihalos}

Having computed the density power spectra for ALPs in a non-periodic potential, in this section we study how the overdense regions, corresponding to the amplified fluctuations of the ALP field, decouple from the Hubble flow and form bound objects. We have already briefly mentioned the formation of long-lived oscillons, which are sustained by the self-interactions. Our focus of this section is on the more important gravitational interaction, and the formation of objects such as halos and solitons. 

Most of the structure formation takes place in the matter-dominated era, when the backreaction of the ALPs on the metric perturbations can no longer be neglected. Our analysis is based on the Press-Schechter (PS) formalism~\cite{Press:1973iz}. We start by reviewing the spherical collapse model from~\citre{Kolb:1994fi} and the Meszaros equation~\cite{Meszaros:1974tb} in \secref{sec:spherical_collapse_model}. Modifications to the formalism, relevant at the length scales associated to fragmentation, are discussed as well \cite{Ellis_Marsh:2020gtq}. In~\secref{sec:Press-Schechter} we apply the formalism to our model and, using the power spectra of the density field from the previous section, compute the halo mass functions. Virialization of the halos is summarised in~\secref{sec:halo_spectrum}, where we also find the halo spectrum.

\subsection{Spherical collapse}
\label{sec:spherical_collapse_model}

We follow~\citre{Kolb:1994fi} and consider a spherical overdense region of matter in the homogeneous background of radiation. The equation of motion for the shell of radius $r$ is given by
\begin{equation}\label{eom_spherical_shell_1}
\ddot{r}=-\frac{8\pi G}{3}\rho_Rr-\frac{GM}{r^2}\,,
\end{equation}
where $\rho_R$ refers to the energy density of the radiation component and $M$ is the conserved mass of the matter inside the shell. 

As it was done in~\citre{Kolb:1994fi}, this equation is investigated by switching to conformal time $d\eta = dt/a$ and introducing the following parameterisation of $r$:
\begin{equation}\label{r_terms_of_deceleration_and_a}
r\equiv a(\eta)\xi(\eta)R\,,
\end{equation}
where $R$ is the comoving size of the spherical region and $\xi(\eta)$ is the \textit{deceleration parameter}. The latter is set to $\xi = 1$ initially and describes how an overdense region decouples from the Hubble flow. One also expresses the mass in terms of the average matter density $\rho$ and the initial over-density parameter $\delta_{\ini}$, 
\begin{equation}\label{eq:spherical_mass}
M=\frac{4\pi}{3}\rho_{\eq}a_{\eq}^3(1+\delta_{\ini})R^3\,,
\end{equation}
Note that the term $(\rho a^3)$ remains constant as soon as the ALPs become non-relativistic and, in the above expression, it is evaluated at the time of matter-radiation equality, as denoted by the index $\rm EQ$. For simplicity the universe is taken to be flat, consisting only of the matter and radiation components. The scale factor can then be written as $a(\eta)=a_{\eq}\left[2(\eta/\eta_*)+(\eta/\eta_*)^2\right],$
where $\eta_*^{-2}=2\pi G \rho_{\eq} a_{\eq}^2/3$~\cite{Kolb:1994fi}. Inserting everything into \eqref{eom_spherical_shell_1} and switching from the time variable to the scale factor $x = a/a_{\rm EQ}$, one arrives at a dimensionless equation for the deceleration parameter~\cite{Kolb:1994fi}:
\begin{equation}\label{eom_spherical_shell_3}
x(1+x)\frac{d^2\xi}{dx^2}+\left(1+\frac{3}{2}x\right)\frac{d\xi}{dx}+\frac{1}{2}\left(\frac{1+\delta_{\ini}}{\xi^2}-\xi\right)=0\,.
\end{equation}

Solving this equation numerically, with $\xi=1$ and $d\xi/dx=0$ as initial conditions, one can find the scale factors at the \textit{turnaround} $x_{\ta}$, after which the shell starts to shrink, and the collapse scale factor $x_{\coll}$. The condition for the turnaround is given by $\dot{r}=0$ which, using the definition (\ref{r_terms_of_deceleration_and_a}), can be expressed as $\dot{\xi}=-\xi H$. In terms of the dimensionless variable $x$ this reads as
\begin{equation}\label{collapse_condition}
\left.\frac{d\xi}{dx}\right\lvert_{x_{\ta}}+\left.\frac{\xi}{x}\right\lvert_{x_{\ta}}=0\,.
\end{equation}
We find numerically that this is fulfilled at 
\begin{equation}\label{condtion_turnaround}
\delta_{\ini}\cdot x_{\rm turn}\approx0.71\equiv C_x\,,
\end{equation}
where we will use the last definition later. We also evaluated the product of $\delta_{\ini}$ with the numerically extracted collapse scale factor $x_{\coll}$, defined by $r(x_{\coll}) \rightarrow 0$ or $\xi(x_{\coll}) \rightarrow 0$. This gives
\begin{equation}\label{condition_collapse}
\delta_{\ini}\cdot x_{\coll}\approx1.13\,.
\end{equation}

\bigskip

\textbf{Linear growth:} The actual over-density parameter of the spherical region $\delta$ will grow during the collapse. It is important to investigate how this parameter evolves in time, under the assumption of it being small, $\delta\ll1$. Since the mass inside the collapsing shell remains constant at value given by~\eqref{eq:spherical_mass}, one has 
\begin{equation}
(1+\delta_{\ini})=\xi^3(x)(1+\delta(x))\,.
\end{equation}
If we now assume $\delta_{\ini}\ll 1$, the left-hand side in the above equation can be set to unity. Solving this for $\xi$ gives $\xi=\left(1/(1+\delta)\right)^{1/3}$, which can be expanded to $\xi=1-\delta/3+\mathcal{O}(\delta^2)$. Inserting $\xi=1-\delta/3$ into \eqref{eom_spherical_shell_3} we arrive at the \textit{Meszaros equation}~\cite{Meszaros:1974tb} for $\delta$:
\begin{equation}\label{eom_spherical_shell_4_linear}
x(1+x)\frac{d^2\delta}{dx^2}+\left(1+\frac{3}{2}x\right)\frac{d\delta}{dx}-\frac{3}{2}\delta=0\,.
\end{equation}

For a given positive initial over-density $\delta_{\ini}$ and initial condition $\dot{\delta}_i\approx 0$, the analytical solution of the Meszaros equation is
\begin{equation}
    \delta(x)=\delta_{\ini}\left(1+\frac{3}{2}x\right).
\end{equation}
Plugging \eqref{condition_collapse} into this result, and assuming that $\delta_{\ini}$ is small, such that $1.1/x_\coll\ll1$, gives the well-known value for the over-density in the linearized theory at which the collapse happens:
\begin{equation}\label{eq:condition_linear_coll}
    \delta_c\equiv\delta(x_\coll)=\frac{1.1}{x_\coll}\left(1+\frac{3}{2}x_\coll\right)\approx1.69.
\end{equation}

\bigskip

\textbf{Jeans length and scale-dependent growth:} The linear growth of density perturbations in \eqref{eom_spherical_shell_4_linear} is scale independent, which is true for CDM perturbations. However, in the case of ALP DM, the growth is modified at high momenta, above a certain Jeans length. To account for these effects due to the gradient pressure, the Meszaros equations can be modified to \cite{Eroncel:2022efc}
\begin{equation}
\label{eom_spherical_shell_4_linearJ}
\boxed{
x(1+x)\frac{d^2\delta_k}{dx^2}+\left(1+\frac{3}{2}x\right)\frac{d\delta_k}{dx}-\left(\frac{3}{2} - \frac{\tilde k^4}{x} \right) \delta_k=0.}
\end{equation}
where $\tilde k$ is defined in \eqref{eq_define_tilde}. Modes with $\tilde k\ll 1$ behave as in the case of CDM, while modes with $\tilde k\gg 1$ start growing only after $x\sim \tilde k^4$, resulting in a suppression of the power spectrum at high momenta.

\subsection{The Press-Schechter formalism and halo mass function}
\label{sec:Press-Schechter}

The standard Press-Schechter formalism has been introduced in \citre{Press:1973iz} and was modified since then (see \citre{Sheth:1999mn}, \citre{Zentner:2006vw} and references therein). The original Press-Schechter formalism was developed for dark matter that has been pressureless for the whole history of the universe, with scale independent power spectrum that has been imprinted by inflation.

In the PS formalism one introduces the smoothed over-density field $\delta_R(\xv)$,
\begin{equation}\label{smooted_density_contrast}
\delta_R(\xv)=\int d^3x'W_R(\xv-\xpv)\delta(\xpv)\,,
\end{equation}
where $W_R(\xv)\equiv W(\xv,R)$ is a \textit{window function} that goes to zero very fast if $|\xv|>R$. An intuitive choice would be the so-called \textit{spherical top-hat}, $W_{\mathrm{sph.\,TH}}(\xv,R)=3/(4\pi R^3)\Theta(R-\xv),$ where $\Theta(x)$ is the Heaviside step function. $R$ is called the \textit{smoothing scale}. 

One can evaluate the variance of the smoothed density contrast, $\sigma_R^2$. A quick calculation gives
\begin{equation}\label{sigma_R}
    \sigma_R^2\equiv\frac{1}{\Vol}\int d^3x \expval{\delta_R(\xv)^2} = \frac{1}{2\pi^2}\int dk\,k^2 \abs{\tilde{W}_R(\kv)}^2P(k)\,,
\end{equation}
where we used $\Vol^{-1}P(k)\equiv\expval{\delta(k)^2}$ and introduced the Fourier transform of the window  function $\tilde{W}_R(\kv)\equiv\tilde{W}(\kv,R)=\int d^3xW(\xv,R)e^{-i\kv\cdot\xv}$.

Under the assumption of initially small over-densities, one can relate the smoothing scale $R$ directly to a halo mass $M$ by setting $M=\rho V_{R}R^3$, where $V_{R}$ is a volume factor that depends on the window function, where for the spherical top-hat it is given by $4\pi/3$. This implies that $\sigma_R=\sigma(R)$ basically becomes a function of $M$, hence we can write $\sigma(M)$.

Assuming that $\delta_R$ follows a Gaussian distribution (for any value of $R$) and identifying $\sigma_R$ as its variance, the probability to find a fluctuation in the range $[\delta_R,\delta_R+d\delta_R]$ is \cite{Press:1973iz}

\begin{equation}\label{smoothed_over-density_f}
F_{\mathrm{sm}}(\delta_R;R)=\frac{1}{\sqrt{2\pi\sigma_R^2}}\exp\left(-\frac{\delta_R^2}{2\sigma_R^2}\right)\,.
\end{equation}

The fraction of collapsed objects with mass $>M$ is then given by
\begin{equation}\label{PS_err_func}
F(>M)=\frac{1}{\sqrt{2\pi}}\int_{\delta_c/\sigma_R}^{\infty}e^{-t^2/2}dt\equiv\mathrm{erfc}[\delta_c/\sigma_R]\,,
\end{equation}
where $\mathrm{erfc}$ denotes the \textit{error function} and $\delta_c\approx 1.69$ is the critical density at which collapse happens.
Due to the one-to-one correspondence between $R$ and $M$, $\sigma_R$ can be understood as a function of $M$. Thus one can calculate:
\begin{equation}
\frac{dF}{dM}=-\frac{1}{\sqrt{2\pi}}\frac{\delta_c}{\sigma_R^2}\frac{d\sigma_R}{dM}\exp\left(\frac{-\delta_c^2}{2\sigma_R^2}\right)\,.
\end{equation}
This is the fraction of collapsed objects in the range $[M,M+dM]$. The fractional energy density in these objects is then $\rho_0\cdot\frac{dF}{dM}$, while finally the number density of objects inside the given mass range is $\rho_0\cdot\frac{dF}{dM}/M$. We arrive at \cite{Press:1973iz}
\begin{equation}\label{PS_number_density}
\frac{dn}{dM}=-\frac{\rho_0}{M}\sqrt\frac{2}{\pi}\frac{\delta_c}{\sigma_R^2}\frac{d\sigma_R}{dM}\exp\left(\frac{-\delta_c^2}{2\sigma_R^2}\right)\,.
\end{equation}
In this result, a seemingly ad hoc factor of 2 is introduced. This is done because the Gaussian distribution of $\delta_R$ also includes negative over-densities. This region is not accounted for when evaluating \eqref{PS_err_func}, but nevertheless these under-densities will be included in collapsed halos since over-dense regions will be surrounded by under-dense regions which are dragged into the halo by the higher gravitational potential in the collapsing regions. The factor of 2 can be rigorously derived in the excursion set formalism \cite{Zentner:2006vw}. 

\begin{figure}[t!]
	\centering
	\includegraphics[width=0.49\textwidth]{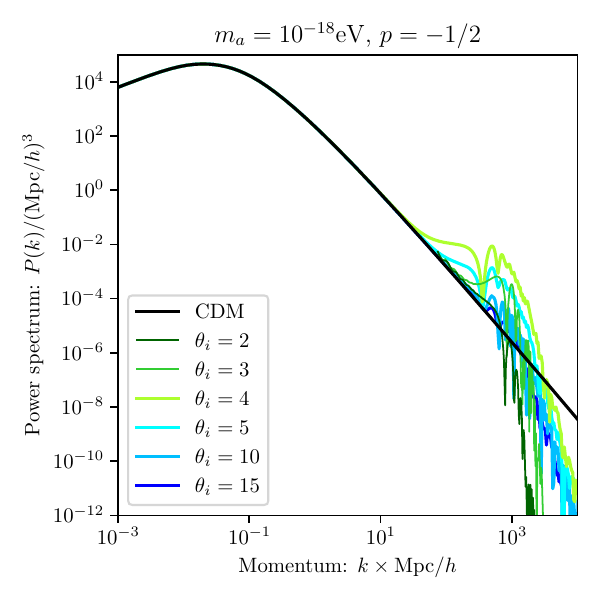}
 	\includegraphics[width=0.49\textwidth]{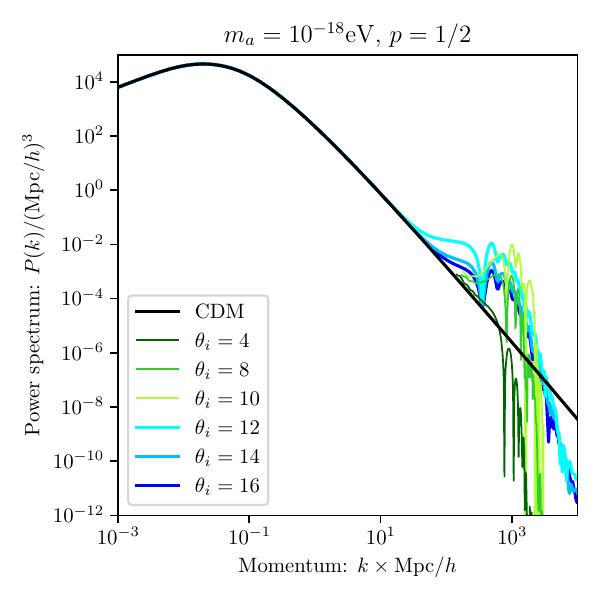}
	\caption{Matter power spectrum today for CDM (black line) vs ALP DM of mass $m_a=10^{-18}\mathrm{eV}$ with $p=-1/2$ (left) and $p=1/2$ (right). Different colours correspond to different values of the initial misalignment angle $\theta_i = \phi_i/f_a$. The power spectra agree with CDM on large scales, are enhanced on scales at which fluctuations were amplified, and are suppressed on small scales due to wave effects.}
	\label{fig:MPS}
\end{figure}

To compute the halo mass function, we start with the density contrast power spectra that were discussed in \secref{sec:the_spectra}. These serve as initial conditions for the Meszaros equation including the Jeans length \eqref{eom_spherical_shell_4_linearJ} and are evolved until today. For the non-linear regime we use the density power spectrum without the oscillon contribution, as it was described in the previous section, together with $d \delta_k/dx =0$ for the initial conditions. The resulting power spectra are shown in \figref{fig:MPS} for $m_a = 10^{-18}\mathrm{eV}$, both $p=-1/2$ and $p=1/2$, and several values of the misalignment angle. The black line corresponds to the CDM power spectrum linearly evolved until today \cite{Maggiore:2018sht},
\begin{equation}\label{Power_spectrum_CDM_eq}
    P_{\mathrm{CDM}}(k)=2\pi\delta_H^2\frac{k^{n_s}}{H_0^{n_s+3}}T^2(k)\,.
\end{equation}
Here $\delta_H$ is a normalising constant we set to $1.9\times 10^{-5}$, $n_s$ is the spectral index of primordial perturbations, which we set to 1 and $T(k)$ is the BBKS transfer function, given by~\cite{1986ApJ...304...15B}
\begin{equation}
    T( k )=\frac{\ln(1+0.171\kappa)}{0.171\kappa}\left(1+0.284\kappa+(1.18\kappa)^2+(0.399\kappa)^3+(0.490\kappa)^4\right)^{-1/4}\,,
\end{equation}
where $\kappa=k/k_{\eq}$ and $k_{\eq}=0.073\Omega_{m,0}h^2\,\mathrm{Mpc}^{-1}$.

\begin{figure}[!t]
	\centering
	\includegraphics[width=0.49\textwidth]{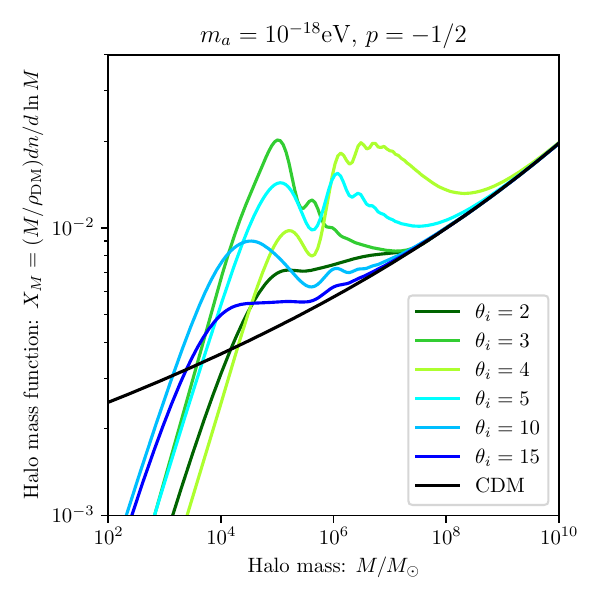}
 	\includegraphics[width=0.49\textwidth]{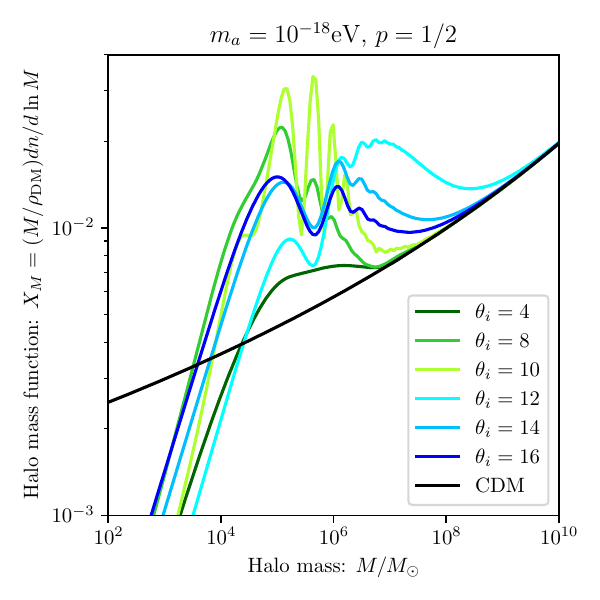}
	\caption{The halo mass function $X_M = (M/\rho_{\mathrm{DM}}) d n/d \ln M$ for ALP DM with several values of the misalignment angle $\theta_i = \phi_i/f_a$. We set $m_a = 10^{-18} \mathrm{eV}$ using both $p=-1/2$ (left) and $p=1/2$ (right). The mass function is enhanced on intermediate masses compared to the CDM mass function, shown as the black line, due to the enhancement of fluctuations. At small masses the halo mass function is correspondingly suppressed compared to CDM.}
	\label{fig:HMF}
\end{figure}

We then compute the variance of the filtered field $\sigma_{R}(a)$, where in this work we always use the spherical top-hat window function. The halo mass function today is then calculated using \eqref{PS_number_density}. It is shown in \figref{fig:HMF}, where we plot the function $X_M\equiv(M/\rho_{\mathrm{DM}}) dn/d\ln M = dF/d\ln M$ for $m_a = 10^{-18}\mathrm{eV}$, both $p=-1/2$ and $p=1/2$, and the same values of the misalignment angle as in \figref{fig:MPS}. The black line again corresponds to the CDM halo mass function today. 

\subsection{Halo spectrum}
\label{sec:halo_spectrum}

While $\xi=0$ is a useful condition to identify the collapse time from~\eqref{eom_spherical_shell_3}, the condition itself is not physical. Instead of collapsing into a singularity, the overdensity is expected to virialize into a halo, in which the potential and the kinetic energies are settled to a stable equilibrium~\cite{Press:1973iz, Kolb:1994fi,Zentner:2006vw,Ellis_Marsh:2020gtq}. In this section we discuss the properties of such halos.

The radius of the virialized halo is half of  the turnaround radius, $r_{\rm vir} = r_{\rm ta}/2$, which follows directly from the virial theorem and energy conservation. Hence the final virial density $\rho_{\rm vir}$ is eight times the density at turnaround \cite{Kolb:1990vq}. $\rho_{\rm vir}$ can also be expressed in terms of the background energy density of ALPs. Remembering that the total mass in the halo stays constant,
\beq
M=\frac{4\pi}{3}\rho_{\eq}a_{\eq}^3(1+\delta_{\ini})R^3=\frac{4\pi}{3}\rho_{\rm ta}r_{\rm ta}^3=\frac{4\pi}{3}\rho_{\rm ta}a_{\rm ta}^3\xi^3_{\rm ta}R^3,
\eeq
the virial density can be expressed as
\begin{equation}
\rho_{\vir}= 8\rho_{\rm ta}= 8C_{\rho} \rho_{\eq}(1+\delta_{\ini})\delta_{\ini}^3,
\end{equation}
where we defined $C_{\rho}=1/(\xi_{\rm ta}^3C_x^3)$ and used $C_x\equiv x_{\rm ta}\delta_{\ini}$ from above. Numerically we find that $C_\rho\approx 17$. Inserting this into the above equation we arrive at~\cite{Kolb:1994fi,Ellis_Marsh:2020gtq}
\begin{equation}\label{virial_density}
\rho_{\rm vir}\approx136\rho_{\eq}(1+\delta_{\ini})\delta_{\ini}^3\,.
\end{equation}
Using $\delta_\ini\approx 1.1/x_{\rm coll}$ we can estimate the virial density of the halo right after collapse as
\begin{equation}
  \label{eq:449}
  \rho_{\rm vir}(x_{\rm coll})\approx 136\overline{\rho}_{m,\rm EQ}\qty(1+\frac{1.1}{x_{\rm coll}})\frac{1.1^3}{x_{\rm coll}^3}=136 \overline{\rho}_{\rm coll}1.1^3\qty(1+\frac{1.1}{x_{\rm coll}}).
\end{equation}
For small fluctuations collapse happens during matter domination, so the $x_{\rm coll}^{-1}$ term can be neglected. Then
\begin{equation}
  \label{virial_density_matter_smalldelta}
  \rho_{\rm vir}(x_{\rm coll})\approx 136 \overline{\rho}_{\rm coll}1.1^3\approx 181 \overline{\rho}_{\rm coll},\quad \delta_\ini\ll 1,
\end{equation}
For collapse during radiation domination we obtain
\begin{equation}
  \label{eq:451}
  \rho_{\rm vir}(x_{\rm coll})\approx 136\overline{\rho}_{m,\rm EQ}\frac{1.1^4}{x_{\rm coll}^4}\approx 200 \overline{\rho}_{R,\rm EQ}\frac{a_{\rm EQ}^4}{a_{\rm coll}^4}\approx 200\overline{\rho}_{\rm coll},\quad \delta_\ini\gg 1.
\end{equation}
The conclusion is that the virialized density of the halo right after collapse is approximately $200$ times the mean density of the universe at collapse, regardless of the cosmological era in which the collapse happens.

It has been confirmed, by means of $N$-body simulations, that CDM forms halos with a characteristic, red-shift dependent profile, dubbed \textit{Navarro-Frenk-White}-(NFW)-profile. It is given by \cite{Navarro:1995iw}
\begin{equation}\label{eq:NFW-profile_eq}
\rho(r,z)=\frac{4\rho_s}{(r/r_s)(1+r/r_s)^2}\,,
\end{equation}
where $\rho_s\equiv\rho(r_s)$ is the \textit{scale density}, and $r_s$ is the \textit{scale radius} defined by $\left.d\ln\rho(r)/d\ln r\right|_{r_s}=-2$ that determines how strongly the density is concentrated within the halo. The numerical simulations show that $\rho_s$ depends on the red-shift at which the halos forms, but it stays constant after the halo is formed. For the remainder of the section we review an approach to find $\rho_s$ for a given collapsing region.

An equivalent definition of the NFW-profile is
\begin{equation}\label{eq:NFW-profile_eq_2}
\rho(r,z)=\frac{\delta_{\mathrm{char}}\rho_c(z)}{(r/r_s)(1+r/r_s)^2}\,,
\end{equation}
where $\rho_c$ is the critical density of the universe and $\delta_\mathrm{char}$ is the characteristic density. 

The mass inside a certain radius $r$ can be found by integrating \eqref{eq:NFW-profile_eq} up to $r$, which gives \cite{Blinov:2021axd}
\begin{equation}
\label{eq:scale-mass}
M(r)=4\pi\int_0^{r}dr'\,r'^2\rho(r')=16\pi\rho_sr_s^3\tilde{f}(r/r_s)\,,
\end{equation}
with 
\begin{equation}
\label{eq:fx-def}
\tilde{f}(x)=\ln(1+x)-\frac{x}{x+1}\,.
\end{equation}
The mass integrated up to $r_s$ is called the scale mass $M_s$.
We discussed that the density of a virialized halo is roughly 200 times the mean density of the universe at the time of formation. This motivates the definition of $r_{200}$, which is the radius at which the average density of the enclosed halo is 200 times the critical density:
\begin{equation}\label{eq:r_200}
    200\rho_c=\left(\frac{4\pi}{3}r_{200}^3\right)^{-1}4\pi\int_0^{r_{200}}dr\,r^2\rho(r),
\end{equation} 
such that at the time of formation, this corresponds roughly to the virial density. The mass inside this region is defined to be
\begin{equation}
\label{eq:m200-def}
    M_{200}=\frac{4\pi}{3}(200\rho_{\mathrm{crit}})r_{200}^3.
\end{equation}
Since we assume that the mass of the halo is conserved during the collapse, the spherical mass $M$ from \eqref{eq:spherical_mass} can be identified with $M_{200}$, which is constant by definition.

One now defines the concentration parameter $c_{200}$ by
\begin{equation}
    c_{200}\equiv\frac{r_{200}}{r_s}.
\end{equation}

We can use \eqref{eq:r_200} to find the characteristic density as a function of the concentration parameter:
\begin{equation}\label{eq:char_dens}
    \delta_{\mathrm{char}}=\frac{200}{3}\frac{c_{200}^3}{\ln(1+c_{200})-c_{200}/(1+c_{200})}.
\end{equation}
From the definitions, this equation of $\delta_{\mathrm{char}}$ leads to a time independent $\rho_s$. We now use the formalism from~\citres{Navarro:1996gj_concentration1,Ellis_Marsh:2020gtq}  and use these definitions to relate the quantities of the collapsing ALP regions to the empirical results of the scale density of halos formed at different redshifts.

Let us assume we identify a halo at redshift $z_{\mathrm{id}}$, which most of the time will be today, $z_{\mathrm{id}}=0$. We now assign the \textit{formation redshift} $z_\form(M_{200},\mathfrak{f})$ which is defined as the redshift at which half the mass of the halo was first contained in progenitors more massive than
some fraction $\mathfrak{f}<1$ of the mass $M_{200}$.\footnote{We note that $z_\form$ is distinct from $z_\coll$ discussed in the spherical collapse model.} We can find this with the Press-Schechter formalism by
\begin{equation}\label{eq:z_form}
    \mathrm{erfc}\left[X(z_\form) - X(z_{\mathrm{id}})\right]=\frac{1}{2},
\end{equation}
with
\begin{equation}
    X(z)=\frac{\delta_c(z)}{\sqrt{2\left[\sigma^2(\mathfrak{f}M_{200},z)-\sigma^2(M_{200},z)\right]}},
\end{equation}
where $\delta_c(z)$ is the red-shift dependent critical density for collapse, and $\sigma^2(M_{200},z)$ is the variance at redshift $z$ evaluated with a spherical tophat as window function for the smoothing scale
\begin{equation}
    R=\left(\frac{3M_{200}}{4\pi\rho_{m,0}}\right)^{1/3},
\end{equation}
assuming here that the ALPs make up all of dark matter and ignoring the baryonic contribution to $\rho_m$. 

The connection to the numerical simulations is now made by a redshift dependent proportionality factor $\mathfrak{C}$ which relates the critical density of the universe at formation to the scale density:
\begin{equation}
    \rho_s=\mathfrak{C}(z_\form)\rho_c(z_\form).
\end{equation}
To summarise this section: We use the variances found in \secref{sec:the_spectra} and plug them into \eqref{eq:z_form}, choose $\mathfrak{f}=10^{-2}$, to find $z_\form$ for given smoothing scale, which sets the halo mass. We use the fit function from the appendix C of \citre{Eroncel:2022efc} to find $\mathfrak{C}$ for given $z_\form$. The relation $\rho_s=\delta_{\mathrm{char}}(z)\rho_c(z)$ immediately gives the characteristic density today, leading to the concentration parameter $c_{200}$. Since we know $r_{200}$ from the identification of $M_{200}$ with the mass of the halo, we find $r_s$ and $M_s$ from $c_{200}$. Scale mass $M_s$ and $\rho_s$ are the two parameters we are looking for in the halo spectrum and which we are going to use to find observational prospects in the next chapter. 

\bigskip

\begin{figure}[t!]
	\centering
	\includegraphics[width=0.49\textwidth]{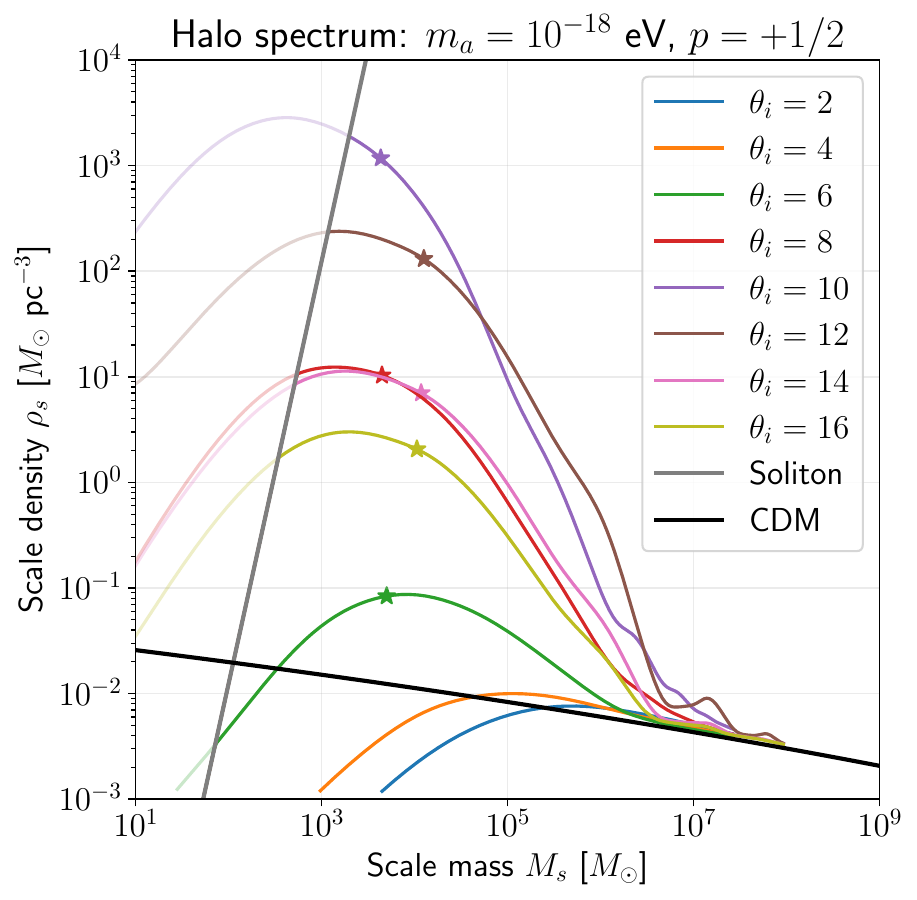}
	\includegraphics[width=0.49\textwidth]{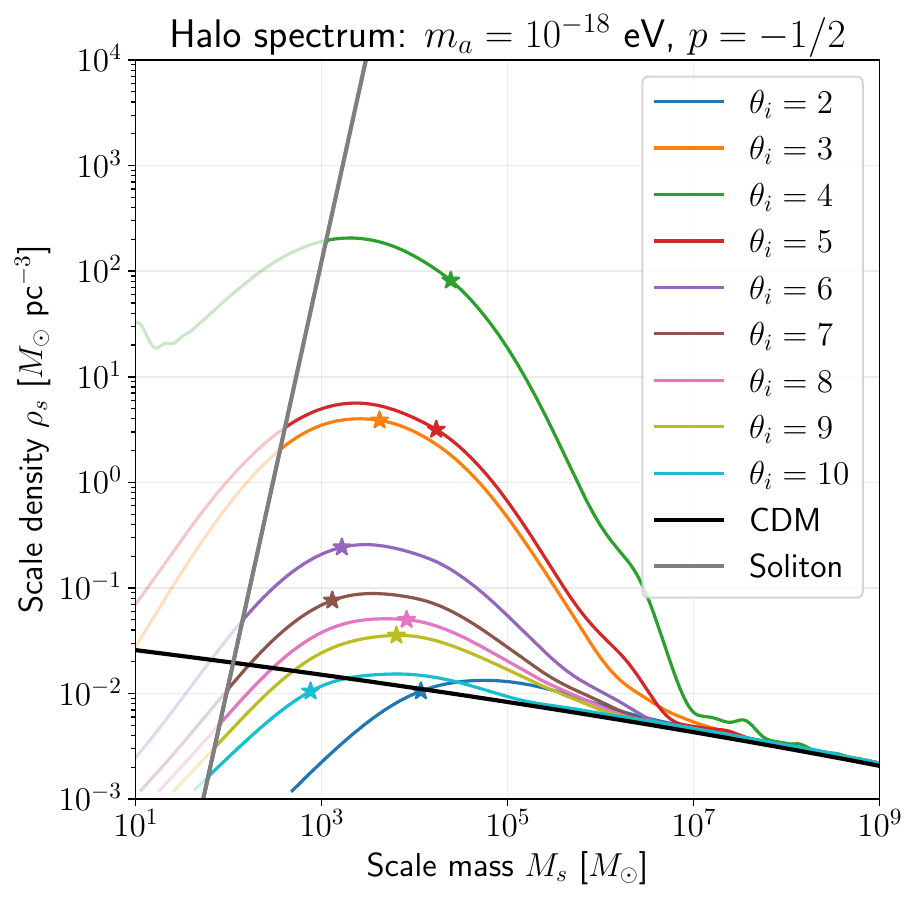}
	\caption{Halo spectra, showing the scale density $\rho_s$ as function of the scale mass $M_s$, for axion-like particles with mass $m_a=10^{-18}\eV$ and potential \eqref{eq:monodromy_potential} with $p=1/2$ (left) and $p=-1/2$ (right). We compare different initial field values $\theta_i$. The variance $\sigma_R$ has been calculated with the spherical window function. The soliton line from \eqref{eq:soliton-line} for this mass is shown as straight grey line. This line acts as an upper bound for the scale density rendering densities above the soliton line as unphysical. This means the actual halo spectrum will follow the line that corresponds to the minimum of soliton bound and the calculated scale densities. We show the parts of the calculation of the scale density that exceed the soliton line as transparent. The stars indicate where the halo mass function, shown in \figref{fig:HMF}, peaks. Also shown is the CDM-line in black.}
	\label{fig:rho_vir_M_iso_sigmatilde_0.5_m_1e-18}
\end{figure}

\textbf{Solitons}: The NFW profile fails to describe the halos formed by ALP dark matter at small scales due to the wavelike nature of the ALP. It has been shown numerically that the central profiles of ALP halos do not obey a ``cuspy'' $r^{-1}$ profile \cite{Schive:2014hza}, but form a \textit{(pseudo-) soliton}, being the eigenstate of the time-independent Schr\"odinger-Poisson equation. A fit for the soliton profile is given by \cite{Schive:2014dra}:
\begin{equation}
    \rho_\mathrm{sol}(r)=\rho_{\mathrm{sol},0}\left[1+(2^{1/8}-1)\left(\frac{r}{r_{1/2}}\right)^2\right]^{-8},
\end{equation}
where 
\begin{equation}
    \rho_{\mathrm{sol},0}\approx1.9\times\left(\frac{10^{-17}\eV}{m_a}\right)^2\left(\frac{\mathrm{pc}}{r_{1/2}}\right)^4\,M_\odot\mathrm{pc}^{-3}
\end{equation}
is the core density and $r_{1/2}$ is the radius at which the density is half of the core density. From this form of the profile, one can calculate, how the scale mass and scale density of the solitons are related:
\begin{equation}
    \label{eq:soliton-line}
    \rho^{\mathrm{sol}}_s\approx1.2\times10^{-4}\left(\frac{m_a}{10^{-17}\eV}\right)^6\left(\frac{M_s^{\mathrm{sol}}}{M_\odot}\right)^4M_\odot\mathrm{pc}^{-3}.
\end{equation}
Far away from the center of the halo, the profile is expected to still follow the NFW-profile. We will neglect a possible interpolation between the soliton core at small scales and the NFW-profile further out. Therefore, we are going to make the assumption that for a given scale mass $M_s$ the scale density of solitons $\rho_{\mathrm{sol}}(m_a,M_s)$ is an upper bound on the scale density and that denser halos cannot form for that mass. Halo spectra for $p=1/2$ and $p=-1/2$ at a benchmark ALP mass and for different initial field values can be seen in \figref{fig:rho_vir_M_iso_sigmatilde_0.5_m_1e-18}.

\section{Observational prospects from gravitational interactions}
\label{sec:obs_prospects}

Now that we have described the ALP halo spectra for a certain mass and given initial conditions, we want to proceed by investigating in which part of the parameter space in the $(m_a,f_a)$-plane gravitational interactions could actually lead to observational signatures of the ALP halos. 

When the halos form, they are much denser than their surrounding. This remains true when they become part of larger dark matter halos. However, if, at later times, they enter galaxies, much denser objects like stars will cause tidal forces on the halos which could potentially disrupt them. Since part of the observational signatures rely on the fact that the halos are inside of the Milky Way or other galaxies, we need to make sure that they actually survive until today. For halos smaller than the typical star separation, $r_s<1\,\mathrm{pc}$, passing a star could actually disrupt the halo completely, depending on its density. For larger halos, interaction with single stars cannot disrupt the halo completely, but here one has to take into account the tidal forces induced by the gravitational potential of the Galaxy. To understand all of the halo population, in a full analysis one would need to find the sub-halo mass function within the Galaxy, as has been done e.g. in \citres{Lee:2021zqw,Shen:2022ltx}. However, we are only interested if the densest halos survive. Hence, without going into details we take the estimate from \citre{Arvanitaki:2019rax} that a halo smaller than $1\,\mathrm{pc}$ must have $\rho_s>10^3\rho_\odot$, with $\rho_\odot\equiv1.1\times10^{-2}M_{\odot}\mathrm{pc}^{-3}$ being the local dark matter density. Halos much larger than $1\,\mathrm{pc}$ only need to be slightly denser than the local average, so we take $\rho_s>11\rho_\odot$. Both of these estimates are rather conservative. For $r_s\approx1\,\mathrm{pc}$, we would actually need to interpolate between the two regimes, but this goes beyond the scope of our work. We show the parameter space in which halos inside the Galaxy likely did not survive until today as light grey area in \figref{fig:observational_prospects_03}.

\subsection{Signatures of ALP minihalos}
\label{ssec:signatures}

There are several halo phenomena that  could lead to observable effects. Halos could act as gravitational lenses or they could affect the motion of stars when they pass them on their movement through the Galactic halo. In the following we discuss the different effects these gravitational interactions could cause \cite{Arvanitaki:2019rax}.

\bigskip

\textbf{Local Gravitational Perturbations:} One distinguishes three different cases of a similar phenomenon: When dark matter subhalos traverse the Galaxy, they can change the 6D phase space distribution of stars in the Milky Way disk \cite{Feldmann:2013hqa_disk_detection} or of stars in the stellar halo \cite{Buschmann:2017ams_stellar_halo_detection}. The third effect that is heavily investigated is the disruption of \textit{stellar streams} by dark matter subhalos \cite{Ibata:2001iv_stellar_streams_detection1,Carlberg:2011xj_stellars_streams_detection2,Johnston:2001wh_stellar_streams_detection3}. These streams are the result of tidal disruption of globular clusters (spherical collections of stars) in the Milky Way halo, resulting in narrow leading and trailing arms of stars, since the mass loss in the clusters is deposited into orbits with slightly higher and lower orbital energies \cite{Bovy:2015mda_stellars_streams_explanation, Johnston:1997fv}. These stellar streams have been experimentally confirmed for example in \citres{SDSS:2000zid_stellar_streams_experiment1, Grillmair:2006bd_stellars_streams_experiment2}.

\begin{figure}[t!]
	\centering
	\includegraphics[width=\textwidth]{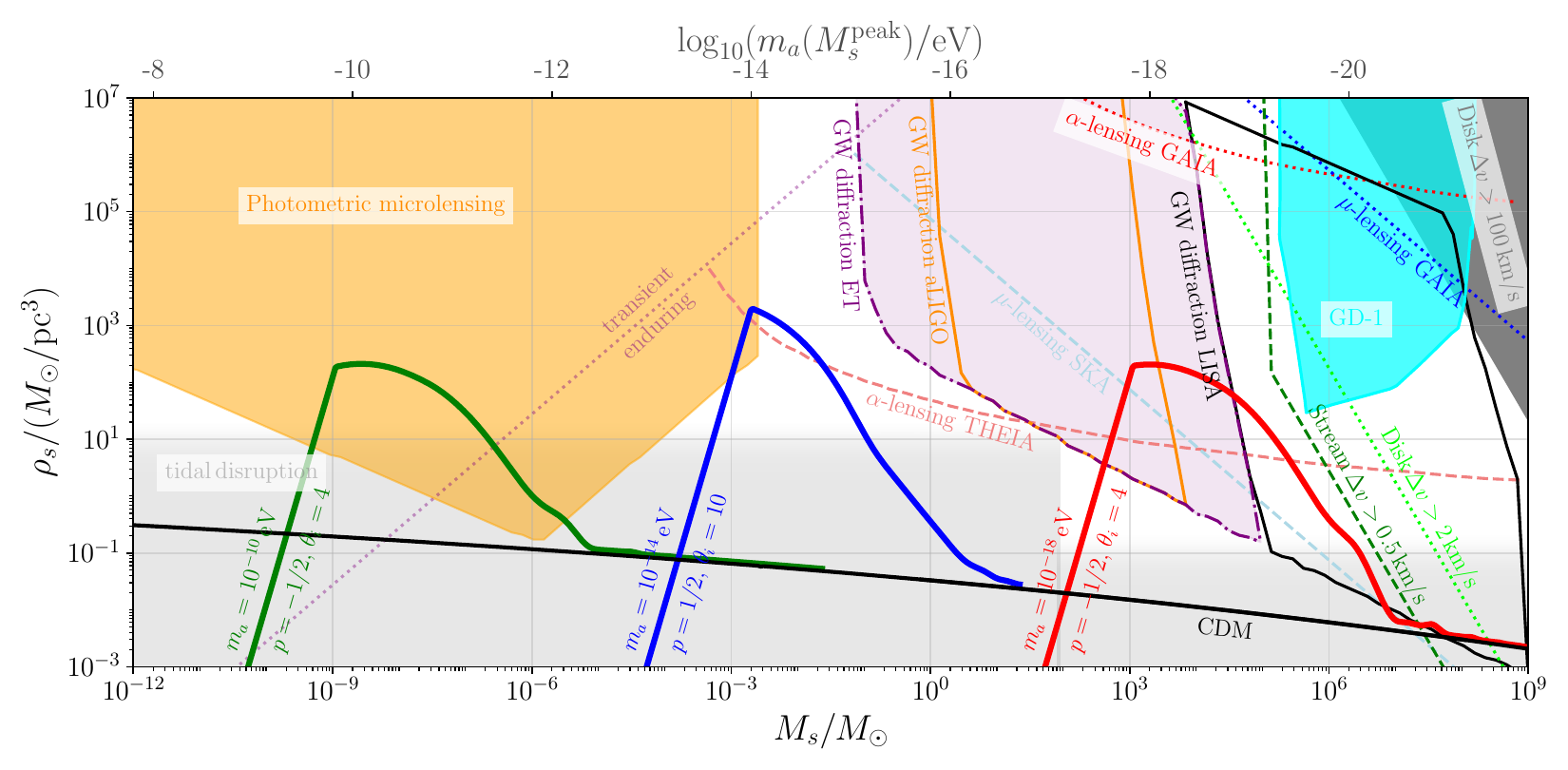}
	\caption{The thick lines show three of the halo spectra with the densest characteristic halos, for $p=-1/2$, $\theta_i=4$ and a mass of  $m_a=10^{-18}\,\mathrm{eV}$ in red, and for a mass of $m_a=10^{-10}\,\mathrm{eV}$ in green, and for $p=1/2$, for a mass of $m_a=10^{-14}\,\mathrm{eV}$ and $\theta_i=10$ in blue. The CDM line is given in black. The light grey area depicts the region in which halos in the Galaxy likely have been tidally disrupted until today.
 We also show the following parameter regions in which halos could be observable by their gravitational interactions: \textbf{Local Gravitational Perturbations:} The dashed green (dotted lime-green) line depicts the parameter space above which halos could significantly change the velocity of stars in stellar streams (the Galaxy's disk), while the grey area shows where dark matter halos could lead to ejection of stars from the disk or even the Galactic halo. The cyan area shows the best-fit parameter space that could explain features in the GD-1 stream.  \textbf{Astrometric Weak Gravitational Lensing:} The dotted purple line shows the boundary between transient effects (to the left) and enduring effects (to the right), where the following phenomena can only test the enduring regime. Above the dotted (dashed) blue line the test statistic $\mathcal{T}_{\mu}$ predicts that local effects in the angular velocity $\mu$ of stars due to lensing effects are observable by GAIA (SKA), while above the dotted (dashed) red lines, corresponding to the global test statistic $\mathcal{C}_{\alpha}$, correlations of the angular acceleration $\alpha$ due to enduring gravitational lensing  should be observable by GAIA (THEIA). \textbf{Photometric Microlensing:} The orange region shows where photometric microlensing, leading to irregularities in the light curve of highly magnified stars that transit a caustic curve, may be observable. \textbf{Diffraction of Gravitational Waves:} The region in which we expect distortions in the gravitational wave signals from BH-BH mergers caused by dark matter halos that could be observable by aLIGO (LISA) [ET] is shown by the dark-orange line (black line) [purple region with dashed-dotted boundary]. On the top we show the corresponding ALP mass $m_a$ that produces a halo spectrum peaking at given $M_s$, see \eqref{eq:ms-peak-fit}.}
	\label{fig:observational_prospects_03}
\end{figure}

\Citre{Feldmann:2013hqa_disk_detection} gives as an approximate formula for the velocity kick a star gets from the passing of the halo:
\begin{equation}\label{velocity_kick}
\Delta\vec{v}\approx-\hat{b}\,0.5\,\mathrm{km~s}^{-1}\left(\frac{M(b)}{10^7M_{\odot}}\right)\left(\frac{166\,\mathrm{km~s}^{-1}}{V}\right)\left(\frac{\mathrm{kpc}}{b}\right)\,,
\end{equation}
where $V$ is the relative velocity between the subhalo and the star we will assume to be $166\,\mathrm{km\,s}^{-1}$, $b$ is the impact distance, and $M(b)$ is the mass of the subhalo that is enclosed within that impact distance. Such a velocity kick could be observed by the space telescope GAIA, whose on-going mission is set to make the most detailed and largest map of stars in the Milky Way to date \cite{Gaia:2018ydn}. We follow \citre{Arvanitaki:2019rax} and calculate the parameter region in the $(M_s,\rho_s)$-plane where the halo can give stars in the Galactic disk velocity kicks that are larger than $2\,\mathrm{km~s}^{-1}$, which we present as dotted light green line in \figref{fig:observational_prospects_03}, while the region where the stars could be ejected from the disk, or even the Galaxy, i.e.~$\Delta v\gtrsim 100\, \mathrm{km\,s}^{-1}$, is shown as grey area in the upper right corner. To find these lines, we take the minimum velocity kick $\Delta\vec{v}$ we want to achieve, insert the given mass of the halo into \eqref{velocity_kick} as $M(b)$ and then find the corresponding $b$ that still allows for the velocity kick we want to achieve. This velocity kick will only be possible if all the mass of the halo \textit{actually} is contained inside the impact distance $b$, so we impose that the scale radius must fulfil $r_s<b/3$ which for given mass automatically determines the minimum density $\rho_s$ such that all of the halo lies withing the impact distance. We also present the parameter space for which the halos could generate velocity kicks large enough to disrupt a very cold stream, resulting in change in speed of stars in the stream of $\Delta v\gtrsim 0.5\, \mathrm{km\,s}^{-1}$ for impact distances and scale radii with $\max\{b, r_s\} \gtrsim 10\,\mathrm{pc}$. This area is shown as dashed dark-green line in \figref{fig:observational_prospects_03}.

One promising stellar stream, which shows characteristic features (including a gap and an off-stream spur of stars) that could be explained by an encounter with a dense dark matter subhalo is the stream GD-1. \Citre{Bonaca:2018fek_GD1} claims that such an encounter is the best explanation, since encounters with known globular clusters, dwarf galaxies or molecular clouds in the Milky Way disk are unlikely to explain the observations. The densities of the dark matter halos needed to explain the features are in a mass range of about $10^6\,M_{\odot}$ to $10^8\,M_{\odot}$ and would need to be more dense than the CDM halo spectrum predicts. However, the halos predicted by ALPs with non-period potential could fall into the region that could explain the features in the GD-1 stream. We copied this region from \citre{Bonaca:2018fek_GD1} and show it as cyan area in \figref{fig:observational_prospects_03}.

We note that these observational prospects do not assume any halo mass function or value of energy density in halos of certain mass. For these observations, encounters with one halo would suffice, while of course a small number density would diminish the possibility to actually observe these velocity kicks or make the encounter of a halo with the GD-1 stream less likely. The results from the halo mass function (see \secref{sec:Press-Schechter}) imply that the number density of dense halos is large enough to make the possibility of these encounters significant.

\bigskip

\textbf{Astrometric Weak Gravitational Lensing:} If the compact halos cross the line-of-sight between Earth and stars or other luminous objects, they can induce apparent motion of the latter through gravitational lensing, without the need for multiple images or magnification \cite{Arvanitaki:2019rax}. This effect, called astrometric weak lensing, was first considered in \citre{Boden:1998me_astrometry1,Belokurov:2001vh_astrometry2,Erickcek:2010fc_astrometry3,Li:2012qha_astrometry4} for point-like objects and cuspy minihalos. We use the results of \citre{VanTilburg:2018ykj_astrometry_program} who proposed a program of searches for compact subhalos using the astrometric possibilities of the space telescope GAIA \cite{Gaia:2018ydn}, working since 2013, of the \textit{Square Kilometer Array (SKA)} \cite{SKA:2018ckk}, a radio interferometer whose construction has started in 2022, and the proposed space telescope \textit{Telescope for Habitable Exoplanets and Interstellar/Intergalactic Astronomy (THEIA)} \cite{Malbet:2021rgr,Theia:2017xtk_Theia}.

We have to distinguish two categories of astrometric lensing signatures: If the minimum impact parameter between the lensed object and the line-of-sight towards the lens is smaller than the change in impact parameter within a typical multi-year astrometric survey, we expect \textit{transient} effects, while \textit{enduring} effects correspond to the case where the change of the impact parameter is small during the observation time.
This means in the transient regime we would expect to actually see a change in a gravitational lensing event, while in the enduring regime, the change of the lensing happens so slow that the instantaneous angular deflection is in practice unobservable since the survey does not record the true position of the sources. 
In \citre{VanTilburg:2018ykj_astrometry_program}, the local velocity dispersion of dark matter $v=166\,\mathrm{km\,s}^{-1}$ is picked as a typical rate of change in the impact parameter. Together with an observation time $\tau$ of roughly 5 years, this results in $v\tau\sim\mathcal{O}(10^{3}\mathrm{pc})$ as the size of the impact parameter that marks the boundary between the transient and enduring regime. Since usually the deflection angle is maximised for impact parameters near the scale radius, i.e.~$b\sim r_s$, a subhalo can produce a gravitational lensing transient only if \cite{Arvanitaki:2019rax}
\begin{equation}
\rho_s\approx2\times10^{10}\rho^{\odot}_{\mathrm{DM}}\left(\frac{M_s}{M_{\odot}}\right)\left(\frac{10^{-3}\,\mathrm{pc}}{v\tau}\right)^{3}\,,
\end{equation}
where again $\rho^{\odot}_{\mathrm{DM}}\approx1.1\cdot10^{-2}\,M_{\odot}/\mathrm{pc}^3$ is the assumed local dark matter density in the Solar System. The line that marks the boundary between these two regions, with transient (enduring) effects possible for smaller (larger) masses at fixed density is shown in purple in \figref{fig:observational_prospects_03}, where we took $v\tau=2\times10^{-3}$.

Contrary to what one would maybe expect at first sight, it seems more promising to observe enduring effect than transient effects \cite{Arvanitaki:2019rax}. Since the density of a halo, for given mass, has to be very large to produce a transient event, while we do not expect extraordinarily large densities for smaller halos, the expected angular deflection would be too small to be detectable for current astrometric observatories.
As already mentioned, in the enduring regime the angular deflection is in principle unobservable. Nevertheless, \citre{VanTilburg:2018ykj_astrometry_program} studied, how the apparent angular velocity $\mu$ and angular acceleration $\alpha$ could hint to dark matter halos. We present two of their core results regarding dark matter halos: Stellar location and velocity data can be tested for consistence with the existence of a lens. This is done by assuming the velocity of a star in the absence of a lens $\dot{\vec{\theta}}$ is a Gaussian variable with known variance, while the local, multi-source test statistic $\mathcal{T}_\mu$, which takes stellar position and velocity data as input, is taken to be the sum over all stars in the dataset of the scalar product of $\dot{\vec{\theta}}$ with $\vec{\mu}$, where $\vec{\mu}$ is a template for the predicted angular velocity in presence of a dark matter halo acting as lens. The predicted sensibility, i.e.~the maximal signal to noise ratio, in \citre{VanTilburg:2018ykj_astrometry_program} is
\begin{equation}\label{SNR_mu_lensing}
\left<\max\mathrm{SNR}_{\mathcal{T}_{\mu}}\right>\approx0.4\,f^{1/3}\left[\frac{M_s}{10^7M_{\odot}}\right]^{2/3}\left[\frac{10\,\mathrm{pc}}{r_s}\right]\left[\frac{N_0}{10^7}\right]^{1/2}\left[\frac{0.01}{\Delta\Omega}\right]^{1/6}\left[\frac{200\mu\mathrm{as\,y}^{-1}}{\sigma_{\mu,\eff}}\right],
\end{equation}
where, here as well as for the following observational prospects, it is assumed that the halo mass function is a delta peak around $M_s$. This means we can write 
\begin{equation}\label{eq:monochromatic_X_M}
    X_M(M)=fM\delta(M-M_s),
\end{equation} where $f$ is the fraction of all dark matter that is assumed to be in these halos. This is called the monochromatic approximation. 
 
$\Delta\Omega$ is the solid angle covered by the survey, $N_0$ is the  product of the angular number density of stars and $\Delta\Omega$, and $\sigma_{\mu,\eff}$ is the average angular velocity noise, i.e.~the effective standard error for the velocities  \cite{VanTilburg:2018ykj_astrometry_program}. 
For the GAIA observations of the Large and Small Magellanic Clouds we followed \citre{VanTilburg:2018ykj_astrometry_program} and took $N_0=10^7$, $\Delta\Omega=0.01$ and $\sigma_{\mu,\eff}=200\mu\mathrm{as\,y}^{-1}$, for quasar observations with the SKA we took $N_0=10^8$, $\Delta\Omega=4\pi$ and $\sigma_{\mu,\eff}=1\mu\mathrm{as\,y}^{-1}$.
The projected curves for halo parameters where the resulting signal-to-noise ratio is larger than unity, i.e.~$ \left<\max\mathrm{SNR}\tau_{\mu}\right>>1$, for all densities above the curve, where we took $f=0.3$, are shown in \figref{fig:observational_prospects_03} in blue, where the dotted line shows the result for the GAIA observations while the dashed line shows the projected region for the SKA (or comparable radio observatories). \Eqref{SNR_mu_lensing} tells us that the smallest observable density scales as $\rho_s\propto f^{-1}$.

The global, multi-sourced test statistic $\mathcal{C}_\alpha$ measures small-angle correlations in angular accelerations that gravitational lensing induces on a field of stellar motions \cite{VanTilburg:2018ykj_astrometry_program}. The results for this are shown in \figref{fig:observational_prospects_03} in red, where the dotted line is a projection for GAIA observations of stars in the Galaxy, while the dashed line is a projection for THEIA observations of stars. Since the calculation of the curves is rather involved, we graphically copied the results given in \citre{VanTilburg:2018ykj_astrometry_program} and note that also here the result scales with $\rho_s\propto f^{-1}$ \cite{VanTilburg:2018ykj_astrometry_program,Arvanitaki:2019rax}.

\bigskip

\textbf{Diffraction of Gravitational Waves:} If the gravitational wave signal of a merger of two black holes \cite{LIGOScientific:2016aoc_first_gravitational_wave} passes through or near a dark matter halo on its way to Earth, the signature of the lensing may allow for another probe of the halo mass distribution. This would be possible even if the lens is not strong enough to produce multiple images of the same merger, which would then be detected as temporally distinct events. Instead, the lens would imprint characteristic distortions in the waveform as well as in the amplitude of the gravitational wave signal \cite{Dai:2018enj_GW_diffraction1}.

The dimensionless parameter $\omega$ characterises the strength of the distortions, and it is given as a function of frequency and mass enclosed in the lens by \cite{Arvanitaki:2019rax}
\begin{equation}
\omega\cong1.3(1+z_L)\left(\frac{f_{\mathrm{GW}}}{10^2\,\mathrm{Hz}}\right)\left(\frac{M_{\mathrm{enc}}}{100\,M_{\odot}}\right)\,,
\end{equation}
where $z_L$ is the redshift of the lens, $f_{\mathrm{GW}}$ is the frequency of the gravitational wave, and $M_{\mathrm{enc}}$ is the enclosed mass within the impact parameter of the lens. The equation is valid if the effective lens distance is of $\mathcal{O}(\mathrm{Gpc})$ and if the effective velocity dispersion is of $\mathcal{O}(\mathrm{km\,s}^{-1})$. \Citre{Dai:2018enj_GW_diffraction1} explains how \textit{advanced LIGO} (\textit{aLIGO}) \cite{LIGOScientific:2014pky} will be able to probe black hole mergers with $\omega\sim\mathcal{O}(1)$ out to distances of a Gigaparsec. Since aLIGO operates in frequencies of $(10-10^3)\,\mathrm{Hz}$, this makes it possible to detect halos of masses $\sim\mathcal{O}(10-10^3)\,M_{\odot}.$ The probability that a BH-BH merger is passing behind a halo with impact parameter $b$ is roughly given as \cite{Dai:2018enj_GW_diffraction1,Arvanitaki:2019rax}:
\begin{equation}\label{eq:prob_gw_diffraction}
P(b)\sim0.045f\left[\frac{1+z_L}{2}\right]^3\left[\frac{D_{\mathrm{BH}}}{5\mathrm{\,Gpc}}\right]\left[\frac{10^5M_{\mathrm{\odot}}}{M_s}\right]\left[\frac{b}{1\mathrm{\,pc}}\right]^2,
\end{equation}
where $D_{\mathrm{BH}}$ is the distance to the merger. To find observable regions, we followed \citre{Arvanitaki:2019rax} and set $D_{\mathrm{BH}}=5\mathrm{\,Gpc}$ and took the redshift to the lens to be $z_L=0.3$. We then found for every scale mass the corresponding $b_{\mathrm{min}}$ at which $P(b)$ is $1\%$. If the resulting $b_{\mathrm{min}}$ was smaller as the so-called NFW smoothing scale $2\pi/(2m_av_s)$ with $v_s=\sqrt{16\pi\ln(2)G\rho_sr_s^2}$, then this smoothing scale was taken as $b_{\mathrm{min}}$ instead. It was then checked if there is a $b>b_{\mathrm{min}}$ such that the mass enclosed in a cylinder within the impact parameter $b$ leads to a distortion in the range $0.5<\omega<5$ for some frequency in the range of aLIGO, $(10-10^3)\,$Hz, the planned space interferometer LISA \cite{LISA:2017pwj}, $(10^{-4}-10^{-1})\,$Hz or the proposed Einstein Telescope (ET) \cite{Punturo:2010zz_ET}, $(1-10^4)\,$Hz. The last condition was that the $b$ we found was not larger as ten times the so-called Einstein radius of the lens, which was taken to be $r_E=0.1\mathrm{\,pc}\left[M_{\mathrm{enc}}/(100M_{\odot})\right]$ \cite{Dai:2018enj_GW_diffraction1}. We show the regions  where this calculation predicts observable effects for aLIGO, LISA and ET in \figref{fig:observational_prospects_03} as dark-orange, black and purple dashed-dotted line, respectively. 
We note that while this prediction depends on $f$, again using the monochromatic approximation \eqref{eq:monochromatic_X_M}, a decrease in $f$ makes denser halos unobservable first, because while they induce strong lensing, the probability that they cross our line of sight to a binary gets smaller, i.e. the smaller $f$ in \eqref{eq:prob_gw_diffraction} can be compensated by a larger $b$. The low density cutoff for the observable region is instead given by the fact that for halos with small density, $b$ will be larger than the Einstein radius.
\bigskip

\textbf{Photometric Microlensing:} Photometric lensing, gravitational lensing events that change the apparent brightness of a celestial object, has been proposed in \citre{Paczynski:1985jf_photometry_proposal}. How such a phenomenon could be observable with the help of stars in the Milky Way, the Magellanic Clouds or Andromeda in the case where the lensing objects are MACHO's, e.g. planetary size objects or primordial black holes, has been investigated in \citres{MACHO:2000qbb_photometry_dense1,EROS-2:2006ryy_photometry_dense2,Niikura:2017zjd_photometry_dense3,Griest:2013aaa_photometry_dense4,Zumalacarregui:2017qqd_photometry_dense5}. At small sizes the halos we discuss here are not dense enough to have the same effects as the dense objects discussed in the references above. \Citre{Dai:2019lud_photometric_halos} discusses, how such halos could nevertheless be observed via \textit{photometric microlensing}. The general idea is the following: At specific distances behind a gravitational lens there are locations of highest possible magnification. The connection of all these locations is called caustic curve, while the projection of such a caustic to the sky is called the critical curve. Recently, observations of stars highly magnified to around a factor of $\mathcal{O}(10^3)$ by this mechanism found near a caustic curve have been reported, e.g. in \citre{Kelly:2017fps}. In these cases the lenses are clusters of galaxies and the highly magnified star is expected to be close to the critical curve of the lensing cluster. \Citre{Dai:2019lud_photometric_halos} proposes to look for small variations to the magnifying event to find the footprint of dark matter minihalos within the lensing cluster. Single stars inside the cluster lens can disrupt the cluster's critical curve into a network of micro-critical curves (projections of micro-caustics) on angular scales of $\sim 10\,\mu\mathrm{as}$, while dark matter minihalos with their sub-planetary masses and solar system sizes are expected to produce even finer magnification variations scales of $\sim 10^2 - 10^3\,\mathrm{nas}$ \cite{Dai:2019lud_photometric_halos}. In such a microlensing event, when a highly magnified star crosses a micro-caustic, the observed light curve would be altered by surface density fluctuations caused by dark matter minihalos inside the cluster. \Citre{Dai:2019lud_photometric_halos} predicts that such an event could be monitored through observations with the most powerful optical and infrared telescopes based in space or on Earth. 

We follow \citre{Dai:2019lud_photometric_halos} and \citre{Blinov:2021axd} to find the parameters of halos that could be observable by this effect. 
The density power spectrum of halos on non-linear scales can be found from
\begin{equation}
    P_{\rho}=\overline{\rho}\int\frac{dM}{M^2}\frac{dF}{d\ln M}\left|\tilde{\rho}^h(q;M)\right|^2,
\end{equation} 
where $\overline{\rho}$ is the averaged density in some region of the cluster and $\tilde{\rho}^h(q;M)$ is the Fourier transform of the halo profile. One now assumes that slices of the lensing cluster all have the same fractional mass distribution and only vary by their averaged density, with which the power spectrum of the surface density becomes an integral along the line of sight:
\begin{equation}
    P_{\Sigma}(q_\perp)=\overline{\Sigma}\int\frac{dM}{M^2}\frac{dF}{d\ln M}\left|\tilde{\rho}^h(q_\perp;M)\right|^2,
\end{equation}
where $q_{\perp}$ is the surface Fourier mode and $\overline{\Sigma}=\int dL\overline{\rho}(L)$ is the mean surface density. It is assumed that inside the cluster it is dominated by the cluster's surface density, i.e.~$\overline{\Sigma}=\overline{\Sigma}_{\cl}$. 

It is convenient to define the lensing convergence
\begin{equation}
    \kappa=\frac{\Sigma}{\Sigma_{\crit}},
\end{equation}
with $\Sigma_{\crit}=c^2/(4\pi G D_{\eff})$ being the critical surface density for an isolated lens, at which it produces multiple images, where we find $D_{\eff}=D_LD_{LS}/D_S$ with $D_L$, $D_{LS}$ and $D_S$, which are the distances to the lens, to the source and from source to lens, respectively. 

The dimensionless convergence power spectrum is now given by
\begin{equation}\label{dimensionless_convergence_power_spectrum}
    \Delta_{\kappa}^2=\frac{q^2P_{\Sigma}}{2\pi\Sigma_{\crit}^2}.
\end{equation}

With the monochromatic approximation, \eqref{eq:monochromatic_X_M}, and using the NFW-profile, \eqref{dimensionless_convergence_power_spectrum} can be written as
\begin{equation}\label{eq:Delta_k_final}
    \Delta_k=\frac{1}{\ln(2/\sqrt{e})}\sqrt{\frac{\Sigma_{\cl}fM_s}{2\pi}}\frac{1}{\Sigma_\crit r_s}\abs{\int_0^\infty d\tilde{r}\sin(qr_s\tilde{r})\frac{1}{(1+\tilde{r})^2}},
\end{equation}
where we have defined $\tilde{r}\equiv r/r_s$ in the Fourier transform of the NFW-profile. The integral in the above equation is a function of $qr_s$ and has a maximum of $\approx0.35$ at $qr_s=0.77$. 
We will now assume that $D_{\eff}=1\,\mathrm{Gpc}$, set $\Sigma_{\cl}=0.8\Sigma_{\crit}$ and introduce the distance the star seems to traverse during the microlensing event $d$, which is given by the product of relative velocity between cluster and star, the event duration and the magnification. As a benchmark value, $d$ is set to $10^3\,\mathrm{AU}$. One assumes that the following four conditions must be fulfilled for possible observations of the halos \cite{Blinov:2021axd}: Firstly, the maximum value of the convergence power spectrum must fulfil $\mathrm{max}\Delta_{\kappa}(q)>10^{-3}$, so we could observe a brightness fluctuation of $\mathcal{O}(1)$ if the star of interest is magnified by a factor of $10^{3}$. Secondly, to ensure statistical treatment, the distance the star seems to traverse should sweep over many halos, i.e.~$f\pi d^2\Sigma_{\cl}/M_s>10$, which leads to a sharp cutoff at a certain mass scale. The third criterion is to assume that the clumps should be smaller than $\sim d$, so the density fluctuations induced during the lensing event are actually significant, so we implement $r_s<2d$. Finally, as fourth criterion we use that the size of the density fluctuations (which is $2\pi/q$) leading to $\Delta_k>10^{-3}$ should not be smaller than $10\,\mathrm{AU}$, so that the fluctuations in the light curve are not washed out by the finite size of the lensed star. To implement this, we evaluate \eqref{eq:Delta_k_final} at $2\pi/q=\max(10\,\mathrm{AU},2\pi r_s/0.77)$. 

We show the region at which this leads to detectable effects of such photometric lensing events for $f=0.3$ in \figref{fig:observational_prospects_03} as orange area. 

\bigskip

\textbf{Observing ALP halos with Pulsar Timing Arrays:}
Pulsar Timing Arrays (PTAs), like NANOGrav \cite{Demorest:2009ex_NANOGrav}, PPTA \cite{Kerr:2020qdo_PPTA} or EPTA \cite{Kramer:2013kea_EPTA} recently became an exciting possibility to test gravitational physics. One observes a set of pulsars, which generally have an extremely stable rotation, making them very accurate astrophysical clocks. Correlation between time residuals, the difference between expected and measured arrival time of the signal, can be analysed to find hints of astrophysical or cosmological gravitational waves, with the possibility that a stochastic GW background has already been observed \cite{NANOGrav:2020bcs}. 

The passing of dark matter halos could be inferred from analysing the pulsar timing data, with a halo/halos passing through the line of sight causing a Shapiro delay, or halos accelerating Earth or the pulsar directly, causing a Doppler effect. The analysis of PTA data for hints of DM halos relies on searching for the expected signal with a filter, where the latter is only easily found for events involving a single halo and a single pulsar (called deterministic), while the stochastic analysis involving the whole halo population and a large number of pulsars is less well understood. This leads to the fact that studies like \citres{Dror:2019twh_transient_regime,Ramani:2020hdo, Lee:2020wfn,Lee:2021zqw} give different results regarding which halo populations could be observable. The observational prospects depend e.g. on the number of observed pulsars, the absolute observation time, how often every single pulsar is observed, the halo mass function of the halos as well their concentration parameters, while the predicted observable regions disagree additionally due to different filters and statistical methods. A simplified analysis with a monochromatic approach for the halo mass function does not suffice for the stochastic scenario, while observation with current and near-future PTAs of single halos is unrealistic for the model we investigate. Therefore we do not show any observational prospects for PTAs and postpone a dedicated analysis of the observational prospects for ALP halos to future work.

\bigskip

{\textbf{Gravitational Waves from ALP fragmentation}:
Axion fragmentation produces a stochastic background of gravitational waves of primordial origin with a peak frequency controlled by the axion mass \cite{Chatrchyan:2020pzh,Madge:2021abk,Eroncel:2022vjg,Kitajima:2018zco}. 
However, the signal is generally suppressed and unobservable by future gravitational-wave observatories when imposing the upper bounds from either the axion dark matter abundance or the axion dark radiation, except at very low frequencies observable by CMB experiments. We will present a quantitative analysis in an upcoming publication.

\subsection{Parameter regions of observable gravitational effects}
\label{ssec:obs_regions_parameter_space}

\begin{figure}[t!]
    \centering
    \includegraphics[width=\textwidth]{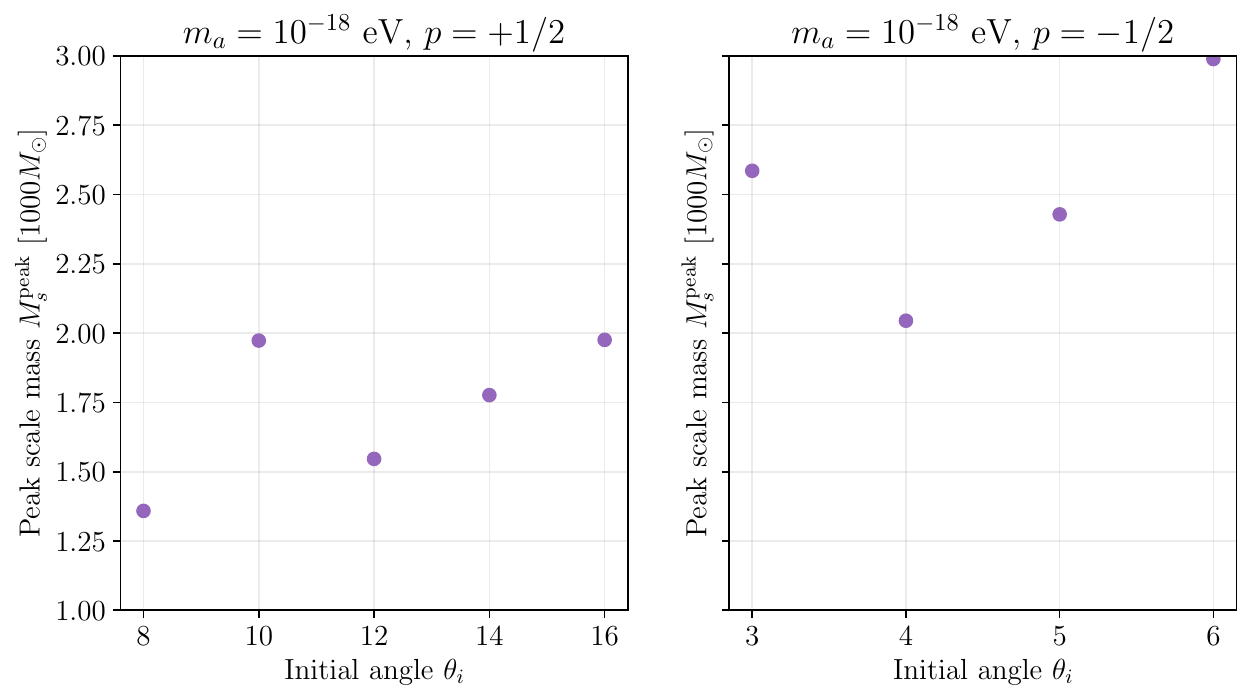}
    \caption{The peak values of the scale mass as a function of the initial angle $\theta_i$ for $m_a=10^{-18}\,\si{\electronvolt}$ obtained by calculating the halo spectrum as described in \secref{sec:halo_spectrum}.}
    \label{fig:ms-peak}
\end{figure}

Our aim is to show in which region of the $(m_a,f_a)$-parameter space the corresponding halos can cause observable effects.. Ideally, this requires calculating the halo spectrum for each point in the $(m_a,f_a)$-parameter space, which is not numerically feasible, especially since we need to run lattice simulations for a sizeable region. Instead, we decided to extrapolate our results for $m_a=10^{-18}\,\si{\electronvolt}$ using a technique which we will now explain. 

We need a simple way to obtain $(M_s^{\rm peak},\rho_s^{\rm peak})$ as a function of model parameters $(m_a,f_a,p)$. In \figref{fig:ms-peak} we show the values of the scale mass at which the halo spectrum gets peaked. We only show the benchmarks at which $\rho_s^{\rm peak}>10^{-1}M_{\odot}\textrm{pc}^{-3}$ since they are relevant for observations. We see that the halo spectra for both $p=1/2$ and $p=-1/2$ are peaked at the scale mass $M_s\sim 2\times 10^3 M_{\odot}$. To find the scaling of the scale mass with respect to the ALP mass $m_a$ we can use the following relation for $M_s$ 
\begin{equation}
    M_s=\frac{f(1)}{f(c_{200})}\frac{4\pi}{3}\rho_{m,0}a_0^3 R^3,
\end{equation}
which can be obtained by combining \eqref{eq:scale-mass} and \eqref{eq:m200-def}. The dark matter density today $\rho_{m,0}a_0^3$ is the same for all benchmark points. We can also assume that $f(c_{200})$ at the peak of the halo spectrum is not sensitive to the model parameters. Then we can write
\begin{equation}
    M_s^{\rm peak}\propto R_{\rm peak}^3\propto k_{\rm peak}^{-3},
\end{equation}
where $k_{\rm peak}$ is the comoving momentum around which structure is enhanced. We expect that the enhancement occurs around the same dimensionless momenta
\begin{equation}
    \tilde{k}_{\rm peak}=\frac{k_{\rm{peak}}/a}{\sqrt{2 m_a H}}
\end{equation}
in all of the parameter space. Since $\tilde{k}$ is conserved during radiation era, we have $\tilde{k}_{\rm peak}\propto k_{\rm peak}/\sqrt{m_a}$ which results in 
\begin{equation}
    M_s^{\rm peak}\propto k_{\rm peak}^{-3}\propto m_a^{-3/2}.
\end{equation}
So we can write the following relation for the scale mass:
\begin{equation}
    \label{eq:ms-peak-fit}
    M_s^{\rm peak}\approx 2\times 10^{3} M_{\odot}\qty(\frac{m_a}{10^{-18}\,\si{\electronvolt}})^{-3/2}.
\end{equation}
A similar relation has also been found in \citre{Arvanitaki:2019rax}.

\setlength{\arrayrulewidth}{0.5mm}
\setlength{\tabcolsep}{18pt}
\renewcommand{\arraystretch}{1.5}
\begin{table}[h!]
  \centering
  \begin{tabular}{| c | c | c |}
    \hline
    & $p=1/2$ & $p=-1/2$ \\
    \hline
    $\alpha_{<}$ & $-6.22$ & $-7.18$ \\
    \hline
    $\alpha_{>}$ & $34.6$ & $10.2$ \\
    \hline
    $\beta_{<}$ & $12.7$ & $15.6$ \\
    \hline
    $\beta_{>}$ & $-26.8$ & $-9.20$ \\
    \hline
    $\gamma$ & $8.02$ & $4.67$ \\
    \hline
  \end{tabular}
  \caption{Best fit parameters of the fit function \eqref{eq:rhos-fit} for the peak scale density. We fixed the ALP mass to be $m_a=10^{-18}\,\si{\electronvolt}$.}
  \label{tab:best-fit}
\end{table}

\begin{figure}[t!]
    \centering
    \includegraphics[width=\textwidth]{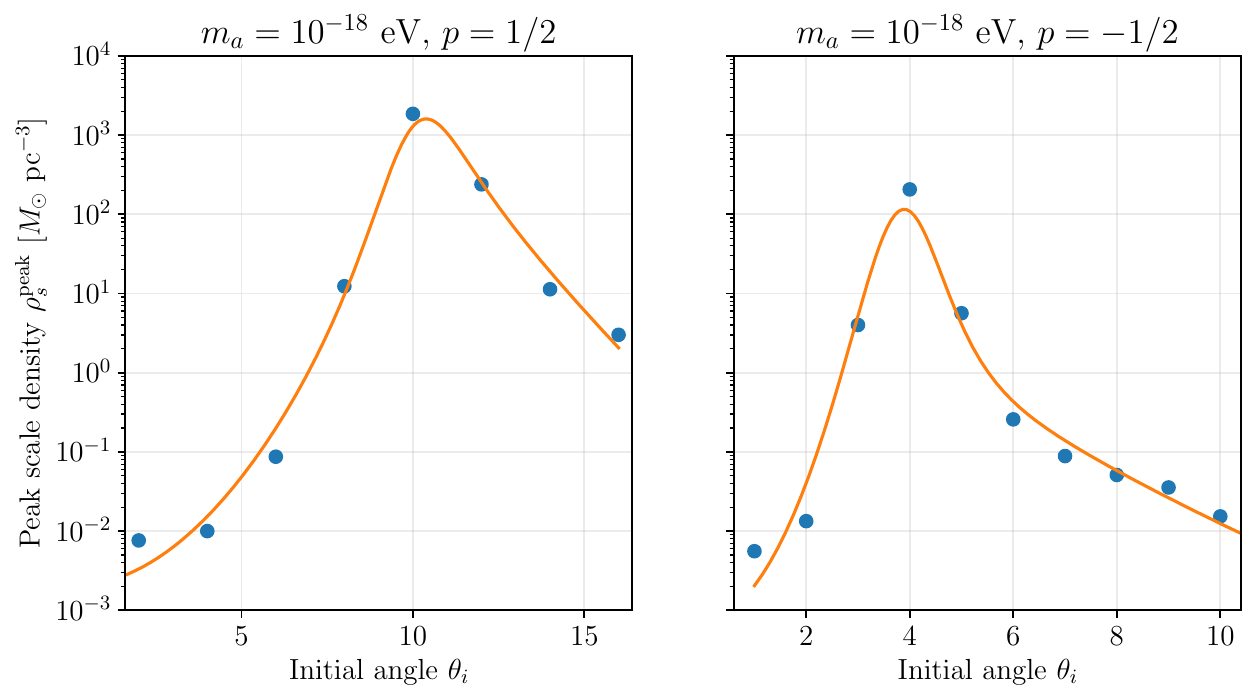}
    \caption{Comparison of the fit function for the peak scale density \eqref{eq:rhos-fit} (solid lines) with the values obtained by the full numerical calculation (circles). The parameters of the fit function are tabulated in Table \ref{tab:best-fit}.}
    \label{fig:rhos-peak}
\end{figure}

For the peak scale density, we found that the following fit function provides a reasonable approximation:
\begin{equation}
\label{eq:rhos-fit}
    \ln \rho_s^{\rm peak}(\theta_i)=\qty(\alpha_{<}+\beta_{<}\,\tilde{\theta}_i^2)h\qty(\gamma(1-\tilde{\theta}_i))+\qty(\alpha_{>}+\beta_{>}\,\sqrt{\tilde{\theta}_i})h\qty(\gamma(\tilde{\theta}_i - 1)),
\end{equation}
where
\begin{equation}
    h(x)=\frac{1}{2}+\frac{1}{2}\tanh{x}\qand \tilde{\theta}_i\equiv \frac{\theta_i}{\theta_i^{\rm peak}}.
\end{equation}
Here $\theta_i^{\rm peak}$ is the value of the initial angle for which the halo spectrum has the strongest peak. We found $\theta_i^{\rm peak}\approx 10$ and $\theta_i^{\rm peak}\approx 4$ for $p=1/2$ and $p=-1/2$, respectively. The best-fit values for the remaining parameters are given in table \ref{tab:best-fit}. A comparison between the numerically obtained peak values and this fit function is given in \figref{fig:rhos-peak}.

\begin{figure}[h!]
	\centering
	\includegraphics[width=0.82\textwidth]{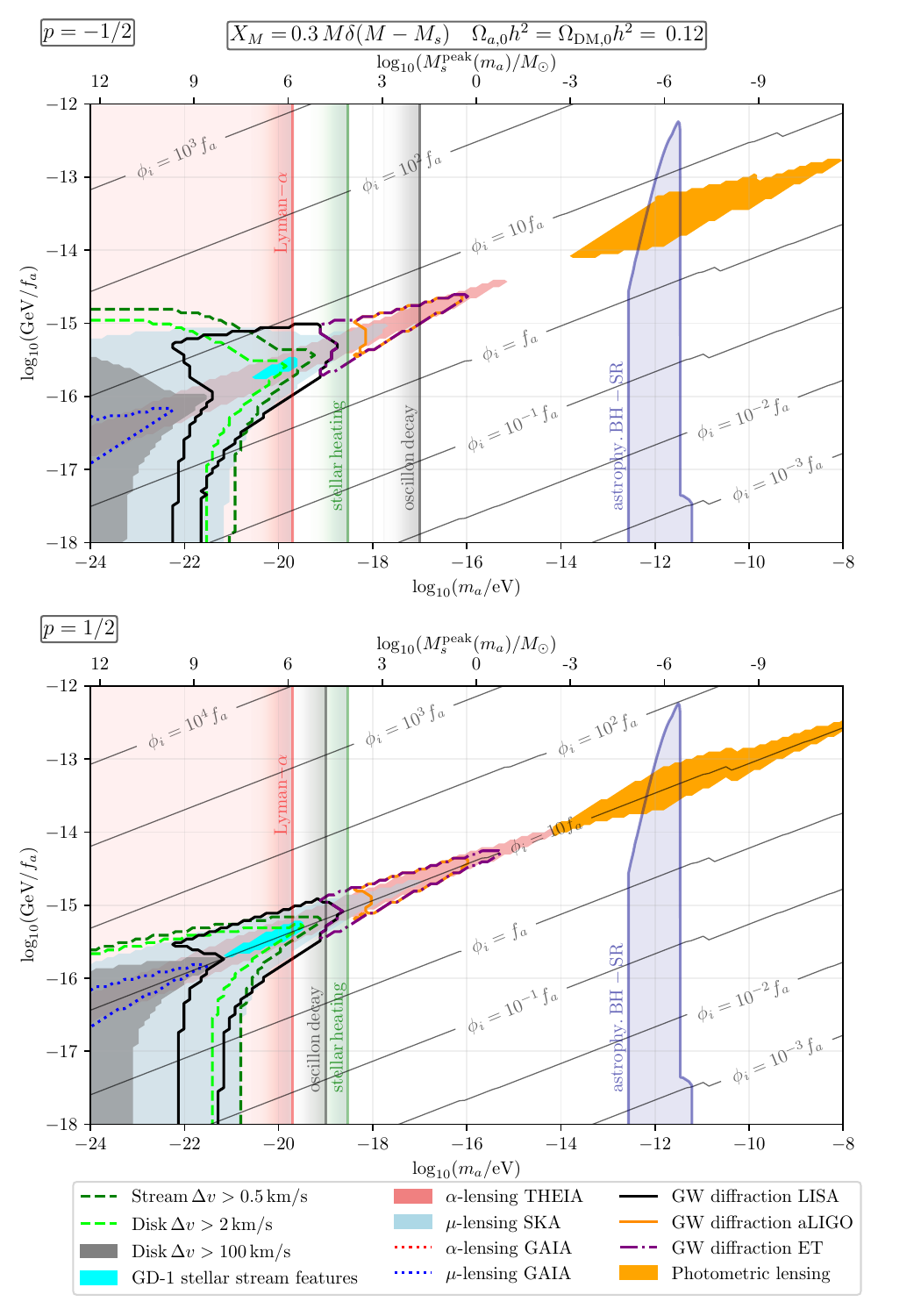}
	\caption{Regions in the $(m_a,f_a)$-plane in which the ALP dark matter can lead to observable gravitational effects. The corresponding initial field values $\theta_i$ are shown by the grey contour lines.
These projections assume that 30\% of all of dark matter is inside of characteristic halos with masses within one decade around $M_s$. The upper plot shows $p=-1/2$, while the lower plot shows $p=1/2$. The labelled regions show the constraint on the minimum mass from the Lyman-$\alpha$ forest (for $p=1$), the minimum mass at which oscillons decay before matter-radiation equality, the constraint from heating of stellar orbits and the constraint from (astrophysical) black hole superradiance.}
	\label{fig:observation_m_f_plane_XM_03}
\end{figure}

\begin{figure}[h!]
	\centering
	\includegraphics[width=0.82\textwidth]{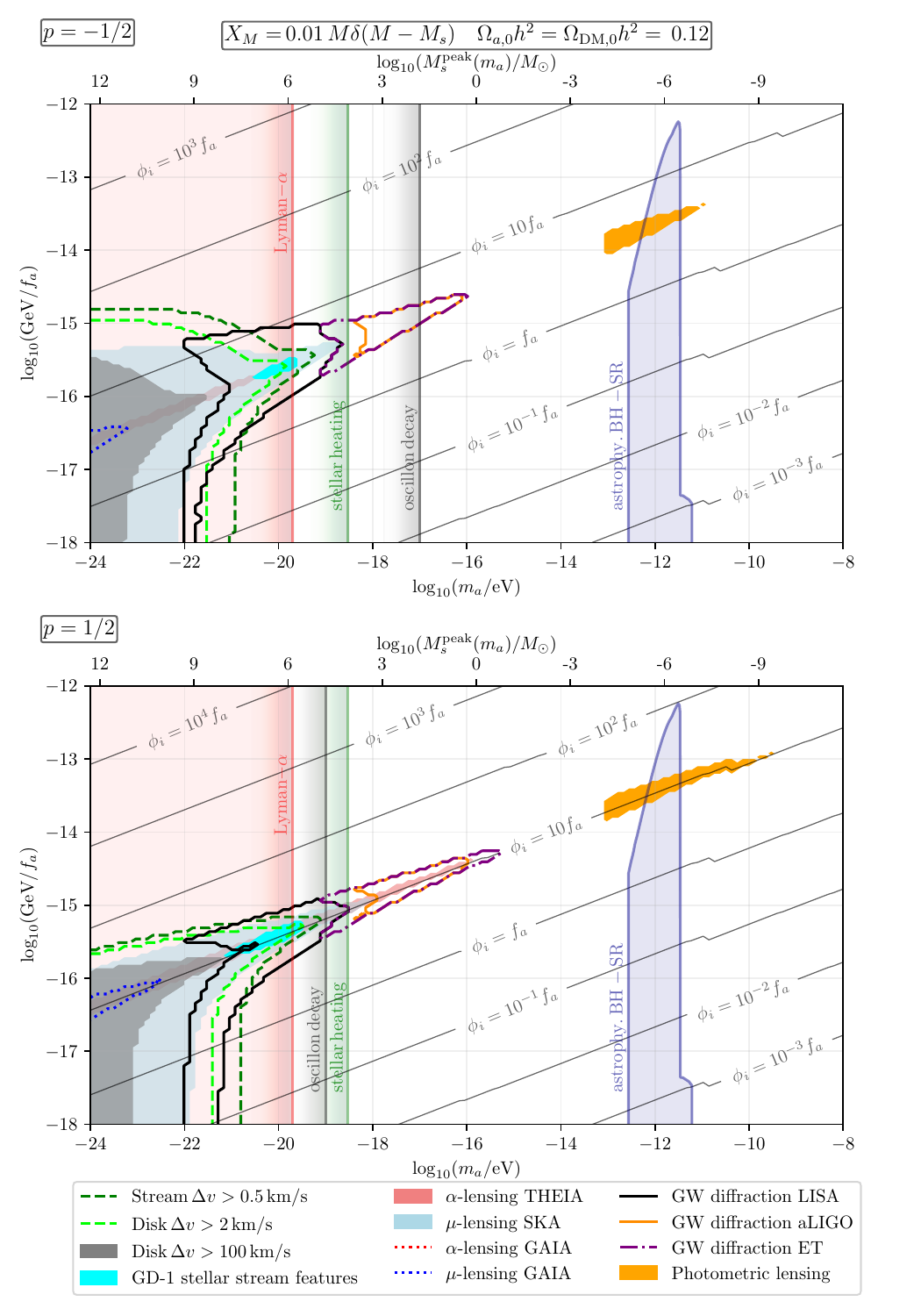}
	\caption{Same as \figref{fig:observation_m_f_plane_XM_03}, but using that 1\% of all dark matter is inside characteristic halos with masses within one decade around $M_s$.}
	\label{fig:observation_m_f_plane_XM_001}
\end{figure}

We can now combine the last two subsections by starting with a given tuple of $m_a$ and $f_a$. When we assume that $\Omega_{a,0}=\Omega_{\mathrm{DM},0}$, the initial tuple ($m_a,f_a$) fixes the initial misalignment $\theta_i=\phi_i/f_a$ to achieve the correct relic abundance by \eqref{eq:relic_density_monodromy} and with this pair of $m_a$ and $\theta_i$, we can find the approximate peak mass and peak density of the characteristic halo via \eqref{eq:ms-peak-fit} and \eqref{eq:rhos-fit}. The parameter space in which we expect observable gravitational effects, as they have been discussed in the previous subsections and presented in \figref{fig:observational_prospects_03}, from these halos can be seen in \figref{fig:observation_m_f_plane_XM_03} for $X_M=0.3M\delta(M-M_s)$ and in \figref{fig:observation_m_f_plane_XM_001} for $X_M=0.01M\delta(M-M_s)$. 
Since the axion-like-particle we investigate could be only gravitationally coupled to the SM, we consider the few constrains that depend only on their self-interactions and coupling to gravity. We added a line to show the currently strongest constraint from the Lyman-$\alpha$ forest at $2\times10^{-20}\,\mathrm{eV}$ \cite{Rogers:2020ltq_ULDM_bounds}\footnote{See \citres{Armengaud:2017nkf,Irsic:2017yje,Kobayashi:2017jcf,Nori:2018pka} for earlier works on constraining the fuzzy dark matter mass via Lyman-$\alpha$ observations.}, for which we point out that it has been derived for ALPs with harmonic potential and could be modified for other potentials~\cite{Leong:2018opi}, but still prospects for smaller masses should be taken with a grain of salt. We also show the constraints from superradiance around astrophysical black holes. In this process, dense clouds of fundamental scalars accumulate around black holes, extracting the latter's rotational energies. For this process to be efficient and the clouds to be long-lived, the scalar's Compton wavelength has to be roughly the size of the black hole horizon. Therefore, observing fast spinning black holes at certain masses excludes the corresponding mass range of the axion-like-particle. Because strong self-interaction quench the cloud, superradiance can only constrain large $f_a$. We show the excluded region found in \citre{Baryakhtar:2020gao}. We also show the mass at which oscillons decay before matter-radiation equality, \eqref{eq:osc_decay_before_eq}, meaning that for lower axion masses the presence of oscillons could influence the evolution of the halos, which we did not include in our analysis. In a recent work, \citre{Dalal:2022rmp} claims that a fuzzy dark matter mass smaller than $3\times10^{-19}\,\mathrm{eV}$ is in conflict with the observed stellar velocity dispersion inside the ultra-faint dwarf galaxies Segue 1 and Segue 2. The main idea is that the wave interference effects produce large fluctuations in the local density and gravitational potential inside the fuzzy dark matter halos. These fluctuations cause gravitational heating of stellar systems~\cite{Hui:2016ltb} and increase the velocity dispersion of the stars. We indicate this bound in our plots with a green band with the label "stellar heating".

As can be seen, the ALP halos typically lead to observable effects in a region of the parameter space where $\theta_i$ is in the range $1$-$100$. For these values of the field, also the change in the equation of state due to the transfer of energy into fluctuations is of $\mathcal{O}(1)$, as it was studied in \citre{Berges:2019dgr} for a similar potential. This justifies neglecting this $\mathcal{O}(1)$-correction in this work.

\section{Comparison between axion fragmentation models}\label{sec:comparison}

The non-periodicity of the potentials we investigated in this work is not a requirement to form ALP miniclusters. Instead, the crucial condition is the delay in the onset of oscillations compared to the common expectation $H_{\osc}\sim m_a$.

In this section, we compare our results with the halo predictions from the Large Misalignment Mechanism (LMM) \cite{Arvanitaki:2019rax} and the Kinetic Misalignment Mechanism (KMM) \cite{Chang:2019tvx,Co:2019jts} 
analyzed in 
\citres{Eroncel:2022vjg,Eroncel:2022efc}. 

We already discussed the LMM in \secref{sec:relic_density_LMM}. Now we shall give a brief review of ALP dark matter in KMM. See \citre{Eroncel:2022vjg} for more details. In KMM, the ALP field is assumed to have an initial kinetic energy parameterised by the "yield" parameter defined by
\begin{equation}
    Y=\frac{f_a \dot{\phi}(T)}{s(T)},
\end{equation}
where $\dot{\phi}(T)$ and $s(T)$ are the ALP velocity and the entropy of the universe at temperature $T$, respectively. At early times, the ALP field is dominated by its kinetic energy so its energy density scales as $\rho_{\phi}\propto a^{-6}$. This implies that $\dot{\phi}\propto a^{-3}$, so the yield parameter is constant if the entropy is conserved. The large initial kinetic energy delays the onset of oscillations from its conventional value $m_a\sim H_{\rm osc}$. Instead, the oscillations start when the kinetic energy drops below the potential energy, i.e.
\begin{equation}
    \frac{1}{2}\dot{\phi}^2(T_{\ast})=2 f_a^2 m_a^2,
\end{equation}
where $T_{\ast}$ is defined by this equation and is called the \emph{trapping temperature}. For a given $m_a$ and $f_a$, the parameters $Y$ and $T_{\ast}$ can be fixed such that the ALP field gives the desired amount of dark matter. In particular, the yield is fixed by
\begin{equation}
    h^2 \Omega_{\phi,0}\approx h^2 \Omega_{\rm DM,0}\,\qty(\frac{m_a}{5\times 10^{-3}\,\si{\electronvolt}})\qty(\frac{Y}{40}),
\end{equation}
where the trapping temperature is given by the relation
\begin{equation}
    \frac{T_{\ast}}{\Lambda_b}=(2\times 10^8)^{1/3}\qty(\frac{g_s(T_{\ast})}{72})^{-1/3}\qty(\frac{\Lambda_b}{\si{\giga\electronvolt}})^{1/3}\qty(\frac{h^2 \Omega_{\phi,0}}{h^2 \Omega_{\rm DM}})^{-1/3},
\end{equation}
where $\Lambda_b=\sqrt{m_a f_a}$. The delay of the onset of oscillations is quantified by $m/H_{\ast}$ where $H_{\ast}=H(T_{\ast})$. The Kinetic Misalignment only occurs when this quantity is larger than the conventional value, i.e. when $m/H_{\ast} \gtrsim 3$. The field is completely fragmented if $m_{\ast}/H_{\ast}\gtrsim 40$.

The $a^{-6}$ scaling of the ALP energy density in Kinetic Misalignment can be problematic if the trapping happens very late. In this case, the ALP field will have too much energy during Big Bang Nucleosynthesis (BBN) so that the constraints from observations of primordial helium-4 and deuterium abundances are violated. This constraint is given by
\begin{equation}
    \label{eq:bbn-bound}
    T_{\ast}\gtrsim 20\,\si{\kilo\electronvolt}\qor \Lambda_b\gtrsim 9\times 10^{-7}\times \qty(\frac{h^2 \Omega_{\phi,0}}{h^2 \Omega_{\rm DM}})^{1/2}.
\end{equation}
Thus, in contrast to an ALP with non-periodic potential or the Large Misalignment Mechanism, the Kinetic Misalignment is incompatible with small ALP masses such as $m_a\lesssim 10^{-16}\,\si{\electronvolt}$.

\setlength{\arrayrulewidth}{0.5mm}
\setlength{\tabcolsep}{8pt}
\renewcommand{\arraystretch}{1.2}
\begin{table}[t]
  \centering
  \begin{tabular}{| p{10em} | c | c | c |}
    \hline
    \textbf{Model} & \textbf{Line style} & $\mathbf{\it f_a}$ \textbf{[GeV]} & \textbf{Model specific}\\
    \hline
    \multirow{3}{10em}{Non-periodic Potential $p=+1/2$ \textcolor{tab:blue}{\textbf{(blue)}}}
     & solid & $9\times 10^{13}$ & $\theta_i=10$\\
     & dashed & $1\times 10^{14}$ & $\theta_i=8$\\
     & dotted & $7\times 10^{13}$ & $\theta_i=12$\\
    \hline
    \multirow{3}{10em}{Non-Periodic Potential $p=-1/2$ \textcolor{tab:purple}{\textbf{(purple)}}}
     & solid & $9\times 10^{13}$ & $\theta_i=5$\\
     & dashed & $1\times 10^{14}$ & $\theta_i=4$\\
     & dotted & $7\times 10^{13}$ & $\theta_i=6$\\
    \hline
    \multirow{3}{10em}{Kinetic Misalignment \textcolor{tab:red}{\textbf{(red)}}}
     & solid & $9\times 10^{13}$ & $m/H_{\ast}\approx 20$\\
     & dashed & $1\times 10^{14}$ & $m/H_{\ast}\approx 12$\\
     & dotted & $7\times 10^{13}$ & $m/H_{\ast}\approx 27$\\
    \hline
    \multirow{3}{10em}{Large Misalignment \textcolor{tab:green}{\textbf{(green)}}}
     & solid & $9\times 10^{13}$ & $\abs{\pi-\theta_i}\approx 1\times 10^{-4}$\\
     & dashed & $1\times 10^{14}$ & $\abs{\pi-\theta_i}\approx 6\times 10^{-3}$\\
     & dotted & $7\times 10^{13}$ & $\abs{\pi-\theta_i}\approx 2\times 10^{-6}$\\
    \hline
    Post-inflationary \newline \textcolor{goldenrod}{\textbf{(yellow)}} & solid & $2\times 10^{14}$ & \\
    \hline
  \end{tabular}
  \caption{The tabulated list of parameters for the benchmark models in \figref{fig:model-comparison}. Here $\theta_i=\phi_i/f_a$ denotes the initial value of the ALP field, and $H_{\ast}$ is the Hubble at the time when the ALP field gets trapped in the Kinetic Misalignment Mechanism.}
  \label{tab:comparison-table}
\end{table}

For the comparison, shown in \figref{fig:model-comparison}, we fix the ALP mass to $m_a=10^{-14}\,\si{\electronvolt}$ and choose several different values for the decay constant $f_a$. Assuming that the ALP makes up all of dark matter, this fixes the initial field value $\phi_i$ in the misalignment model with the non-periodic potential and the LMM, and the yield and the trapping temperature in the KMM. All the model parameters are also tabulated in table \ref{tab:comparison-table}.

In the post-inflationary scenario, the initial angle $\theta_i$ is not a free parameter anymore. Therefore, for a given ALP mass $m_a$, there is single value for the ALP decay constant $f_a$ such that ALPs make all of dark matter. The relic abundance for this case has recently been calculated in \citre{OHare:2021zrq} as\footnote{This result includes contribution from the decay of topological defects such as strings and domain walls. However, due to the computational limitations they were able to simulate up to $\kappa\approx 8$ where $\kappa\equiv\ln(m_{s}/H)$ is the string tension, and $m_s$ is the mass of the radial mode. In realistic scenarios, $kappa$ is much larger, for example $\kappa\sim 70$ for the QCD axion. How the extrapolation should be done is still an open problem. See \citres{Hindmarsh:2019csc,Gorghetto:2020qws,Hindmarsh:2021vih} for more details.}
\begin{equation}
    h^2\Omega_a \simeq 0.019\qty(\frac{g_{\rho}(T_1)}{70})^{3/4}\qty(\frac{g_s(T_1)}{70})^{-1}\qty(\frac{m_a}{\si{\micro\electronvolt}})^{1/2}\qty(\frac{f_a}{10^{12}\,\si{\giga\electronvolt}})^2,
\end{equation}
where $T_1$ is the temperature such that $m_a=(8/5)H(T_1)$. From this expression one finds that for $m_a=10^{-14}\,\si{\electronvolt}$, $f_a\approx 2\times 10^{14}\,\si{\giga\electronvolt}$ gives the correct dark matter density. The spectrum and the halo mass function for this scenario as shown in \figref{fig:model-comparison} are calculated according to the semi-analytical method of \citre{Enander:2017ogx} which is explained in detail in \citre{Eroncel:2022efc}.

\begin{figure}[t!]
    \centering
    \includegraphics[width=\textwidth]{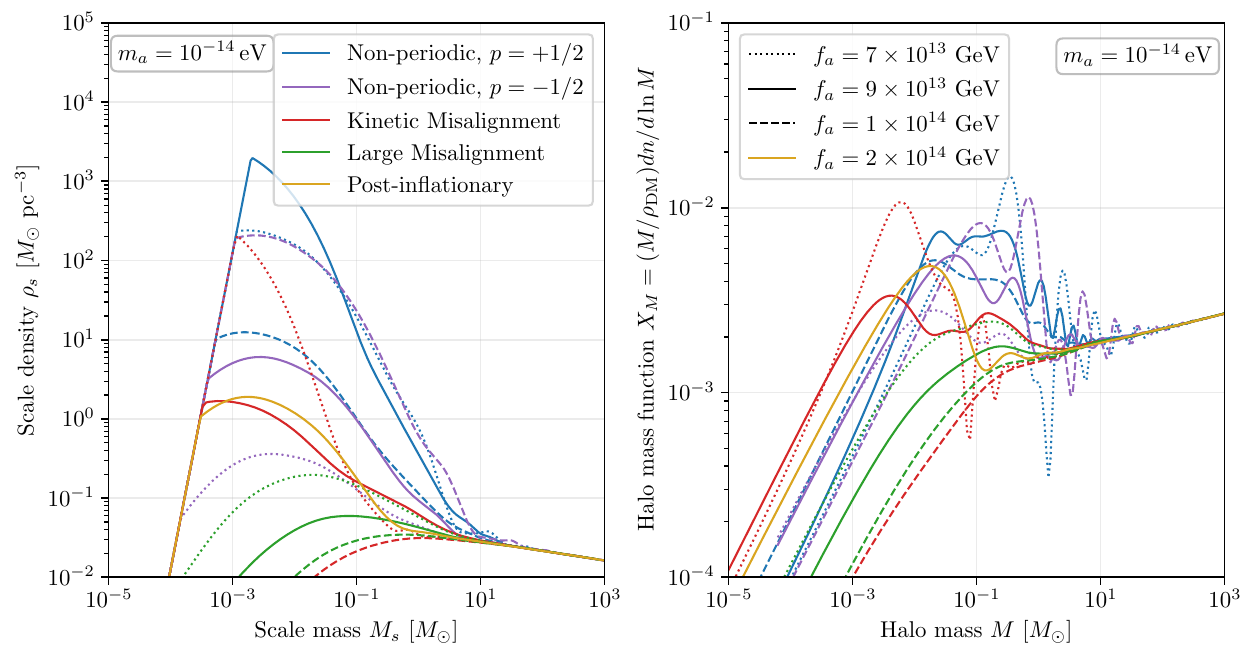}
    \caption{In this figure we compare the halo spectra (left) and the dimensionless halo mass function (right) for different production mechanisms. For all lines the ALP mass is set to $m_a=10^{-14}\,\si{\electronvolt}$. Different colours denote different production mechanisms labelled by the legend on the left plot. Different line styles denote different values of $f_a$ that are labelled by the legend on the right plot. In all the benchmarks, the ALP makes up all of dark matter. We see that the mechanism that predicts denser halos depends on the model parameters, in particular on the value of the ALP decay constant. }
    \label{fig:model-comparison}
\end{figure}

\begin{figure}
    \centering
    \includegraphics[width=0.9\textwidth]{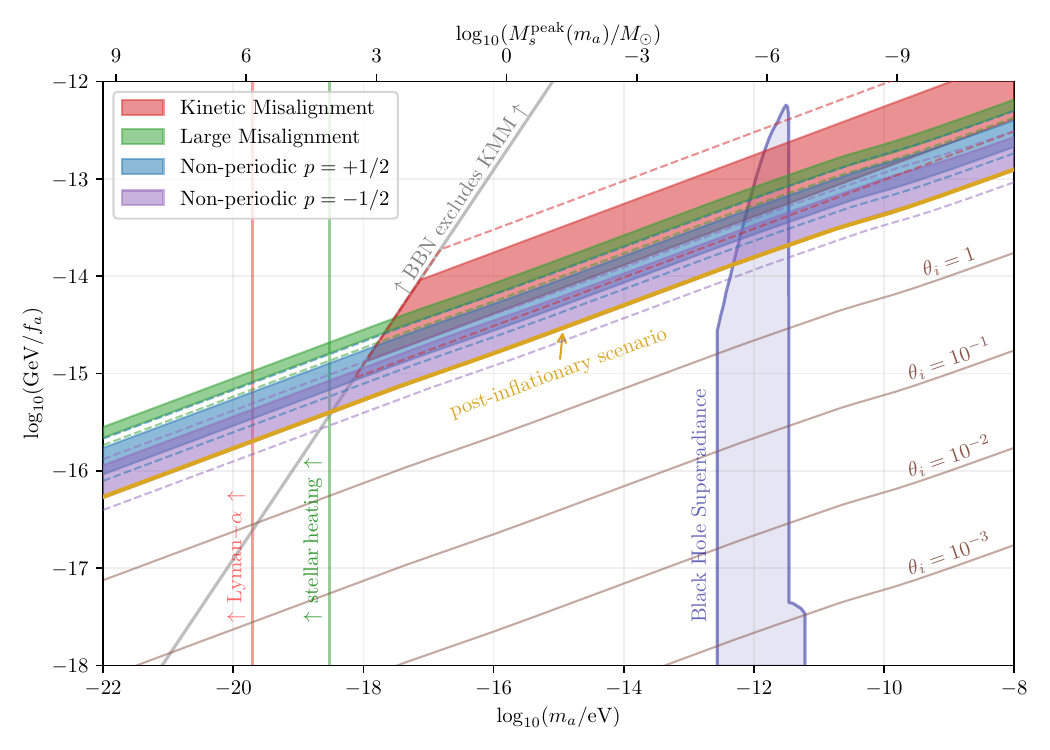}
    \caption{Regions in the ALP parameter space where the parametric resonance might create halos whose scale densities are larger than $\rho_s\gtrsim 10 M_{\odot}\,\textrm{pc}^{-3}$. Such halos  likely survive the tidal stripping, see Section \ref{sec:obs_prospects}, so they can be observable. Different colours show different production mechanisms, and we assumed that ALPs make all of dark matter. The dashed lines indicate how the regions will expand if we impose a smaller bound $\rho_s \gtrsim M_{\odot}\,\textrm{pc}^{-3}$. Above the gray line, the Kinetic Misalignment Mechanism is excluded by BBN due to the bound in \eqref{eq:bbn-bound}. The brown lines show the contours of the initial angle in the standard misalignment mechanism. For these values the initial angle is independent of the shape of the potential as long as it is quadratic around the minimum. Finally, we show the prediction for the case when the ALPs are generated after the inflation via the label "post-inflationary scenario". }
    \label{fig:cluster-region}
\end{figure}

The comparison shows that it depends on the point in the parameter space which model produces the most peaked halo spectrum. This would mean if we observe effects from dark matter halos in the future, and would also find a way to independently determine $f_a$ and the ALP mass, the combination of these observations could make it possible to determine which of the studied mechanisms and models has been realised in nature.

An independent determination of the decay constant might not be possible if the ALP has only gravitational interactions. However, even in this case, observation of dense halos with scale densities $\rho_s\gtrsim 10 M_{\odot}\,\textrm{pc}^{-3}$ can significantly constrain the ALP dark matter parameter space. The reason is that parametric resonance results in dense halos in a small region of the parameter space, and this region does not depend on the production mechanism in a drastic way. We demonstrate this in \figref{fig:cluster-region}. In this plot, the coloured bands show the regions where parametric resonance results in halos whose scale densities are larger than $\rho_s\gtrsim 10 M_{\odot}\,\textrm{pc}^{-3}$. The reason behind our choice for this critical value is that these halos likely survive tidal stripping as we have discussed in \secref{sec:obs_prospects}. Therefore, it is possible that these halos will be observed in the future. We stress that these regions are different from the ones that we show in \figsref{fig:observational_prospects_03}, \ref{fig:observation_m_f_plane_XM_03} and \ref{fig:observation_m_f_plane_XM_001}. Here, we just show the regions where dense halos are predicted, but we remain agnostic about the detection method.

Let us say an experiment has detected dense halos with masses around $M_s \sim M_{\odot}$. Such an observation would not alone reveal the production mechanism, but estimates the ALP mass and the decay constant to be $m_a\sim 10^{-16}\,\si{\electronvolt}$ and $f_a\sim 10^{14}$ -- $10^{15}\,\si{\giga\electronvolt}$, respectively. Then, more observations and detailed simulations might allow us to discover the production mechanism.

When determining the bands in \figref{fig:cluster-region} we use \eqref{eq:rhos-fit} for the non-periodic potentials and we have found that dense halos are predicted when $8\lesssim \theta_i\lesssim 14$ for $p=+1/2$ and $3\lesssim \theta_i\lesssim 5$ for $p=-1/2$. For the Kinetic Misalignment we use the methods of \citre{Eroncel:2022efc} and determined the dense halo region as $30\lesssim m_a/H_{\ast}\lesssim 170$. We should note that the method of \citre{Eroncel:2022efc} does not include a lattice simulation in the non-linear regime, so it is less reliable compared to the analysis we performed for the non-periodic potential. In particular, non-linear effects that can be captured only via a lattice simulation broaden the power spectrum which decreases the peak scale density. Therefore, we expect that a precise analysis of the Kinetic Misalignment will shrink the corresponding band. We also note that a sizable region of the low-mass parameter space in Kinetic Misalignment is excluded due to the BBN bound of \eqref{eq:bbn-bound}. Finally, for Large Misalignment, we have found that dense halos are predicted only in the case of significant tunings $\abs{\pi-\theta_i}\lesssim 10^{-11}$ which is consistent with the findings of \citre{Arvanitaki:2019rax}. For these values, the non-linear effects are expected to be important, but we did not take these into account. Nevertheless, we define the Large Misalignment band as $10^{-15}\lesssim\abs{\pi-\theta_i}\lesssim 10^{-11}$, and show it on the \figref{fig:cluster-region} for completeness.

In \figref{fig:cluster-region}, we also show via the dashed lines how the bands can be expanded if we had taken the weaker bound $\rho_s \gtrsim M_{\odot}\,\textrm{pc}^{-3}$. The brown lines are the contours of the initial angle in the standard misalignment mechanism, where the ALP potential can be approximated by a quadratic one. Finally, via the yellow line we show the prediction in the case where the ALPs are generated after inflation, the so-called post-inflationary scenario\footnote{Recently, it has been claimed in \citre{Amin:2022nlh} that there exists a lower bound on the ALP mass $m_a\gtrsim 10^{-18}\,\si{\electronvolt}$ if it has been produced after inflation.}. 

So far, we have not assumed any coupling of the ALP to the Standard Model besides gravity. If we assume that the ALP also couples to the electromagnetic field or the neutron, like it is the case in many ALP models, via terms like 
\begin{equation}
    \mathcal{L}_{a\gamma\gamma}=g_{a\gamma\gamma}\phi F_{\mu\nu}\tilde{F}^{\mu\nu},\quad \mathcal{L}_{an}=c_{an}\frac{\partial_{\mu}\phi}{2 f_a}\bar{n}\gamma^{\mu}\gamma_5 n,
\end{equation}
where $F_{\mu\nu}$ is the electromagnetic field strength tensor, $\tilde{F}^{\mu\nu}$ is its dual, and $n$ is the neutron, there is an abundance of experiments that could detect these particles. 
We will now assume that the ALP has KSVZ-model-like couplings, i.e. \cite{DiLuzio:2020wdo}
\begin{equation}\label{eq:KSVZ_coupling}
    \abs{g_{a\gamma\gamma,\,\mathrm{KSVZ}}}=\frac{\alpha_{\mathrm{EM}}}{2\pi}\frac{1.92}{f_a}\approx \frac{2.23\times 10^{-3}}{f_a},\quad c_{a n}\approx -0.02.
\end{equation}
Combining all the regions where dense halos are predicted from the different models, and embedding them in the $(m_a,f_a)$-parameter region with all the excluded regions and experimental prospects from past or ongoing experiments probing the electromagnetic couplings above, which have been listed in the appendix D of \citre{Eroncel:2022vjg} and have been collected from \citre{AxionLimits}, we arrive at \figref{fig:dense-halo-region}, which was shown in \secref{sec:introduction}. For this rough sketch, we have ignored the BBN bound on the Kinetic Misalignment Mechanism.

\section{Conclusions}

Amplification of the perturbations of an axion-like field by parametric resonance results in the formation of denser DM halos compared to CDM. In this work we have computed the spectrum of these halos and predicted their observable effects. 

We focused on ALPs with non-periodic potentials. These can be characterised by the mass $m_a$, the decay constant $f_a$ and the parameter $p$. We restricted ourselves to the case when $f_a\gg H_{I}$ such that inflation leaves the ALPs with a homogeneous field value $\phi_i$. Since $m_a$ and $f_a$ of the ALP are independent and $\phi_i$ is not bounded from above, a large part of the parameter space in the $(m_a,f_a)$-plane is allowed. We investigated the parametric resonance semi-analytically using Floquet analysis and numerically using lattice calculation. For large initial field values, $\phi_i \gtrsim f_a$, parametric resonance leads to the fragmentation of the initially homogeneous field, in which all of the energy density is transferred to high momentum modes. This is in contrast to ALPs with a cosine potential where fragmentation requires the tuning of the initial field value to the maximal value $\phi_i = \pi f_a$. 

We then used the Press-Schechter formalism, together with empirical relations obtained from $N$-body simulations, to investigate how the fluctuations, amplified due to resonance effects, decouple from the Hubble flow and collapse into halos. We use density power spectra obtained by linearly evolving the mode functions of the scalar field, as well as, when necessary, by simulating the full non-linear dynamics of the field on a lattice. To our knowledge this is the first work which takes into account such non-linear effects for the computation of the halo mass function following fragmentation. In particular, oscillons are produced in the course of fragmentation and can be very long-lived in certain cases. We used a special procedure to separate the oscillon contribution to the power spectrum and access the late-time power spectrum after the decay of oscillons.

These halos can have observable effects on stars and other astrophysical objects, including actual gravitational perturbations, astrometric and photometric lensing effects and diffraction of gravitational waves. All of these observable effects rely on gravitational interactions, which would be the only possibility to probe an ALP that is completely secluded from the SM. We presented these observational prospects in \secref{ssec:signatures}. Assuming that the ALP makes up all of dark matter, i.e. $\Omega_{a,0}h^2=\Omega_{\mathrm{DM},0}h^2$, given values of $m_a$ and $f_a$ fix the initial field value $\phi_i$ and, therefore, the characteristic halo mass and density. Our main results are summarised in \figsref{fig:observation_m_f_plane_XM_03} and \ref{fig:observation_m_f_plane_XM_001} where we show the parameter region of ALP DM in the $(m_a,f_a)$-plane in which gravitational effects of the halos may be observable. The results show that ALPs with very small masses of around $m_a\sim(10^{-21}-10^{-19})\eV$ would form halos of masses $(10^6-10^8)M_{\odot}$ which could explain the observed features in the GD-1 stellar stream. ALPs with masses smaller than $\sim3\times10^{-20}\eV$ would form halos that are heavy enough to lead to gravitational perturbations in other stellar streams or the Galactic disk. However, we note that there arise stronger and stronger bounds on ultralight ALPs, which would leave just a small, but finite, parameter space to explain e.g. the GD-1 features. Halos from ALPs with masses up to $m_a\sim10^{-16}\eV$ could lead to astrometric weak gravitational lensing that could be found in SKA and THEIA data. Halos from ALPs with masses up to roughly a few $10^{-16}\,\mathrm{eV}$ would lead to halos that could cause observable diffraction in the gravitational wave signals of aLIGO and ET, while halos from ALP masses smaller than $\sim10^{-19}\mathrm{\,eV}$ would leave similar effects in LISA data. Finally, larger ALP masses between $10^{-13}\eV<m_a<10^{-10}\eV$, a range that could be partially excluded by black hole superradiance, would lead to halos that could cause observable photometric microlensing effects of stars that are highly magnified by a galaxy cluster as lens. All of these signatures are observable when the initial field value is of the same order as the critical field value. For this to be the case for smaller masses, $f_a$ would need to be of order $10^{15}\,\mathrm{GeV}$, while for the larger masses observable by photometric lensing, $f_a$ would be of order $10^{13}\,\mathrm{GeV}$.  
For the observational prospects the halo mass function was always assumed to be peaked around the characteristic halo mass, and while this is a good assumption for estimates of the observational prospects \cite{VanTilburg:2018ykj_astrometry_program}, investigating these predictions for a more realistic halo mass function could be strengthening the arguments to search for these halos. 

An important role is played by non-linear effects. These transfer energy towards higher momentum modes and lead to the formation of oscillons which eventually decay. The corresponding length scales are too small and experience a large gradient pressure which prevents them from collapsing gravitationally. We observe that halo formation is most efficient when the field value is comparable to the critical field value determining the transition to the fragmentation regime. For higher field values fluctuations are distributed among higher momentum modes. These modes would collapse later due to the Jeans length and form less dense halos. Hence, the interesting parameter space of the initial field values is concentrated around $\phi_{\mathrm{crit}}$. This is also the reason why we observe qualitatively the same behaviour for the different values of $p$ we investigated, because the critical field value is of the order of $f_a$ for those values. 

In this work we focused on the case when oscillons decay before the onset of structure formation in the matter dominated era. The interesting scenario in which the oscillons decay later or even survive until today, which is possible for smaller ALP masses, is left for future work.

\acknowledgments

 We are thankful to Hyungjin Kim, Alessandro Lenoci and Marcello Musso for helpful discussions about gravitational collapse, to Andrea Mitridate for discussions about PTAs as well as to Jacob M. Leedom, Nicole Righi and Alexander Westphal for insights into axion monodromy. We thank Philip S\o rensen for providing the Mathematica code used to show the experimental constraints and projections in \figref{fig:dense-halo-region}. We are also thankful to Nikita Blinov for clarifications on photometric microlensing. This work is supported by the Deutsche Forschungsgemeinschaft under Germany’s Excellence Strategy – EXC 2121 ,,Quantum Universe“ – 390833306. This work has been produced benefiting from the 2236 Co-Funded Brain Circulation Scheme2 (CoCirculation2)
of The Scientific and Technological Research Council of Turkey T\"UB\.{I}TAK (Project No: 121C404). We also made heavy use of open source software \texttt{Matplotlib}~\cite{Hunter:2007}, \texttt{Numpy}~\cite{harris2020array}, and \texttt{Scipy}~\cite{2020SciPy-NMeth}.

\appendix

\section{UV-completions for ALPs with non-periodic effective potential}\label{App.UV}

\textbf{Axion Monodromy in String Theory:} A strong motivation for the existence of axions and ALPs in general, comes from string theory.
Axion-like particles arise in the 4D low-energy effective field theory after  compactification of the extra-dimensions, 
see e.g. \citre{Ringwald:2012hr_ALP_review}. The number of axions is related to the topology of the compact dimensions, while it is not unusual to expect numerous axion-like particles \cite{Marsh:2015xka}, possibly including the QCD-axion \cite{Witten:1984dg_string_axions1,Conlon:2006tq_string_axions2,Svrcek:2006yi_string_axions3}. The term \textit{string-axiverse} \cite{Arvanitaki:2009fg_String_Axiverse,Cicoli:2012sz_axiverse_pheno} refers to the prediction of many ALPs with masses evenly distributed in logarithmic scale.

ALPs from string theory also respect a shift symmetry, since they arise when a $p$-form gauge potential is integrated over a non-trivial $p$-cycle in the compact manifold. For example \cite{McAllister:2008hb}, in type IIB string theory, axions arise as $b_I=\int_{\Sigma_I^{(2)}}B$, which is the integral of the Neveu-Schwarz (NS) two-form potential $B_{MN}$ over two-cycles ${\Sigma_I^{(2)}}$. Similarly, by integrating the Ramond-Ramond (RR) two-form $C_{MN}$ one finds $c_I=\int_{\Sigma_I^{(2)}}C$. 

In this context, axions are initially massless, and, as usual, a periodic potential for these axions arises from instanton effects. However, if there is a \textit{Dp-brane} wrapping the cycles, it carries a potential energy that increases without bound when the axion value increases. In presence of additional field strengths, i.e.~\textit{fluxes}, the axion can couple non-trivially to them which leads to a non-periodic potential. The ALP experiences \textit{monodromy}, meaning that the potential energy depends on how many times it has circled the compact dimension \cite{McAllister:2016vzi,Silverstein:2008sg,McAllister:2008hb,Dong:2010in_Monodromy3}. For instance, a D5-brane wrapped on a two-cycle $\Sigma^{(2)}$ with size $l\sqrt{\alpha'}$ yields a potential for $b$ of the form
\begin{equation}
    V(b)\sim\frac{1}{g_s(2\pi)^5\alpha'^2}\sqrt{l^2+b^2}.
\end{equation}
For large values of the axion $b$, this potential becomes linear, which motivates the choice of $p=0.5$ for the effective potential in our analysis.

\bigskip

\textbf{Pure Gauge-Fields in the Large $\mathbf{N}$ limit:}
 The ALPs could be coupled to pure Yang-Mills fields of a SU$(N)$ gauge theory described by a Lagrangian like \cite{Nomura:2017zqj}
\begin{equation}
    \mathcal{L}=N\left(-\frac{1}{4\lambda}G_{\mu\nu}G^{\mu\nu}+\frac{1}{32\pi^2}\frac{\phi}{Nf_a}G_{\mu\nu}\tilde{G}^{\mu\nu}\right).
\end{equation} 
As shown by 't Hooft \cite{tHooft:1973alw_Large_N_3} and Witten \cite{Witten:1980sp_Large_N_1,Witten:1998uka_Large_N_2}, in the limit of large $N$, while keeping $\lambda=g^2N$ fixed, the vacuum energy $E$ is a smooth function of $\phi/N$, while it also respects the shift symmetry of $\phi$, i.e. $E(\theta)=E(\theta+2\pi n)$. If $E$ is not constant, this can only be realised if the potential for $\phi$ is  multi-branched, i.e. a multi-valued functional of the field value with a tower of meta-stable states above the vacuum state \cite{Dubovsky:2011tu_Large_N_4}, where the single branches are non-periodic. Such potential can be seen in \figref{fig:multibranched_pot} for $V(\phi)=-m_a^2f_a^2\left[\left(1+(\phi/f_a+2\pi n)^2\right)^{-(1/2)}-1\right]$ with $n\in\mathbb{N}$. The system is still invariant under the shift symmetry $\phi\to\phi+2\pi f_a$, while under adiabatic change the field stays within one non-periodic branch \cite{Dubovsky:2011tu_Large_N_4,Nomura:2017ehb_pure_natural_inflation}. This result can be confirmed from calculations in simplified models in lattice field theory, from realisations of four dimensional gauge-theory in M-theory \cite{Witten:1997sc}, or from the AdS/CFT correspondence \cite{Witten:1998uka_Large_N_2}. The approximate form of the potential can be found from the assumption that we have invariance under CP transformation $\phi\to-\phi$, which implies that $V(\phi)$ is a function of $\phi^2$. Secondly, when $\phi$ increases, the dynamics that generate the potential will become weaker and hence the potential is expected to flatten. Assuming that it is given by a simple power law, which would mean it looks like $\sim1/(\phi^2)^p$, but then regulating it such that it is not singular for $\phi\to 0$, and setting the minimum to $0$, we find \eqref{eq:monodromy_potential} with negative $p$ \cite{Nomura:2017ehb_pure_natural_inflation,Dubovsky:2011tu_Large_N_4,Witten:1998uka_Large_N_2}. It is model dependent under which conditions the field tunnels from branch to branch, but such a transition is usually suppressed if the field does not travel too far from the local minimum \cite{Dubovsky:2011tu_Large_N_4}.

\begin{figure}[t]
    \centering
    \includegraphics[width=0.6\textwidth]{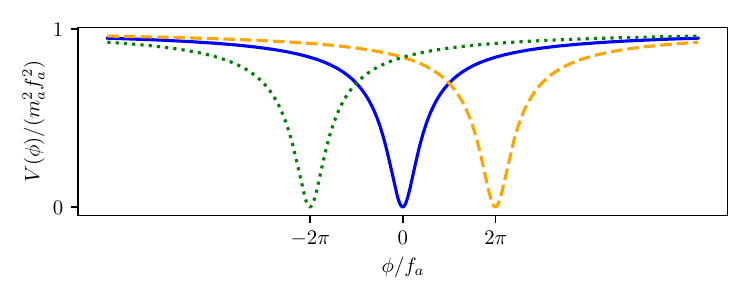}
    \caption{Three branches of the multi-branched potential described in appendix~\ref{App.UV}.}
    \label{fig:multibranched_pot}
\end{figure}

To understand the phenomenological consequences of such ALPs, in this work we assume that these two mechanisms that break the shift symmetry in the low energy limit effectively lead to the potential \eqref{eq:monodromy_potential} with a specific $p$ which we treat as free parameter. We consider general axion-like particles with no fixed relation between symmetry breaking scale $f_a$ and mass $m_a$, such that both can be treated as independent parameters. $m_a$, $f_a$ and $p$ are the parameters that will be determined by the specific UV-completion, as hinted above for $p$.
In general the symmetry breaking scale $f_a$ of ALPs in string theory is non-trivially related to the geometry of the compact manifold \cite{McAllister:2016vzi} and the theories predict a wide range of possible values. It is typically assumed to be near the GUT scale at $f_a\sim10^{16}\GeV$ \cite{Svrcek:2006yi_string_axions3}, while lower values of $f_a\sim10^{10}\GeV-10^{12}\GeV$ are possible \cite{Cicoli:2012sz_axiverse_pheno}. An upper bound is given by the Planck scale, i.e. $f_a<\Mpl$ \cite{Marsh:2015xka,Arkani-Hamed:2006emk_weak_gravity}.

\section{Floquet theorem}\label{App.1}
The Floquet theorem can be found explained in more detail and applied to inflationary cosmology in section~3 of \citre{Amin:2014eta_floquet}, which we follow throughout this appendix.

In order to apply the Floquet theorem we define $f_k(\ttilde)\equiv\frac{k^2}{m_a^2}+\frac{V''(\phi)}{m_a^2}$, which is a periodic function of $\ttilde$. With that \eqref{eom_floquet_fluctuations_no_expansion} becomes \begin{equation}\label{Hills_equation}
\ddot{u}_k(\ttilde)+f_k(\ttilde)\cdot u_k(\ttilde)=0
\end{equation}
which is the so-called \textit{Hills-equation} \cite{10.1007/BF02417081_Hills_equation}. 

One can write \eqref{Hills_equation} in matrix form by introducing $x_k(\ttilde)=\begin{pmatrix}u_k(\ttilde)&\dot{u}_k(\ttilde)\end{pmatrix}^T$, which brings the EOM to the form 
\begin{equation}\label{Hills_equation_matrix_form}
\dot{x_k}(\ttilde)=U_k(\ttilde)x_k(\ttilde)
\end{equation}
with
\begin{equation}
U_k(\ttilde)=\begin{pmatrix}0&1\\-f_k(\ttilde)&0\end{pmatrix}\,,
\end{equation}
where $U_k(\ttilde)$ is periodic, i.e.~$U_k(\ttilde)=U_k(\ttilde+\tilde{T})$ and $\tilde{T}$ is the (dimensionless) period of the homogeneous mode. This period can be found by using energy conservation of a periodic oscillation, $\dot{\theta}(\ttilde)^2+V(\phi(\ttilde))/m_a^2=V(\phi_{\mathrm{max}})/m_a^2$, where $\phi_{\mathrm{max}}$ is the field value corresponding to the maximum of the potential. By separation of variables this can be solved for the period. To do this, one integrates from minimum to maximum of the potential and arrives at
\begin{equation}\label{period_floquet}
\tilde{T}=2\int_{\theta_{\mathrm{min}}}^{\theta_{\mathrm{max}}}\frac{d\theta}{\sqrt{2(V(\phi_{\mathrm{max}})/m_a^2-V(\phi)/m_a^2})}\,,
\end{equation}
which can be evaluated numerically.

We pick $\ttilde_0$ to be the time of the onset of oscillation,
at which the homogeneous mode has the maximal amplitude, and introduce the so-called \textit{fundamental matrix} $\mathcal{O}(\ttilde)$ which is defined to be the solution to 
\begin{equation}\label{eom_fundamental_matrix}
\dot{\mathcal{O}}(\ttilde,\ttilde_0)=U_k(\ttilde)\mathcal{O}(\ttilde,\ttilde_0)
\end{equation}
with initial condition 
\begin{equation}
\mathcal{O}(\ttilde_0,\ttilde_0)=\begin{pmatrix}1&0\\0&1\end{pmatrix}\,.
\end{equation}
To avoid clutter we suppress the $k$-dependence of the fundamental matrix, as well as of $U(\ttilde)$ and $x(\ttilde)$ from here on. 

This matrix can be understood as consisting of (dimensionless) linearly independent solutions for the fluctuations $\begin{pmatrix}x^{(1)}(\ttilde_0)\\\dot{x}^{(1)}(\ttilde_0)\end{pmatrix}=\begin{pmatrix}1\\0\end{pmatrix}$ and $\begin{pmatrix}x^{(2)}(\ttilde_0)\\\dot{x}^{(2)}(\ttilde_0)\end{pmatrix}=\begin{pmatrix}0\\1\end{pmatrix}$ in its two columns. It follows that the fundamental matrix evolves a general initial state $x(\ttilde_0)$ in time:
\begin{equation}
x(\ttilde)=\mathcal{O}(\ttilde,\ttilde_0)x(\ttilde_0)\,,
\end{equation}
The Floquet theorem states that the fundamental matrix can be written as
\begin{equation}\label{Floquet_theorem}
\mathcal{O}(\ttilde,\ttilde_0)=P(\ttilde,\ttilde_0)\exp((\ttilde-\ttilde_0)\tilde{\Lambda}/m_a)\,,
\end{equation}
where $P(\ttilde,\ttilde_0)$ is a periodic function with period $\tilde{T}$, and $\tilde{\Lambda}$ is a time-independent matrix.
The eigenvalues $\mu_k^\pm$ of $\tilde{\Lambda}$ are called \em Floquet exponents \em and indicate if a solution grows exponentially, which it does if the Floquet exponents have non-zero real part, or oscillates, which happens when the Floquet exponents are purely imaginary. Note that $m_a$ appears in the equation since we are working in dimensionless time $\ttilde$.
One can find these eigenvalues by noting that the initial conditions of $\mathcal{O}(\ttilde,\ttilde_0)$ imply $P(\ttilde_0,\ttilde_0)=1$ and one has $P(\ttilde_0+\tilde{T},\ttilde_0)=1$ since $P(\ttilde,\ttilde_0)$ is $\tilde{T}$-periodic. With that one finds
\begin{equation}
\mathcal{O}(\ttilde_0+\tilde{T},\ttilde_0)=\exp(\tilde{T}/m_a\tilde{\Lambda})\,.
\end{equation}
The so-called \em Floquet multipliers \em $\pi^{\pm}_k$ are now defined as the eigenvalues of $\mathcal{O}(\ttilde,\ttilde_0)$ at $\ttilde_0+\tilde{T}$.
Since $\det(\mathcal{O})$ is always unity, these can be calculated via
\begin{equation}
\pi^{\pm}_k=\frac{\Tr(\mathcal{O}(\ttilde_0+\tilde{T}))}{2}\pm\sqrt{\frac{\Tr(\mathcal{O}(\ttilde_0+\tilde{T}))^2}{4}-1}\,.
\end{equation}
\Eqref{Floquet_theorem} implies that the Floquet exponents are given by
\begin{equation}
\mu^{\pm}_k=\frac{m_a}{\tilde{T}}\ln(\pi^{\pm}_k)\,.
\end{equation}

To investigate if there will be exponential growth of fluctuations for given  $k/m_a$ and initial value of the homogeneous mode $\phi_i$ in \eqref{eom_floquet_fluctuations_no_expansion}, one has to calculate the period $\tilde{T}$ with \eqref{period_floquet} for given $\phi_i$, then numerically evolve \eqref{eom_fundamental_matrix} from $\ttilde_0$ to $\ttilde_0+\tilde{T}$ and then compute Floquet multipliers and Floquet exponents as discussed.

The key improvement when using the Floquet theorem is that by calculating the Floquet exponents, the regions of parametric resonance can be found after simulating only one oscillation period, eliminating the need for a full numerical solution over many periods.

\section{Lattice simulation}
\label{App.Lattice}

In this appendix we discuss the numerical set-up for the lattice simulations, which were used to study the dynamics of the ALP field in the non-linear regime.

In such a lattice simulation the classical field equations of motion \eqref{full_eom_axion_unperturbed} are solved on a discretized spatial grid. We use rescaled conformal variables, $d\eta = dt m_a / a$ and $\phi_c = a\phi/f_a$, as well as rescaled spatial coordinates $x m_a$. In terms of these variables the Hubble expansion manifests itself only via the time-dependence of the potential, and the dependence of the equations of motion on $m_a$ and $f_a$ drops out. For the case of the non-periodic potential the corresponding equations of motion are
\beq
(\partial_\eta^2 - \Delta) \varphi_c + a^2 \varphi_c [1+(\varphi_c/a)^2]^{p-1} = 0 .
\eeq
The equations of motion are solved using a C++ program, similar to LATTICEEASY~\cite{Felder:2000hq}, which was also used in \citres{Berges:2019dgr, Chatrchyan:2020pzh}. 
We have used cubic lattices with up to $512^3$ points, with periodic boundary conditions and a fixed comoving volume. A standard leap-frog algorithm was used. We have checked that the results are insensitive to the lattice spacing as well as to the volume.

In classical-statistical simulations a Gaussian initial state can be sampled according to 
\beq
\phi(\mathbf{x}) = \phi_0 + \int \frac{d^3 k}{(2\pi)^3} \sqrt{ \frac{n_{k, 0}+1/2}{\omega_k}}  c_k e^{ikx}, \: \: \: \: 
\pi(\mathbf{x}) = \pi_0 + \int \frac{d^3 k}{(2\pi)^3}  \sqrt{ (n_{k, 0} +1/2) \omega_k } \tilde{c}_k e^{ikx},
\eeq
where $n_{k, 0}$ is the initial occupation number, $\omega_k^2=M^2+k^2$,while $c_k$ and $\tilde{c}_k$ are Gaussian random numbers multiplied by a random complex phase, uncorrelated between each other and satisfying $$\langle c_k \rangle  = \langle \widetilde{c}_k  \rangle  = 0, \: \: \: \:  \: \: \: \: \langle c_k c^{\star}_{k'} \rangle = \langle \tilde{c}_k \tilde{c}^{\star}_{k'} \rangle  = (2\pi)^3\delta(k-k').$$ Such a statistical ensemble generates the following Gaussian initial state
\beq
\nonumber
\begin{split}
\langle \hat \pi(k) \hat \pi^{\star}(k') \rangle_c = (n_{k, 0}+1/2)\omega_k (2\pi)^3\delta(k-k'),\: \: \:  \: \: \:  &\langle \hat \phi(k) \hat \phi^{\star}(k') \rangle_c = \frac{n_{k, 0}+1/2}{\omega_k} (2\pi)^3\delta(k-k'), \\
\langle \hat \phi \rangle = \phi_0, \: \: \:  \: \: \: \: \: \:\langle \hat \pi \rangle = \pi_0, \: \: \: \: \: \: \: \: \: &\langle \hat \phi(k) \hat \pi(k') \rangle_c =0.
\end{split}
\eeq
Here the subscript $c$ denotes the connected part of the two-point function. The occupation number at any time $t$ can be related to the two-point field correlation functions via
\beq
n_k(t) + \frac{1}{2} = \frac{1}{V} \frac{\langle |\pi(t, k)|^2\rangle  + \omega_k^2\langle |\phi(t, k)|^2\rangle}{2 \omega_k}.
\eeq

We first solve the linearized equations of motion,  \eqref{eom_zero_general} and \eqref{eom_with_curv_pert_fourier}, without taking into account the back-reaction of the fluctuations on the background field. The linear simulation is stopped while the total variance of the fluctuations is still small, $\langle \delta\theta^2 \rangle = 10^{-4}$, so that the linear approximation is still valid. After that, the values of the field and its derivative, as well as the occupation number, computed using the mode functions from \eqref{occup_mode}, are then used to initialize the Gaussian random field on the lattice.

\section{Oscillon decay}
\label{App.2}

This appendix is devoted to the decay of oscillons and we demonstrate why it is justified to use \eqref{density contrast power spectrum2} as an approximation to the power spectrum after the decay.

The decay of oscillons can also be modelled as the self-interactions being switched off at $t \approx \tau_{\os}$. In the absence of the binding force, the ALPs, forming the oscillons, simply free stream away, very much in analogy to what would happen after the actual decay (see~\cite{Imagawa:2021sxt} for more details). To force the decay on the lattice, we follow the approach from \citre{Vaquero:2018tib} and simply switch off the non-linearities of the ALP potential at some time. 

\begin{figure}[t!]
	\centering
	\includegraphics[width=0.82\textwidth]{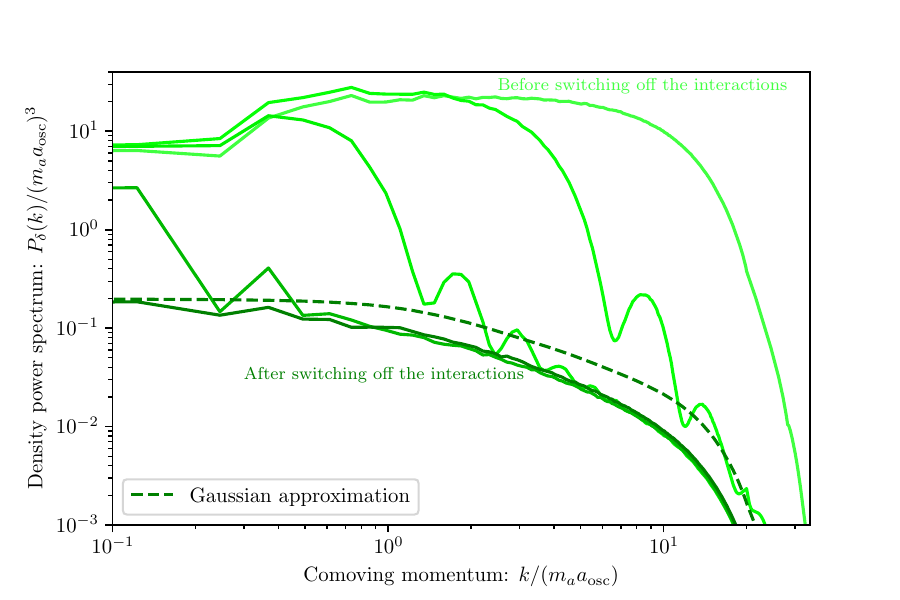}
	\caption{Snapshots of the energy density contrast power spectrum for $\theta_i=6$ and $p=-1/2$ for the scale factors ranging from $a = 20a_{\osc}$ (light green), when the interactions are instantly switched off, to $a = 28a_{\osc}$ (dark green). The dashed green line corresponds to the Gaussian approximation for the power spectrum according to \eqref{density contrast power spectrum2}.}
	\label{fig:energy_ps_WKB_app}
\end{figure}

Switching off the interactions does not affect the occupation numbers much. However, the energy density power spectrum changes significantly, as observed in the numerical simulations. This is shown in \figref{fig:energy_ps_WKB_app}, which contains snapshots of the power spectrum at different scale factors, ranging from the moment when the interactions are switched off, $a=20a_{\osc}$, to $a=28a_{\osc}$, from when the spectrum becomes approximately constant. As can be seen, the free streaming of ALPs ``washes out'' the energy density spectrum, from small to large length scales. The spectrum at small momenta decreases by a factor of around $10$. Which scales of the spectrum are ``washed out'' depends on the free streaming length of the ALPs from this decay.

The dotted green curve in the figure was obtained according to \eqref{density contrast power spectrum2} using the occupation numbers evaluated at the same time as the latest power spectrum from that figure in dark green. As can be seen, the two match quite well. The final power spectrum, obtained from the procedure described above, depends on the time when the oscillons are forced to decay. However, if this decay happens sufficiently late, a time-independent contribution is observed at small momenta of the final power spectrum, coming from the fluctuation (and not the oscillon-related) part of the field power spectrum. Given that the actual lifetimes of the oscillons are very long, this time-independent contribution is expected to describe the final power spectrum more accurately.

\bibliographystyle{JCAP}
\bibliography{references}

\end{document}